\documentclass[%
 reprint, 
superscriptaddress,
 amsmath,amssymb,
 aps,
]{revtex4-2}

\usepackage{graphicx}
\usepackage{dcolumn}
\usepackage{bm}
\newtheorem{lemma}{Lemma}
\newtheorem{thm}{Theorem}
\newtheorem{remark}{Claim}
\newtheorem{cor}{Corollary}
\usepackage{color}
\usepackage{braket}
\usepackage[normalem]{ulem}
\usepackage{tabularx}
\usepackage{hyperref}
\usepackage{comment}

\begin{document}

\preprint{APS/123-QED}

\title{Thermodynamic and Stoichiometric Laws Ruling the Fates of Growing Systems}

\author{Atsushi Kamimura}
\thanks{These two authors contributed equally.}
\affiliation{Institute of Industrial Science, The University of Tokyo, 4-6-1, Komaba, Meguro-ku, Tokyo 153-8505 Japan}
\author{Yuki Sughiyama}
\thanks{These two authors contributed equally.}
\affiliation{Institute of Industrial Science, The University of Tokyo, 4-6-1, Komaba, Meguro-ku, Tokyo 153-8505 Japan}
\affiliation{Graduate School of Information Sciences, Tohoku University, Sendai 980-8579, Japan}
\author{Tetsuya J. Kobayashi}
\email[E-mail:]{tetsuya@mail.crmind.net}
\altaffiliation[Also at ]{Universal Biology Institute, The University of Tokyo, 7-3-1, Hongo, Bunkyo-ku, Tokyo 113-8654, Japan.}
\affiliation{Institute of Industrial Science, The University of Tokyo, 4-6-1, Komaba, Meguro-ku, Tokyo 153-8505 Japan}

\date{\today}

\begin{abstract}
We delve into growing open chemical reaction systems (CRSs) characterized by autocatalytic reactions within a variable volume, which changes in response to these reactions. 
Understanding the thermodynamics of such systems is crucial for comprehending biological cells and constructing protocells, as it sheds light on the physical conditions necessary for their self-replication. 
Building on our recent work, where we developed a thermodynamic theory for growing CRSs featuring basic autocatalytic motifs with regular stoichiometric matrices, we now expand this theory to include scenarios where the stoichiometric matrix has a nontrivial left kernel space. 
This extension introduces conservation laws, which limit the variations in chemical species due to reactions, thereby confining the system's possible states to those compatible with its initial conditions. 
By considering both thermodynamic and stoichiometric constraints, we clarify the environmental and initial conditions that dictate the CRSs' fate—whether they grow, shrink, or reach equilibrium. We also find that the conserved quantities significantly influence the equilibrium state achieved by a growing CRS. These results are derived independently of specific thermodynamic potentials or reaction kinetics, therefore underscoring the fundamental impact of conservation laws on the growth of the system.
\end{abstract}

\maketitle


\section{Introduction}
\label{sec:introduction}

Chemical thermodynamics provides solid physical principles for explaining the energetics and for predicting the fate of chemical reaction processes \cite{callen1985thermodynamics,kondepudi1998modern}. 
Applications of these principles to autocatalytic reactions are essential to elucidate physical constraints on the capability of self-replication \cite{neumann1966automata,KinematicSelfReplication}.
In the self-replication process, both the chemical components and the encapsulating volume of the system have to grow in a coherent manner \cite{de2017mathematical,de2020elementary,muller2021elementary,muller2022elementary}.
This consideration introduces a unique theoretical challenge to establish the thermodynamic consistency between autocatalytic reactions and volume expansion because the conventional chemical thermodynamics is based solely on the density (concentration) of chemicals, presuming a constant volume \cite{horn1972general,qian2005nonequilibrium,beard2008chemical,craciun2009toric,perez2012chemical,polettini2014irreversible,ge2016nonequilibrium,rao2016nonequilibrium,craciun2019,qian2021stochastic}.

We have recently established a thermodynamic theory for growing systems, in which the number of chemicals and the volume are treated based on the rigorous thermodynamic basis \cite{sughiyama2022chemical}. 
Accordingly, the theory generally formulates physical conditions to realize the growth of the system, identifies several thermodynamic constraints for the possible states of the growing system, and derives the form of entropy production and heat dissipation accompanying growth. 
However, this theory can only address systems with regular (full-rank) stoichiometric matrices, thus limiting its applicability to  a set of minimum autocatalytic motifs \cite{blokhuis2020universal}. 
Given that chemical reaction systems (CRSs) are subject to stoichiometric constraints in general, and various biological functions are robustly realized by specific stoichiometric properties \cite{shinar2010structural,hirono2023complete}, it becomes imperative to delve into the influence of stoichiometry to growing systems  for comprehensive understanding of the thermodynamics of self-replication.

In this work, we clarify how stoichiometric conservation laws shape the fate of growing systems.
The stoichiometry gives rise to linear combinations of the numbers of chemicals being conserved during the dynamics of chemical reactions \cite{horn1972general,schuster1991determining, schuster1995information,de2005typical,craciun2009toric,de2009role,de2012neumann,perez2012chemical,polettini2014irreversible,rao2016nonequilibrium,ge2016nonequilibrium,rao2018conservation_jchemphys,rao2018conservation_newjphys,pekavr2021non}.
These conservation laws stringently restrict possible changes in the number of chemicals within the stoichiometric compatibility class, defined by the stoichiometric matrix's left kernel and the initial condition.
In biological contexts, particularly in metabolic networks, these laws are crucial in linking the dynamic variations of different chemicals (metabolites) and also in providing insights into cells' production capabilities \cite{hofmeyr1986metabolic,famili2003convex,imielinski2006systematic,shinar2009sensitivity,martelli2009identifying,de2014identifying,haraldsdottir2016identification,kamei2023raman}. 

This study elucidates the complex interplay between thermodynamics and stoichiometric conservation laws for the growing system and its consequence to the fate of the growing systems. 
By disentangling the geometric relationship among chemical numbers, densities, and potentials, we establish the conditions for the system to grow, shrink, or equilibrate, while simultaneously satisfying the second law of thermodynamics and the conservation laws. 
Furthermore, we show that the existence of conservation laws qualitatively alter the fate of the system and the geometric properties of the equilibrium state. 

These results are derived solely based on thermodynamic and stoichiometric requirements and thus remain independent of specific thermodynamic potentials or reaction kinetics. 
This renders our theory universally applicable, enhancing our understanding of the origins of life and the construction of protocells\cite{hypercycle,kauffman1986autocatalytic,jain1998autocatalytic, segre2000compositional, andrieux2008nonequilibrium,kita2008replication,protocellsmit, noireaux2011development, kurihara2011self,ichihashi2013darwinian,mavelli2013theoretical,ruiz2014prebiotic, himeoka2014entropy,kurihara2015recursive,   shirt2015emergent, pandey2016analytic,serra2017modelling,joyce2018protocells,lancet2018systems, liu2018mathematical,steel2019autocatalytic,blokhuis2020universal,pandey2020exponential,unterberger2022stoechiometric,gagrani2024polyhedral}, and enabling the search for the universal laws of biological cells \cite{scott2010interdependence,scott2011bacterial,scott2014emergence, maitra2015bacterial,reuveni2017ribosomes, barenholz2017design,jun2018fundamental,thomas2018sources, furusawa2018formation,kostinski2020ribosome, dourado2020analytical, lin2020origin, kostinski2021growth,roy2021unifying}.

This paper is organized as follows. We devote Sec. \ref{sec:CRN} to outline our main results accompanying with an illustrative example of a growing CRS. Here, we clarify the environmental and initial conditions that determine the fate of the system: growth, shrinking or equilibration. In Sec. \ref{sec:geometry}, 
we explain the geometric structure of the growing system to characterize the equilibrium state as a maximum of the total entropy function, which the system may achieve. 
In Sec. \ref{sec:entropy}, the form of the total entropy function is further investigated to determine if the system grows or shrinks by following the second law of thermodynamics. 
In Sec. \ref{sec:proof}, we provide a mathematical proof of our main results outlined in Sec. \ref{sec:CRN}.
Finally, we summarize our work with further discussions in Sec. \ref{sec:summary}.
For better readability, we list the symbols and notations in Appendix \ref{appendix_notations}.

\section{Outline of the main results}
\label{sec:CRN}

We outline our main results before presenting their derivations. In Sec. \ref{sec:setup}, we give a thermodynamic setup of a growing chemical reaction system (CRS). By introducing the chemical potential space, we analyze candidates of chemical equilibrium states in Sec. \ref{sec:outline_eq}. By computing accessible regions of the system, 
we characterize the equilibrium state as an intersection of the accessible region and the candidates in Sec. \ref{sec:eq_as_intersection_outline}.  In Sec. \ref{sec:claims}, we present our main claims with the above preparation. We comment our claims for a special class of the growing CRS in Sec. \ref{sec:productivity}. These results are summarized in Table \ref{tab:table1}. Then, we demonstrate the claims by numerical simulations using a simple example of the growing CRS in Sec. \ref{sec:numerical}. Finally, in Sec. \ref{sec:outline_derivations}, we outline the derivations of our results, which will be presented in the subsequent sections. 

\subsection{Thermodynamic setup}
\label{sec:setup}

\begin{figure}
    \centering
    \includegraphics[width=0.5\textwidth]{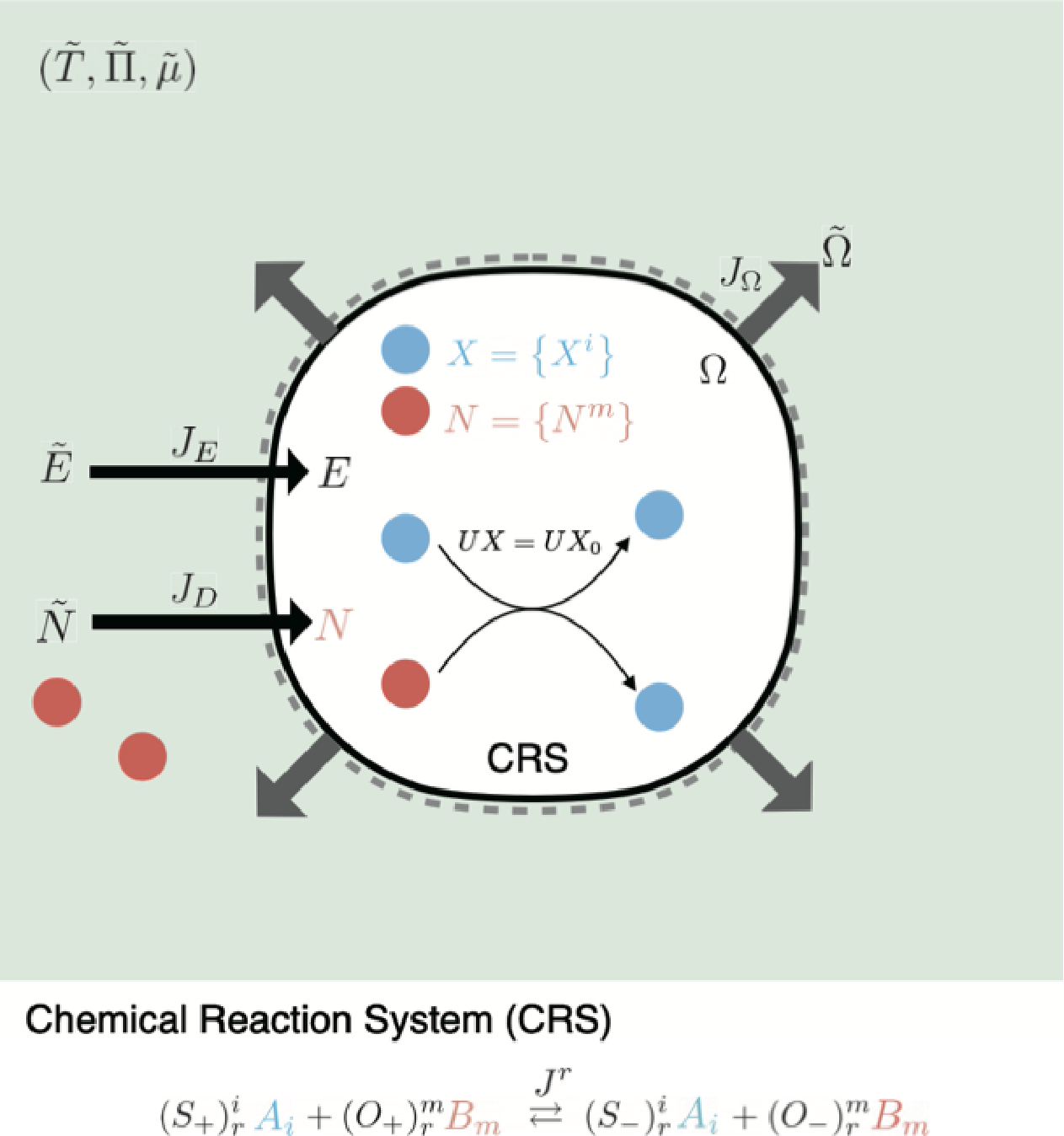}
    \caption{Diagrammatic representation of open CRSs. The chemical reactions occur with the reaction flux $J(t) = \{ J^r(t) \}$, the $r$th reaction of which is represented as the chemical equation at the bottom. Here, $A = \{ A_i \}$ is the label of the confined chemicals, and $B = \{ B_m\}$ is the label of the open chemicals, which can move across the membrane with the diffusion flux $J_D(t) = \{J_D^m(t)\}$. In addition, $J_E(t)$ denotes the energy exchange rate with the environment, and $J_{\Omega}(t)$ is the volume expansion rate. Let $X = \{ X^i \}$ and $N = \{ N^m \}$ denote the numbers of the confined and the open chemicals in the system, respectively. Also, $\left( S_+ \right)^i_r$ and $\left( O_+ \right)^m_r$ denote stoichiometric coefficients of the reactants in the $r$th reaction, whereas $\left( S_- \right)^i_r$ and $\left( O_- \right)^m_r$ are the ones of the products. The stoichiometric matrices are given as $S^i_r = \left( S_- \right)^i_r - \left( S_+ \right)^i_r$ and $O^m_r = \left( O_- \right)^m_r - \left( O_+ \right)^m_r$. For theoretical simplicity, we ignore the tension of the membrane and assume that the membrane never bursts. The stoichiometric matrix $S = \{ S^i_r \}$ has only a left kernel space. By representing the basis matrix of $\mathrm{Ker}[S^T]$ by $U$, $L = U X$ is conserved under the chemical reaction dynamics. }
    \label{fig:1}
\end{figure}

We consider the following thermodynamic setup in the present paper (Fig. \ref{fig:1}). A growing open chemical reaction system
is surrounded by an environment. We assume that the system is always in a well-mixed state (a local equilibrium state), and therefore, we can completely describe it by extensive variables $(E, \Omega, N, X)$. Here, $E$ and $\Omega$ represent the internal energy and the volume of the system. 
$N = \{N^m\} \in \mathbb{R}_{> 0}^{\mathcal{N}_{N}}$ denotes the number of chemicals that can move across the membrane between the system and the environment, which we call open chemicals hereafter. $X = \{X^i\} \in \mathfrak{X} = \mathbb{R}_{> 0}^{\mathcal{N}_{X}}$ is the number of chemicals confined within the system; the indices $m$ and $i$ respectively run from $m=1$ to $\mathcal{N}_{N}$ and from $i=1$ to $\mathcal{N}_{X}$, where $\mathcal{N}_N$ and $\mathcal{N}_X$ are the number of species of the open and the confined chemicals. 
The environment is characterized by intensive variables $(\tilde{T}, \tilde{\Pi}, \tilde{\mu})$, where $\tilde{T}$ and $\tilde{\Pi}$ are the temperature and the pressure; $\tilde{\mu} = \{\tilde{\mu}_m\} \in \mathbb{R}^{\mathcal{N}_{N}}$ is the chemical potential corresponding to the open chemicals. Also, $(\tilde{E}, \tilde{\Omega}, \tilde{N})$ denote the corresponding extensive variables.

In thermodynamics, the entropy function is defined on $(E, \Omega, N, X)$ as a concave and smooth function $\Sigma[E, \Omega, N, X]$. We also write the entropy function for the environment as $\tilde{\Sigma}_{\tilde{T},\tilde{\Pi},\tilde{\mu}}[\tilde{E},\tilde{\Omega},\tilde{N}]$, and the total entropy can be expressed as
\begin{equation}
    \Sigma^{\mathrm{tot}} = \Sigma[E, \Omega, N, X] + \tilde{\Sigma}_{\tilde{T},\tilde{\Pi},\tilde{\mu}}[\tilde{E},\tilde{\Omega},\tilde{N}],
    \label{enttot}
\end{equation}
where we use the additivity of the entropy. Further, the entropy function for the system has the homogeneity. Therefore, without loss of generality, we can write it as
\begin{equation}
    \Sigma[E,\Omega,N,X] = \Omega \sigma[\epsilon,n,x],\label{ent_homogeneity}
\end{equation}
where $\sigma[\epsilon,n,x]$ is the entropy density and $(\epsilon,n,x):=(E/\Omega,N/\Omega,X/\Omega)$. 
In this work, we consider only a situation without phase transition, and therefore, we assume that $\sigma[\epsilon,n,x]$ is strictly concave. 

The dynamics of the number of confined chemicals $X(t)$ is described by
\begin{align}
    \frac{dX^i}{dt} = S^i_r J^r(t),
    \label{system_dynamics_chem}
\end{align}
where $J(t) = \{ J^r(t) \}$ represents the chemical reaction flux; $S = \{ S^i_r\}$ denotes 
the stoichiometric matrix for the confined chemicals (see Fig. \ref{fig:1}). 
The index $r$ runs from $r=1$ to $\mathcal{N}_{R}$, where
$\mathcal{N}_{R}$ is the number of reactions. 
Also, Einstein's summation convention has been employed in Eq. (\ref{system_dynamics_chem}) for notational simplicity.
By integrating Eq. (\ref{system_dynamics_chem}), we obtain 
\begin{align}
    X^i(t) = X^i_0 + S^i_r \Xi^r(t),
    \label{accessible_space_X}
\end{align}
where $X_0 = \{ X^i_0 \}$ denotes the initial condition and $\Xi(t) = \{ \Xi^r(t) \}$ is the integration of $J(t)$, i.e., the extent of reaction. 

In this work, we assume that the stoichiometric matrix $S$ does not have the right kernel space, 
\begin{align}
\dim \mathrm{Ker} [S] = 0.
\label{KerS0}
\end{align} 
By contrast, the left kernel space exists and generates a set of conserved quantities:
\begin{align}
L = \{ L^l = U^l_i X^i_0\}, 
\label{conserved_quantity}
\end{align}
i.e., the vector $L \in \mathcal{L} = \mathbb{R}^{\dim \mathrm{Ker}[S^T]}$. 
Here, $U$ is a basis matrix: $\{ U_i^l \}:= (\textbf{U}^1, \textbf{U}^2, ...)^T$ whose row vectors $\textbf{U}^l$ form the basis of $\mathrm{Ker} [S^T]$, i.e., $U S = 0$, and $l$ runs from $1$ to $\dim \mathrm{Ker}[S^T]$. 
Since we are mainly interested in the growth of the system, the time evolution of $X(t)$ should perpetually increase the number of all confined chemicals while satisfying the conservation laws (Eq. (\ref{conserved_quantity})). In order to realize this situation, the stoichiometric matrix $S$ should satisfy 
\begin{equation}
\mathrm{Im}[S] \cap \mathbb{R}_{> 0}^{\mathcal{N}_{X}} \neq \emptyset,
\label{productiveS}
\end{equation}
which is known as the condition for the productive $S$ \cite{blokhuis2020universal}.
The productive $S$ excludes the mass conservation-type laws, the existence of which trivially precludes the number of confined chemicals to increase continuously.

The number of open chemicals changes through the reactions and the diffusions. By defining the stoichiometric matrix for open chemicals as $O = \{ O^m_r \}$ and the diffusion fluxes as $J_D(t) = \{J^m_D(t)\}$, we have 
\begin{align}
    \frac{dN^m}{dt} = O^m_r J^r(t) + J_D^m(t).
    \label{system_dynamics_open_chem}
\end{align}

\textit{Example 1:} To give an illustrative example, we consider the following chemical reactions:
\begin{align}
    R_1: & A_1 + A_3 + B_1 \rightleftarrows 2A_2, \notag \\
    R_2: & A_2 \rightleftarrows A_1 + A_3 + B_2. \label{Reactions_example}
\end{align}
Here, three confined chemicals $A = (A_1, A_2, A_3)$ and two open chemicals $B = (B_1, B_2)$ are involved in two reactions $R_1$ and $R_2$ \footnote{One may feel that the order of the reactions in Eq. (\ref{Reactions_example}) is high. The reason why we consider this example is just because it is straightforward to visualize our geometric representation in Sec. \ref{sec:geometry}. Since $\mathcal{N}_{X} = \dim \mathrm{Ker}[S^T] + \dim \mathrm{Im}[S]$ and we can visualize up to three dimensional space ($\mathcal{N}_{X}=3$), relevant cases are when $\dim \mathrm{Im}[S]$ is one, two or three. The present example corresponds to the case in which $\dim \mathrm{Im}[S]$ is two. In Sec. I and II of the Supplementary material, we also consider other examples with which $\dim \mathrm{Im}[S]$ is one and three, respectively.}. 
The stoichiometric matrices $S$ and $O$ for the confined and open chemicals, respectively, are 
\begin{align}
    S = 
    \bordermatrix{     & R_1 & R_2 \cr
               A_1 & -1 & 1 \cr
               A_2 & 2 & -1 \cr
               A_3 & -1 & 1 \cr
            },\mbox{  } 
    O = 
    \bordermatrix{     & R_1 & R_2 \cr
               B_1 & -1 & 0\cr
               B_2 & 0 & 1 \cr
            }.
            \label{S_O_example}
\end{align}

Since $\mathcal{N}_R = 2$ in this case, we can represent the number of confined chemicals by introducing $\Xi = ( \Xi^1, \Xi^2 )^T$ as  
\begin{equation}
\left(
\begin{array}{c}
X^1 \\
X^2 \\
X^3
\end{array}
\right) - 
\left(
\begin{array}{c}
X_0^1 \\
X_0^2 \\
X_0^3
\end{array}
\right) = 
S \Xi = 
\left(
\begin{array}{c}
-1 \\
2 \\
-1
\end{array}
\right)
\Xi^1 + 
\left(
\begin{array}{c}
1 \\
-1 \\
1
\end{array}
\right)
\Xi^2
.\label{V1V2}
\end{equation}

For this example, $\dim \mathrm{Ker}[S^T]$ is one, and $U = (-1, 0, 1)$ satisfies $US = 0$. 
Thus, the reactions have the conserved quantity $L = UX = -X^1+X^3$.  
We can intuitively see the conserved quantity from the chemical equations in Eq. (\ref{Reactions_example}) as follows. The numbers of $A_1$ and $A_3$, i.e., $X^1$ and $X^3$, change in the same manner: 
each chemical is consumed by one molecule in the forward reaction of $R_1$ and is produced by one in that of $R_2$.
Thus, the difference between $X^1$ and $X^3$ is conserved when each of the reactions occurs. 

\rightline{$\blacksquare$}

In this paper, we assume that the time scale of the chemical reactions is much slower than that of the others, i.e., $J \ll J_E, J_{\Omega}, J_D$. 
This means that we consider the isothermal and isobaric process with a balance of chemical potentials for open chemicals.
This assumption allows us to separately analyze the slow dynamics of chemical reactions from the fast dynamics of the energy, the volume, and the chemical diffusion. As shown in Appendix \ref{appendix_derivations}, our dynamics is effectively governed by Eq. (\ref{system_dynamics_chem}), and the other variables $(E, \Omega, N)$ are rapidly relaxed to the values at the equilibrium state of the fast dynamics. As a result, $(E, \Omega, N) = (E(X), \Omega(X), N(X))$ is obtained as a function of $X$. 

With the above assumptions, we obtain the following expression for the total entropy (see Eq. (\ref{appendix_Enttot_partialGP}) in Appendix \ref{appendix_derivations}), 
\begin{align}
    \Sigma^{\mathrm{tot}}(\Xi) = &- \frac{1}{\tilde{T}} \left\{ \Omega(X) \varphi \left( \frac{X}{\Omega(X)} \right) + \Omega(X) \tilde{\Pi} + \tilde{\mu}_m O^m_r \Xi^r \right\} \notag \\
    &+ \mathrm{const}, 
    \label{Enttot_partialGP}
\end{align}
where $X = X_0 + S\Xi$. 
Here, $\varphi(\cdot)$ is the partial grand potential density, which is given by a Legendre transformation of $\sigma\left[ \epsilon, n, x\right]$: 
\begin{align}
\varphi(x) := \varphi \left[ \tilde{T}, \tilde{\mu}; x \right] = \min_{\epsilon, n} \left\{ \epsilon -  \tilde{T} \sigma\left[ \epsilon, n, x\right] - \tilde{\mu}_m n^m \right\}.  \label{varphi_x}
\end{align}
Note that $\varphi(x)$ is strictly convex because $\sigma$ is strictly concave. 
The volume $\Omega(X)$ is variationally determined under isobaric conditions as 
\begin{equation}
    \Omega(X) = \arg \min_{\Omega} \left\{ \Omega \varphi\left(\frac{X}{\Omega} \right) + \Omega \tilde{\Pi} \right\}.\label{volume}
\end{equation}

\textit{Example 2:}
If we assume that the system and the environment are composed of ideal gas or solution, 
the partial grand potential density $\varphi(x)$ is expressed as 
\begin{align}
    \varphi(x) = x^i \nu_i^o\left(\tilde{T}\right) &+ R \tilde{T} \sum_i \left\{ x^i \log x^i - x^i \right\} \notag \\
    & - R\tilde{T} \sum_m e^{\left\{\tilde{\mu}_m - \mu_m^o \left( \tilde{T} \right) \right\}/R\tilde{T}}, \label{partialGP_ideal}
\end{align}
where $R$ is the gas constant, and $\nu^o(\tilde{T}) = \{ \nu_i^o(\tilde{T}) \}$ and $\mu^o(\tilde{T}) = \{ \mu^o_m(\tilde{T}) \}$ are the standard chemical potentials of the confined and the open chemicals, respectively\cite{sughiyama2022hessian}. 

Since we assume that the environment also consists of the ideal gas, the chemical potential $\tilde{\mu}$ in Eq. (\ref{partialGP_ideal}) is represented as
\begin{align}
\tilde{\mu}_m = \mu_m^o(\tilde{T}) + R \tilde{T} \log \tilde{n}^m,\label{chemical_potential_idealgas}
\end{align}
where $\tilde{n} = \{ \tilde{n}^m\}$ is the density of the open chemicals in the environment.  
Furthermore, Eq. (\ref{volume}) gives the following expression for the volume $\Omega$,  
\begin{align}
    \Omega(X) = \frac{R\tilde{T} \sum_i X^i}{\tilde{\Pi}- R\tilde{T} \sum_m \tilde{n}^m},\label{EqState}
\end{align}
which corresponds to the equation of state. 

\rightline{$\blacksquare$}

From Eq. (\ref{volume}), the density $x \in \mathcal{X} = \mathbb{R}_{>0}^{\mathcal{N}_{X}}$ is obtained by a nonlinear function $\rho_{\mathcal{X}}(X)$ of $X$ as
\begin{align}
\rho_{\mathcal{X}}(X) := x(X) = X/\Omega(X) \in \mathcal{X}.\label{map_density_of_x}
\end{align}

Since $\Omega(X)$ is homogeneous, we have $\rho_{\mathcal{X}}(\alpha X) = \rho_{\mathcal{X}}(X)$ for $\alpha > 0$. This fact implies that $\rho_{\mathcal{X}}$ induces one-to-one correspondence between rays in the number space $\mathfrak{X}$ and points in the density space $\mathcal{X}$. 
Here, we write the corresponding ray to a density $x$, i.e., the fiber of $x$ for $\rho_{\mathcal{X}}$, as
\begin{align}
\mathfrak{r}^{\mathcal{X}}(x) :&= \Set{ X | \rho_{\mathcal{X}}(X) = x, X > 0 } \subset \mathfrak{X},\notag \\
&= \Set{ X | X = \alpha x, \alpha > 0 }. \label{ray_x}
 \end{align}

\subsection{Candidates of chemical equilibrium states}
\label{sec:outline_eq}
We define the full grand potential density $\varphi^*(y) = \varphi^*[\tilde{T}, \tilde{\mu}; y]$ by the Legendre transformation of $\varphi(x)$ in Eq. (\ref{varphi_x}): 
\begin{equation}
    \varphi^* (y) := \max_{x} \left\{ y_i x^i - \varphi(x) \right\}.\label{varphi_y}
\end{equation}
We mention that $\varphi^*(y)$ is strictly convex. 
Because of the one-to-one correspondence of the Legendre transformation by $\varphi(x)$ and $\varphi^*(y)$, a state of the system is equivalently 
specified either by the density $x \in \mathcal{X} = \mathbb{R}_{>0}^{\mathcal{N}_{X}}$ of the confined chemicals or by its Legendre dual variable $y = \partial \varphi(x) \in \mathcal{Y} = \mathbb{R}^{\mathcal{N}_{X}}$ \footnote{The inverse transformation is given by $x = \partial \varphi^*(y)$. The existence of one-to-one correspondence between $\mathbb{R}_{>0}^{\mathcal{N}_{X}}$ and $\mathbb{R}^{\mathcal{N}_{X}}$ is physically reasonable (See Sec. \ref{sec:proof1}).}. 
The thermodynamic interpretation of $y$ is the corresponding chemical potential to $x$. In addition, $\varphi^*(y)$ can be interpreted as the pressure of the system at the state $y$. 
Recalling the one-to-one correspondence by $\rho_{\mathcal{X}}$ between rays in $\mathfrak{X}$ and points in $\mathcal{X}$, we notice that there exists the one-to-one correspondence between rays in $\mathfrak{X}$ and points in $\mathcal{Y}$. Here, we write the corresponding ray to a chemical potential $y$ as
\begin{align}
\mathfrak{r}^{\mathcal{Y}}(y) := \Set{ X | \partial \varphi \circ \rho_{\mathcal{X}}(X) = y, X > 0 } \subset \mathfrak{X}. 
\label{ray_y}
 \end{align}

The chemical equilibrium states are given by the balance of chemical potentials between reactants and products \cite{sughiyama2022hessian}:
\begin{equation}
y_i S^i_r + \tilde{\mu}_m O^m_r = 0.\label{simultaneous_equation_outline_new}
\end{equation}
Thus, we can obtain the candidates of chemical equilibrium states by the solutions to Eq. (\ref{simultaneous_equation_outline_new}):
\begin{equation}
    \mathcal{M}^{\mathcal{Y}}_{\mathrm{EQ}}(\tilde{\mu}) := \Set{ y| y_i S^i_r + \tilde{\mu}_m O^m_r = 0 }, \label{eq_manifold_y_outline}
\end{equation}
which we term the equilibrium manifold.
Here, we note that $\mathcal{M}^{\mathcal{Y}}_{\mathrm{EQ}}(\tilde{\mu}) \neq \emptyset$, because we have assumed $\dim \mathrm{Ker} [S] = 0$ in Eq. (\ref{KerS0}), i.e., $\dim \mathrm{Im}[S^{T}]=\mathcal{N}_{R}$.

\subsection{An equilibrium state as the intersection}
\label{sec:eq_as_intersection_outline}

The reaction dynamics in Eq. (\ref{system_dynamics_chem}) with its conserved quantities $L$ restricts the accessible region in $\mathfrak{X}$ as
\begin{equation}
    \mathcal{M}^{\mathfrak{X}}_{\mathrm{STO}}(L) := \Set{ X | U^l_i X^i = L^l, X > 0 }.\label{stoichiometric_manifold}
\end{equation}
This subspace is known as the stoichiometric compatibility class\cite{feinberg2019foundations}. 
By using the mappings $\rho_{\mathcal{X}}$ in Eq. (\ref{map_density_of_x}) and $\partial \varphi$, the accessible regions in the density space $\mathcal{X}$ and the chemical potential space $\mathcal{Y}$ are respectively obtained as 
\begin{align}
\mathcal{M}^{\mathcal{X}}_{\mathrm{STO}}(L) &= \rho_{\mathcal{X}}(\mathcal{M}^{\mathfrak{X}}_{\mathrm{STO}}(L)), \label{stoichiometric_x_claim}\\
\mathcal{M}^{\mathcal{Y}}_{\mathrm{STO}}(L) &= \partial \varphi(\mathcal{M}^{\mathcal{X}}_{\mathrm{STO}}(L)).\label{stoichiometric_y_claim}
\end{align}

Since the system evolves only on $\mathcal{M}^{\mathcal{Y}}_{\mathrm{STO}}(L) \in \mathcal{Y}$, 
the equilibrium state to which the system converges is represented by the intersection:
\begin{align}
\mathcal{M}^{\mathcal{Y}}_{\mathrm{STO}}(L) \cap \mathcal{M}^{\mathcal{Y}}_{\mathrm{EQ}}(\tilde{\mu}),\label{intersection_outline}
\end{align}
if it is not empty. 
If the intersection is empty, the system never converges to the equilibrium state and the growth of the volume may happen. 

\subsection{Main claims of the present paper}
\label{sec:claims}

In this subsection, we state our two main claims of this work before presenting their derivations in the following sections.  

Our first claim provides the conditions to determine the fate of the system, i.e., growth, shrinking or equilibration. 
To formulate our claim, we define $y^{\mathrm{min}}$ as 
\begin{align}
y^{\mathrm{min}} := \arg \min_y \Set{ \varphi^*(y) | y \in \mathcal{M}^{\mathcal{Y}}_{\mathrm{EQ}}(\tilde{\mu}) }.\label{y_min_outline}
\end{align}
Note that the productivity of $S$ in Eq. (\ref{productiveS}) guarantees the existence and the uniqueness of $y^{\mathrm{min}}$ (see the latter half of Appendix \ref{appendix:birch}).
This $y^{\mathrm{min}}$ gives the chemical equilibrium state
whose pressure $\varphi^{*}( y^{\mathrm{min}})$ is the minimum in the candidates, $\mathcal{M}^{\mathcal{Y}}_{\mathrm{EQ}}(\tilde{\mu})$.

\begin{remark} 
The fate of the system is classified by $y^{\mathrm{min}}$ and the conserved quantities $L$ in Eq. (\ref{conserved_quantity}) as follows (see also Table \ref{tab:table1}):

\begin{enumerate}
    \item If $\varphi^{*}\left( y^{\mathrm{min}} \right)-\tilde{\Pi} > 0$, the system grows and finally diverges for any $L$.
    \item If $\varphi^{*}( y^{\mathrm{min}} ) - \tilde{\Pi} = 0$, the system converges to an 
    equilibrium state when $L = 0$, whereas it grows when $L \neq 0$.
    \item If $\varphi^{*}( y^{\mathrm{min}} ) - \tilde{\Pi} < 0$, the system shrinks and finally vanishes when $L = 0$, whereas it converges to an equilibrium state when $L \neq 0$.
\end{enumerate}
\label{claim1}
\end{remark}

Here, $L = 0$ represents the case in which $L^l = 0$ for all the components of $L = \{ L^l \}$ in Eq. (\ref{conserved_quantity}), and $L \neq 0$ indicates that at least one of the components is not zero. 

We first notice that one of the conditions in the claim is given by the sign of $\varphi^*(y^{\mathrm{min}}) - \tilde{\Pi}$. Here, $\varphi^{*}\left( y^{\mathrm{min}} \right)$ represents the minimum pressure of the system within $\mathcal{M}^{\mathcal{Y}}_{\mathrm{EQ}}(\tilde{\mu})$ (see Eq. (\ref{y_min_outline})), whereas $\tilde{\Pi}$ is the pressure of the environment. 
The growth of the system occurs independently of the value of $L$ when $\varphi^*(y^{\mathrm{min}})>\tilde{\Pi}$, because the environmental pressure can not prevent the system from growing by counteracting any internal pressures $\varphi^*(y)$ for $y \in \mathcal{M}^{\mathcal{Y}}_{\mathrm{EQ}}(\tilde{\mu})$.

The equilibration becomes possible only when $\varphi^*(y^{\mathrm{min}}) - \tilde{\Pi} \leq 0$ is fulfilled. 
For the singular case of $L=0$, the equilibration occurs only when the minimum internal pressure perfectly balances with the external one, namely, $\varphi^*(y^{\mathrm{min}}) - \tilde{\Pi} = 0$. When $\varphi^*(y^{\mathrm{min}}) - \tilde{\Pi} < 0$, the system shrinks and $X(t)$ approaches $0$ as $t \to \infty$ because the external pressure dominates the internal one at the equilibrium state. For the generic $L\neq 0$, the fate of the system is qualitatively altered. The system equilibrates even if $\varphi^*(y^{\mathrm{min}}) - \tilde{\Pi} < 0$ owing to the non-singular conservation laws, which prevent $X(t)$ from approaching $0$, i.e., shrinking. In fact, the equilibrium state $y \in \mathcal{M}^{\mathcal{Y}}_{\mathrm{EQ}}(\tilde{\mu})$ can have a higher pressure in this case than $\varphi^*(y^{\mathrm{min}})$, and it balances with the external pressure $\tilde{\Pi}$. 
In addition, the system grows for $\varphi^*(y^{\mathrm{min}}) - \tilde{\Pi} = 0$.

From claim \ref{claim1}, the system converges to an equilibrium state when (1) $\varphi^{*}( y^{\mathrm{min}} ) - \tilde{\Pi} = 0$ and $L=0$, and (2) $\varphi^{*}( y^{\mathrm{min}} ) - \tilde{\Pi} < 0$ and $L \neq 0$. For these cases, our second claim describes how the equilibrium states appear in the number space $\mathfrak{X}$. 

\begin{remark}
In the number space $\mathfrak{X}$, the equilibrium states appear as follows (see also Table \ref{tab:table1}):
    \begin{enumerate}
        \item When $\varphi^{*}( y^{\mathrm{min}} ) - \tilde{\Pi} = 0$ and $L=0$, the equilibrium state to which the system converges depends on the initial condition $X_0 \in \mathcal{M}^{\mathfrak{X}}_{\mathrm{STO}}(L=0)$ and the functional form of the reaction flux $J(t)$. Such an equilibrium state lies on the ray $\mathfrak{r}^{\mathcal{Y}}(y^{\mathrm{min}})$ in Eq. (\ref{ray_y}).
        \item When $\varphi^{*}( y^{\mathrm{min}} ) - \tilde{\Pi} < 0$ and $L \neq 0$, the equilibrium state is uniquely determined, irrespective of the initial condition $X_0 \in \mathcal{M}^{\mathfrak{X}}_{\mathrm{STO}}(L\neq0)$ and the form of the reaction flux $J(t)$.  
    \end{enumerate}
    \label{claim2}
\end{remark}

\begin{table}
\caption{\label{tab:table1}%
The fate of the system is classified by the sign of $\varphi^*(y^{\mathrm{min}})-\tilde{\Pi}$ and the conserved quantities $L = \{ L^l \}$. Equilibration is indicated by $\mathrm{EQ(D)}$ and $\mathrm{EQ(I)}$. For $\mathrm{EQ(D)}$, the equilibrium state depends on the initial conditions and the functional form of the reaction flux $J(t)$. For $\mathrm{EQ(I)}$, it is independent for a given $L$. Here, $y^{\mathrm{min}} = \{ y^{\mathrm{min}}_i \}$ is defined in Eq. (\ref{y_min_outline}) and it is identical with $y^{\mathrm{EQ}}_i = -\tilde{\mu}_m O^m_r (S^{-1})^r_i$ for regular $S$.
}
\begin{ruledtabular}
\begin{tabular}{lccr}
\textrm{}&
$L\neq 0$&
\multicolumn{1}{c}{$L = 0$}&
\textrm{Regular $S$}\\
\colrule
$\varphi^*(y^{\mathrm{min}}) - \tilde{\Pi} > 0$ & Growing & \textrm{Growing} & Growing \\
$\varphi^*(y^{\mathrm{min}}) - \tilde{\Pi} = 0$ & Growing & EQ (D) & EQ (D)\\
$\varphi^*(y^{\mathrm{min}}) - \tilde{\Pi} < 0$ & EQ (I) & Shrinking & Shrinking \\
\end{tabular}
\end{ruledtabular}
\end{table}

\subsection{Regular stoichiometric matrix $S$}
\label{sec:productivity}

The fates of the system for the singular case of $L=0$ are basically the same as 
the case that $S$ is regular; no conservation laws exist so that $\dim \mathrm{Ker} [S^T] = 0$, and $\dim \mathrm{Ker} [S] = 0$ from Eq. (\ref{KerS0}) \footnote{Note that regular $S$ is productive in Eq. (\ref{productiveS}).} (see Sec. \ref{sec:proof_corollaries} for the reason why regular $S$ can be interpreted as $L=0$). 
Thus, the two claims clarify that the existence of conservation laws is a fundamental determinant of the fate of the growing system.
We investigated the case of regular $S$ in our previous work \cite{sughiyama2022chemical} (see also \footnote{In Sec. II of the Supplementary material, we show an example when $S$ is regular in which the dimension of $\mathrm{Ker}[S^T]$ is zero and that of $\mathrm{Im}[S]$ is three.}). 
For comparison, we summarize the results here (see the cases for $L=0$ and regular $S$ in Table \ref{tab:table1}). 

For a regular $S$, the candidates of chemical equilibrium states, $\mathcal{M}^{\mathcal{Y}}_{\mathrm{EQ}}(\tilde{\mu})$ in Eq. (\ref{eq_manifold_y_outline}), consist of precisely one point $y_i^{\mathrm{EQ}} = - \tilde{\mu}_m O^m_r (S^{-1})^r_i$. Here, we note that $y^{\mathrm{min}}$ is identical with $y^{\mathrm{EQ}}$ by Eq. (\ref{y_min_outline}).
Then, we obtain
\begin{cor}
    \label{cor1}
    When the stoichiometric matrix $S$ is regular, the fate of the system is classified by $y_i^{\mathrm{EQ}} = - \tilde{\mu}_m O^m_r (S^{-1})^r_i$ as follows (see also Table \ref{tab:table1}):
    \begin{enumerate}
    \item If $\varphi^{*}\left( y^{\mathrm{EQ}} \right)-\tilde{\Pi} > 0$, the system grows and finally diverges.
    \item If $\varphi^{*}( y^{\mathrm{EQ}} ) - \tilde{\Pi} = 0$,
    equilibrium states form the ray $\mathfrak{r}^{\mathcal{Y}}(y^{\mathrm{EQ}})$ in Eq. (\ref{ray_y}).
    The system converges to one of them, depending on the initial condition $X_0 \in \mathfrak{X}$ and the functional form of the reaction flux $J(t)$. 
    \item If $\varphi^{*}( y^{\mathrm{EQ}} ) - \tilde{\Pi} < 0$, the system shrinks and finally vanishes.
\end{enumerate}
\end{cor}
This statement corresponds to claim 1 in \cite{sughiyama2022chemical}.

\subsection{Numerical demonstrations}
\label{sec:numerical}

In this subsection, we demonstrate our claims by using the specific reactions in \textit{Example 1} with the ideal gas case (see \textit{Example 2}).

We first calculate the equilibrium manifold $\mathcal{M}^{\mathcal{Y}}_{\mathrm{EQ}}(\tilde{\mu})$ in Eq. (\ref{eq_manifold_y_outline}) to obtain $y^{\mathrm{min}}$ in Eq. (\ref{y_min_outline}). 
For the specific reactions in Eq. (\ref{Reactions_example}) and the stoichiometric matrices in Eq. (\ref{S_O_example}),
the simultaneous equations in Eq. (\ref{simultaneous_equation_outline_new}) are written as
\begin{align}
    \begin{pmatrix}
    -y_1 + 2 y_2 - y_3 - \tilde{\mu}_1  \\
    y_1 - y_2 + y_3 + \tilde{\mu}_2 
    \end{pmatrix} = 
    \begin{pmatrix}
    0 \\
    0
    \end{pmatrix}
    .
    \label{simultaneousEQ_example_new}
\end{align}
The solutions $y = \{ y_i \}$ are expressed as 
\begin{align}
    \begin{pmatrix}
    y_1 \\
    y_2 \\
    y_3 
    \end{pmatrix}
    = 
    \begin{pmatrix}
    -h \\
    \tilde{\mu}_1 - \tilde{\mu}_2 \\
    \tilde{\mu}_1 - 2\tilde{\mu}_2 +  h 
    \end{pmatrix},\label{example_solution_y_new}
\end{align}
where $h \in \mathbb{R}$ is the coordinate of $\mathrm{Ker}[S^T]$. The set of solutions in Eq. (\ref{example_solution_y_new}) represents the equilibrium manifold $\mathcal{M}^{\mathcal{Y}}_{\mathrm{EQ}}(\tilde{\mu})$. 

Second, using Eqs (\ref{partialGP_ideal}) and (\ref{varphi_y}), the full grand potential density $\varphi^*(y)$ is obtained for the ideal gas or solution as 
\begin{align}
    \varphi^*(y) = R\tilde{T} \sum_i e^{\left\{ y_i - \nu_i^o \left( \tilde{T} \right) \right\}/R\tilde{T}} + R\tilde{T} \sum_m e^{\left\{ \tilde{\mu}_m - \mu_m^o \left( \tilde{T} \right) \right\}/R\tilde{T}}.\label{fullGP_ideal}
\end{align}
Using the solutions in Eq. (\ref{example_solution_y_new}), $\varphi^*(y)$ is represented on $\mathcal{M}^{\mathcal{Y}}_{\mathrm{EQ}}(\tilde{\mu})$ as  
\begin{align}
    \bar{\varphi}^*(h) &= R\tilde{T} \biggl[ e^{\left\{ -h - \nu_1^o \left( \tilde{T} \right) \right\}/R\tilde{T}} + e^{\left\{ \tilde{\mu}_1 - \tilde{\mu}_2 - \nu_2^o \left( \tilde{T} \right) \right\}/R\tilde{T}} \notag \\
&+ e^{\left\{ \tilde{\mu}_1 - 2\tilde{\mu}_2 + h - \nu_3^o \left( \tilde{T} \right) \right\}/R\tilde{T}} \biggr] \notag \\ 
    &+ R\tilde{T} \left[ e^{\left\{ \tilde{\mu}_1 - \mu_1^o \left( \tilde{T} \right) \right\}/R\tilde{T}} + e^{\left\{ \tilde{\mu}_2 - \mu_2^o \left( \tilde{T} \right) \right\}/R\tilde{T}} \right].\label{fullGP_ideal_example}
\end{align}

\begin{figure}
    \centering
    \includegraphics[width=0.5\textwidth]{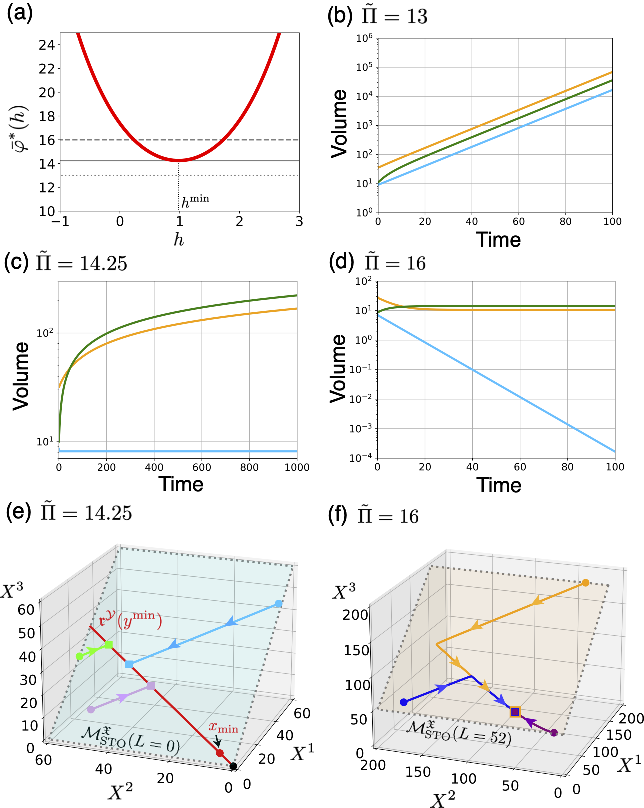}
    \caption{(a) The pressure $\bar{\varphi}^*(h)$ in Eq. (\ref{fullGP_ideal_example}) is shown by the red curve. The coordinate $h$ gives each element $y \in \mathcal{M}^{\mathcal{Y}}_{\mathrm{EQ}}(\tilde{\mu})$ in Eq. (\ref{example_solution_y_new}). The unique $h^{\mathrm{min}}$ for the minimum pressure is obtained as $h^{\mathrm{min}} = (-\nu^o_1+\nu^o_3-\tilde{\mu}_1+2\tilde{\mu}_2)/2 = 0.981$. The corresponding $y$ is $y^{\mathrm{min}}= (-0.981, -0.288, 1.099)$ and $\bar{\varphi}^*(h^{\mathrm{min}}) = \varphi^*(y^{\mathrm{min}}) = 14.25$. The values at $\bar{\varphi}^*(h) = 13, 14.25$, and $16$ are shown in dotted, solid, and dashed gray lines, respectively.  
    (b-d) Numerical plot of the time evolution for the reactions in Eq. (\ref{Reactions_example}) with the assumptions of ideal gas and mass-action kinetics. The volume $\Omega(X)$ is shown for three different pressures (b) $\tilde{\Pi} = 13$, (c) $\tilde{\Pi} = 14.25$, and (d) $\tilde{\Pi} = 16$. In (b-d), different colors indicate different initial conditions: $(X_0^1, X_0^2, X_0^3)= (48,1,48)$ (light blue), $(148,25,200)$ (orange), $(90,5,20)$ (green). Here, the initial conditions correspond to the conserved quantities $L = 0$, $L = 52$ and $L = -70$, respectively, where $L$ is calculated as $L = UX_0 = -X^1_0 + X^3_0$. 
    (e,f) The time evolution of $X$ is shown in the number space $\mathfrak{X}$ for three initial conditions. (e) We set that all the initial conditions have $L=0$ and are given by $(X_0^1, X_0^2, X_0^3)= (48,1,48)$ (light blue), $(25,60,25)$ (light green), $(10,50,10)$ (light purple). Each initial condition is indicated by a circle in the corresponding color. From each initial condition, the curve is constrained in $\mathcal{M}^{\mathfrak{X}}_{\mathrm{STO}}(L=0)$ and finally converges to the square when $\tilde{\Pi} = 14.25$. All the squares are located on the red ray $\mathfrak{r}^{\mathcal{Y}}(y^{\mathrm{min}})$. The black and red circles on this ray indicate the origin $X=0$ and $x^i_{\mathrm{min}}=\partial^i \varphi^*(y^{\mathrm{min}}) = \exp\{ (y^{\mathrm{min}}_i-\nu^o_i)/R\tilde{T}\}$, respectively. (f) We set that the three initial conditions have $L=52$ and are given by $(X_0^1, X_0^2, X_0^3)= (148,25,200)$ (orange), $(10,175,62)$ (blue), and $(1,10,53)$ (purple), respectively. Each of them is indicated by a circle in the corresponding color. The curve is constrained in $\mathcal{M}^{\mathfrak{X}}_{\mathrm{STO}}(L=52)$ and finally converges to the unique point (square). 
    The parameters are fixed as $R = \tilde{T} = 1$, $x^1_o = 8$, $x^2_o = 7$, $x^3_o = 1$, $n^1_o = 2$, $\tilde{n}^1 = 1$, $n^2_o =3$, $\tilde{n}^2 = 2$. Given the parameters, the rate constants $\hat{w}^r_+$ and $\hat{w}^r_-$ are determined by Eq. (\ref{localdetailedbalance3}) with $\hat{w}^r_- = 1$ in the Appendix \ref{appendix_numerical}. 
    }
    \label{Timeevolution}
\end{figure}

In Fig. \ref{Timeevolution}(a), we numerically plot the pressure $\bar{\varphi}^*(h)$ on $\mathcal{M}^{\mathcal{Y}}_{\mathrm{EQ}}(\tilde{\mu})$. 
Here, we have a unique $h^{\mathrm{min}} = (-\nu^o_1+\nu^o_3-\tilde{\mu}_1+2\tilde{\mu}_2)/2$ that attains the minimum $\bar{\varphi}^*(h)$. Substituting it into Eq. (\ref{example_solution_y_new}), we obtain $y^{\mathrm{min}}$ and the minimum pressure $\varphi^*(y^{\mathrm{min}})$. 
With the parameters given in the caption, $\varphi^*(y^{\mathrm{min}}) = 14.25$.


To demonstrate that the fate of the system is classified by $y^{\mathrm{min}}$ and the conserved quantities $L$ in claim 1, we show numerical simulations of the dynamics, Eq. (\ref{system_dynamics_chem}).
From the solution $X(t)$, the volume $\Omega(X)$ of the system is determined by the equation of state, Eq. (\ref{EqState}). 
To numerically solve Eq. (\ref{system_dynamics_chem}) with the stoichiometric matrix $S$ in Eq. (\ref{S_O_example}), we employ the mass-action kinetics with the local detailed balance condition to specify the flux function $J(t)$ (see Appendix \ref{appendix_numerical}). 

\textit{Claim 1:} In Fig. \ref{Timeevolution}(b-d), we show numerical trajectories of the volume $\Omega(X)$ from three initial conditions for different pressures $\tilde{\Pi}$. The color of curves corresponds to each initial condition. 
The fate of the system changes with $\tilde{\Pi}$ as follows.
When $\tilde{\Pi} = 13 < \varphi^*(y^{\mathrm{min}}) = 14.25$  (Fig. \ref{Timeevolution}(b)), the volume of the system increases with time from any initial conditions. 
As $\tilde{\Pi}$ increases to $\tilde{\Pi} = \varphi^*(y^{\mathrm{min}}) = 14.25$ (Fig. \ref{Timeevolution}(c)), the trajectory from an initial condition with $L=0$ in light blue achieves an equilibrium state, whereas trajectories from the other two with $L\neq 0$ continue growing.
As $\tilde{\Pi}$ increases further, $\tilde{\Pi} = 16 > \varphi^*(y^{\mathrm{min}}) = 14.25$ (Fig. \ref{Timeevolution}(d)), the trajectory with $L=0$ in light blue decreases and the system shrinks. By contrast, those from the other two with $L\neq 0$ achieve equilibrium states. 
These numerical results demonstrate the claim 1. 

\textit{Claim 2:} In Fig. \ref{Timeevolution}(e,f), we demonstrate the time evolution of the number $X$. When $\tilde{\Pi} = \varphi^{*}( y^{\mathrm{min}}) = 14.25$ and $L=0$, the equilibrium state to which the system converges depends on the initial conditions in $\mathcal{M}^{\mathfrak{X}}_{\mathrm{STO}}(L=0)$. Furthermore, it indeed lies on the ray $\mathfrak{r}^{\mathcal{Y}}(y^{\mathrm{min}})$, which is given by Eq. (\ref{ray_y}) (see Fig. \ref{Timeevolution}(e)). In our numerical simulation, we compute it as follows. From $y^{\mathrm{min}}$, we obtain the corresponding density $x^i_{\mathrm{min}}= \partial^i \varphi^*(y^{\mathrm{min}}) = \exp \{(y^{\mathrm{min}}_i - \nu_i^o)/R\tilde{T}\}$ by using Eq. (\ref{fullGP_ideal}). The ray $\mathfrak{r}^{\mathcal{Y}}(y^{\mathrm{min}})$ is plotted by the one which includes $x_{\mathrm{min}}$. When $\tilde{\Pi} =16 > \varphi^{*}( y^{\mathrm{min}})$ and $L = 52 \neq0$, the system indeed converges to the unique equilibrium state, irrespective of the initial conditions in $\mathcal{M}^{\mathfrak{X}}_{\mathrm{STO}}(L=52)$ (see Fig. \ref{Timeevolution}(f)).

\subsection{Outline of the derivations of claims \ref{claim1} and \ref{claim2}, and Corollary \ref{cor1}}\label{sec:outline_derivations}

The fates of the system stated in claims 1 and 2, and Corollary \ref{cor1} are summarized in Table \ref{tab:table1}.
Before closing this section, we outline their derivations, which will be presented in the subsequent sections. They will be derived in two steps. In the first step, we investigate the existence of an equilibrium state, that is the intersection $\mathcal{M}^{\mathcal{Y}}_{\mathrm{STO}}(L) \cap \mathcal{M}^{\mathcal{Y}}_{\mathrm{EQ}}(\tilde{\mu})$ in Eq. (\ref{intersection_outline}). In Sec. \ref{sec:geometry}, we will show that the intersection is not empty in the case $2$ for $L=0$ and in the case $3$ for $L\neq 0$ of claim \ref{claim1}, and is empty for the other cases. This will be summarized in Theorem \ref{thm1_new}. 
Following Theorem \ref{thm1_new}, we derive claim \ref{claim2} from the correspondence in Eq. (\ref{ray_y}) between the number space $\mathfrak{X}$ and the chemical potential space $\mathcal{Y}$. 
In the second step, when the intersection is empty, we investigate the landscape of the total entropy function $\Sigma^{\mathrm{tot}}$ to classify whether the system grows or shrinks in Sec. \ref{sec:entropy}. We will show that the system has two possibilities for $L=0$: it grows when $\Sigma^{\mathrm{tot}}$ is not bounded above, or it shrinks when the supremum of $\Sigma^{\mathrm{tot}}$ is at the origin $X=0$. By contrast, the system always grows for $L\neq 0$ when the intersection is empty because $\Sigma^{\mathrm{tot}}$ is not bounded above. 
This will be summarized in Theorem \ref{thm2_new}. Mathematical proofs of Theorems \ref{thm1_new} and \ref{thm2_new}, and Corollary \ref{cor1} will be presented in Sec. \ref{sec:proof}.

\section{Geometric representation of equilibrium states}
\label{sec:geometry}

For the derivation of claims \ref{claim1} and \ref{claim2}, we 
analyze the accessible regions in the number space $\mathfrak{X}$, the density space $\mathcal{X}$ and the chemical potential space $\mathcal{Y}$ by using a geometric technique. 
In Sec. \ref{sec:proof1}, we give the assumptions to the full grand potential density $\varphi^*(y)$ for the following mathematical treatment. In Sec. \ref{sec:isobaric}, we show that the accessible region is restricted by the isobaric condition, which we term the isobaric manifold. This manifold is further partitioned by the conservation laws in Sec. \ref{sec:conservation_law}.
After defining the equilibrium manifold as the candidates of chemical equilibrium states in Sec. \ref{sec:thermodynamics}, we characterize the equilibrium state to which the system converges as an intersecting point of the accessible region and the equilibrium manifold in Sec. \ref{sec:eq_as_intersection}. 
Furthermore, by using numerical simulations, we identify the conditions for the existence of the intersecting point, which we state as Theorem \ref{thm1_new}.

\subsection{Assumptions to $\varphi^*(y)$}
\label{sec:proof1}

The thermodynamic property of the system is encoded in the full grand potential density $\varphi^*(y)$ (see Eq. (\ref{varphi_y})). Here, $y \in \mathcal{Y} = \mathbb{R}^{\mathcal{N}_{X}}$ is the chemical potential of the confined chemicals, and $\varphi^*(y)$ represents the pressure of the system at the state $y$. 
In this work, we assume that $\varphi^*(y)$ is a smooth and strictly convex function on $\mathcal{Y} = \mathbb{R}^{\mathcal{N}_{X}}$, and the image of the associated Legendre transformation,
\begin{align}
    \partial \varphi^*(y): y \in \mathcal{Y} \mapsto \partial \varphi^*(y) = \left\{ \partial^i \varphi^* \right\} = \left\{ \frac{\partial \varphi^*}{\partial y_i} \right\}, 
\end{align}
is equal to $\mathcal{X} = \mathbb{R}_{>0}^{\mathcal{N}_{X}}$. In addition, we assume the following properties; (1) $\varphi^*(y)$ strictly increases with $y_i$ for an arbitrary fixed $\{ y_j \}_{j\neq i}$. (2) $\partial^i \varphi^*(y) \rightarrow 0$ when $y_i \rightarrow - \infty$.
They are satisfied in most thermodynamic systems such as the ideal gas in Eq. (\ref{fullGP_ideal}). 

Furthermore, we assume that the mapping $\partial \varphi^*: \mathbb{R}^{\mathcal{N}_{X}} \rightarrow \mathbb{R}_{>0}^{\mathcal{N}_{X}}$ is bijective, and we consider that its inverse map is given by $\partial \varphi : \mathbb{R}_{>0}^{\mathcal{N}_{X}} \rightarrow \mathbb{R}^{\mathcal{N}_{X}}$ (see Eq. (\ref{varphi_x}) for $\varphi(x)$). By using this, the above property (2) is rephrased as $y_i = \partial_i \varphi(x) \rightarrow - \infty$ when $x^i \rightarrow 0$.

Finally, we assume that the pressure of the environment $\tilde{\Pi}$ is greater than $\Pi_{\mathrm{min}}$, to guarantee that the volume $\Omega(X)$ is uniquely determined (see Appendix \ref{appendix_min_pressure}). Here, $\Pi_{\mathrm{min}}$ denotes the minimum pressure that the system can take as 
\begin{align}
    \Pi_{\mathrm{min}} := \inf_{y\in \mathcal{Y}} \varphi^*(y) = \lim_{\{y_i\} \rightarrow \{ - \infty \}} \varphi^*(y).\label{min_pressure}
\end{align} 
The second equality holds from the above property (1).

\subsection{The isobaric manifold and the projection $\rho_{\mathcal{X}}$}
\label{sec:isobaric}

We remind that the density $x$ is given by the mapping $\rho_{\mathcal{X}}$ in Eq. (\ref{map_density_of_x}):
\begin{align}
\rho_{\mathcal{X}}:= X \in \mathcal{X} \mapsto \rho_{\mathcal{X}}(X) = x(X) = \left\{ \frac{X^i}{\Omega(X)} \right\} \in \mathcal{X}.
\end{align}
The range of the mapping  $\rho_{\mathcal{X}}$ describes possible states of the density $x$ under the isobaric condition.
To calculate the range, we compute the volume $\Omega$ by solving the minimization in Eq. (\ref{volume}). 
Its critical equation is given by 
\begin{align}
    \varphi\left( \frac{X}{\Omega} \right) - \frac{X^i}{\Omega} \partial_i \varphi\left( \frac{X}{\Omega} \right) + \tilde{\Pi} = 0.\label{isobaric_x}
\end{align}
Therefore, the possible region in $\mathcal{X}$ under the isobaric condition is constrained in  
\begin{align}
\mathcal{I}^{\mathcal{X}}\left(\tilde{\Pi},\tilde{\mu}\right):= \Set{x|\varphi\left(x\right)-x^i \partial_i \varphi(x) + \tilde{\Pi} = 0, x > 0 },
\label{isobaric_manifold_x}
\end{align}
which we call the isobaric manifold.
Since the range of $\rho_{\mathcal{X}}$ is represented in Eq. (\ref{isobaric_manifold_x}), we find that the mapping $\rho_{\mathcal{X}}(X)$ is a projection of $X \in \mathfrak{X}$ into $\mathcal{I}^{\mathcal{X}}(\tilde{\Pi},\tilde{\mu})$ along the ray which includes $X$.

\begin{figure}
    \centering
    \includegraphics[width=0.5\textwidth]{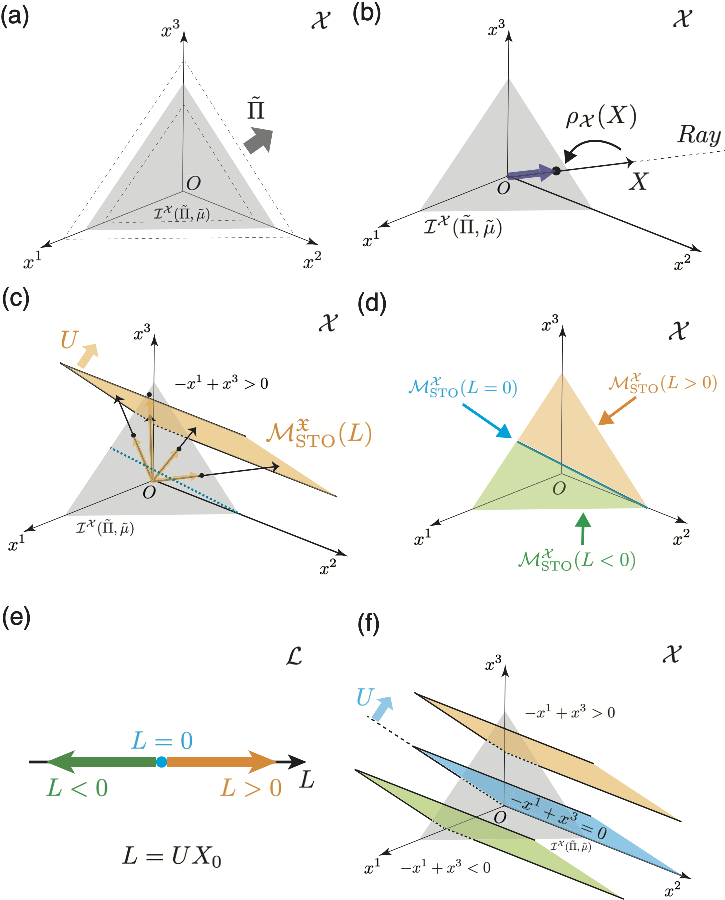}
    \caption{
    (a) For the ideal gas case, the isobaric manifold $\mathcal{I}^{\mathcal{X}}(\tilde{\Pi},\tilde{\mu})$ is represented by a simplex in gray. 
    (b) The mapping $\rho_{\mathcal{X}}(X)$ gives the projection of $X \in \mathfrak{X}$ into $\mathcal{I}^{\mathcal{X}}(\tilde{\Pi},\tilde{\mu})$ along the ray which includes $X$. The thick arrow is pointing the corresponding density $x = \rho_{\mathcal{X}}(X)$.
    (c) For the reactions in Eq. (\ref{Reactions_example}), the stoichiometric manifold $\mathcal{M}^{\mathfrak{X}}_{\mathrm{STO}}(L)$ in Eq. (\ref{stoichiometric_manifold2}) represents a two dimensional plane with $U = (-1, 0, 1)$. By applying the mapping $\rho_{\mathcal{X}}(X)$ to each point $X \in \mathcal{M}^{\mathfrak{X}}_{\mathrm{STO}}(L)$, we get the corresponding manifold, $\mathcal{M}^{\mathcal{X}}_{\mathrm{STO}}(L)$ (see the upper region of the simplex in (d) colored by orange).
    (d) In the present case, the isobaric manifold $\mathcal{I}^{\mathcal{X}}(\tilde{\Pi},\tilde{\mu})$ is partitioned by $\mathcal{M}^{\mathcal{X}}_{\mathrm{STO}}(L>0)$, $\mathcal{M}^{\mathcal{X}}_{\mathrm{STO}}(L<0)$, and $\mathcal{M}^{\mathcal{X}}_{\mathrm{STO}}(L=0)$. Here, $\mathcal{M}^{\mathcal{X}}_{\mathrm{STO}}(L > 0)$ and $\mathcal{M}^{\mathcal{X}}_{\mathrm{STO}}(L < 0)$ are given by the orange and the green regions, respectively, and their interface is given by $\mathcal{M}^{\mathcal{X}}_{\mathrm{STO}}(L = 0)$ in the light blue line.
    (e) The stoichiometric manifold $\mathcal{M}^{\mathcal{X}}_{\mathrm{STO}}(L)$ is characterized by a ray $\mathfrak{r}$ in the space $\mathcal{L}$ of the conserved quantities (see the line below Eq. (\ref{conserved_quantity}) for $\mathcal{L}$). Since $\dim \mathrm{Ker}[S^T] = 1$ in the present example, we have three characterizations $L = 0$, $L >0$ and $L<0$. In Sec. I of the Supplementary material, we investigate another example with a higher dimension of $\mathrm{Ker}[S^T]$. 
    (f) Given a ray $\mathfrak{r}$, any plane  $\mathcal{M}^{\mathfrak{X}}_{\mathrm{STO}}(L)$ for $L \in \mathfrak{r}$ is mapped to the same region on the simplex by the mapping $\rho_{\mathcal{X}}$. Any plane with $L > 0$ (e.g. in orange) is mapped to the upper region of the simplex (see (d)), whereas, any plane with $L < 0$ (e.g. in green) is mapped to the lower region.
    For $L = 0$, the plane in light blue is mapped to the interface between the regions for $L > 0$ and $L < 0$ (see the light blue line in (d)). 
    }
    \label{fig:3D}
\end{figure}

\textit{Example 3:}
For the ideal gas case, by substituting $\varphi(x)$ in Eq. (\ref{partialGP_ideal}) into Eq. (\ref{isobaric_manifold_x}), we obtain  
\begin{align}
\mathcal{I}^{\mathcal{X}}\left(\tilde{\Pi},\tilde{\mu}\right) = \Set{x |
\sum_i x^i = \frac{\tilde{\Pi}}{R\tilde{T}} - \sum_m \tilde{n}^m, x>0
    }, 
\label{EqState_iso}
\end{align}
where $\tilde{n}^m$ is the density of the open chemicals in the environment (see Eq. (\ref{chemical_potential_idealgas})).
Thus, the isobaric manifold in $\mathcal{X}$ represents a simplex for the ideal gas case (Fig. \ref{fig:3D}(a)).

The mapping $\rho_{\mathcal{X}}$ is the projection of $X \in \mathfrak{X} = \mathbb{R}^{\mathcal{N}_{\mathcal{X}}}_{>0}$ into this simplex (see Fig. \ref{fig:3D}(b)).

\rightline{$\blacksquare$}

Finally, by using the mapping $\partial \varphi$, the isobaric manifold in $\mathcal{Y}$ is represented as
\begin{align}
\mathcal{I}^{\mathcal{Y}}\left(\tilde{\Pi},\tilde{\mu}\right):= \partial \varphi\left(\mathcal{I}^{\mathcal{X}}\right) = \Set{y|\varphi^{*}\left(y\right)-\tilde{\Pi} = 0},
\label{isobaric_manifold_y}
\end{align}
which is straightforward to interpret that the pressure of the system $\varphi^*(y)$ at the chemical potential $y$ is balanced with $\tilde{\Pi}$ in this manifold.

\subsection{The stoichiometric manifold as partition of the isobaric manifold}
\label{sec:conservation_law}

The accessible subspace of the state $X(t)$ from an initial condition $X_0$ in Eq. (\ref{accessible_space_X}) is  
\begin{equation}
    \mathcal{M}^{\mathfrak{X}}_{\mathrm{STO}}(L) := \Set{ X | U^l_i X^i = U^l_i X^i_0 = L^l, X > 0 },\label{stoichiometric_manifold2}
\end{equation}
which we call the stoichiometric manifold in $\mathfrak{X}$.

By applying the mapping $\rho_{\mathcal{X}}$ to $\mathcal{M}^{\mathfrak{X}}_{\mathrm{STO}}(L)$, we have the stoichiometric manifold in the density space $\mathcal{X}$ as
\begin{align}
    \mathcal{M}^{\mathcal{X}}_{\mathrm{STO}}(L) := &\left. \biggl\{ x \right| \varphi(x) - x^i \partial_i \varphi(x) + \tilde{\Pi} = 0, \notag \\
    &U^l_i x^i = \frac{L^l}{\alpha}, x>0 \biggr\},
\label{stoichiometric_manifold_x}
\end{align}
where $\alpha > 0$ is arbitrary.
Using the mapping $\partial \varphi$, the stoichiometric manifold in the chemical potential space $\mathcal{Y}$ is given by 
\begin{equation}
    \mathcal{M}^{\mathcal{Y}}_{\mathrm{STO}}(L) := \Set{ y | \varphi^*(y) - \tilde{\Pi} = 0,  U^l_i \partial^i \varphi^*(y) = \frac{L^l}{\alpha} }. \label{stoichiometric_manifold_y}
\end{equation}

\textit{Example 4:}
For the specific stoichiometric matrix in Eq. (\ref{S_O_example}), $\mathcal{M}^{\mathcal{X}}_{\mathrm{STO}}(L)$ is schematically depicted in Fig. \ref{fig:3D}(c,d) by applying the mapping $\rho_{\mathcal{X}}$ to $\mathcal{M}^{\mathfrak{X}}_{\mathrm{STO}}(L)$.
\rightline{$\blacksquare$}

For further investigations, we define the following manifold in $\mathcal{X}$ as 
\begin{align}
\hat{\mathcal{M}}^{\mathcal{X}}_{\mathrm{STO}}\left(\hat{L}\right) := \Set{ x | U^l_i x^i = \hat{L}^l, x > 0 },\label{sto_isochoric_x}
\end{align}
and that in $\mathcal{Y}$ as 
\begin{align}
\hat{\mathcal{M}}^{\mathcal{Y}}_{\mathrm{STO}}\left( \hat{L} \right) &:= \partial \varphi \left( \hat{\mathcal{M}}^{\mathcal{X}}_{\mathrm{STO}}\left( \hat{L} \right) \right) \notag \\ &= \Set{ y | U^l_i \partial^i \varphi^*(y) = \hat{L}^l }.
\label{sto_isochoric_y}
\end{align}
These manifolds correspond to the stoichiometric manifolds with conserved quantities $\hat{L}$ under the \textit{isochoric} condition \cite{sughiyama2022hessian}. Hereafter, we term the \textit{isochoric} stoichiometric manifolds.

Using these manifolds, the stoichiometric manifold under isobaric condition is represented as follows: 
\begin{align}
\mathcal{M}^{\mathcal{X}}_{\mathrm{STO}}(L) = \bigcup_{\alpha > 0} \hat{\mathcal{M}}^{\mathcal{X}}_{\mathrm{STO}}\left( \frac{L}{\alpha} \right) \cap \mathcal{I}^{\mathcal{X}}(\tilde{\Pi}, \tilde{\mu}),
\label{stoichiometric_manifold_x_isobaric_isochoric}
\end{align}
and
\begin{align}
\mathcal{M}^{\mathcal{Y}}_{\mathrm{STO}}(L) = \bigcup_{\alpha > 0} \hat{\mathcal{M}}^{\mathcal{Y}}_{\mathrm{STO}}\left( \frac{L}{\alpha} \right) \cap \mathcal{I}^{\mathcal{Y}}(\tilde{\Pi}, \tilde{\mu}).
\label{stoichiometric_manifold_y_isobaric_isochoric}
\end{align}
For arbitrary $c > 0$, we find $\mathcal{M}^{\mathcal{X}}_{\mathrm{STO}}(c L) = \mathcal{M}^{\mathcal{X}}_{\mathrm{STO}}(L)$ and, correspondingly,  $\mathcal{M}^{\mathcal{Y}}_{\mathrm{STO}}(c L) = \mathcal{M}^{\mathcal{Y}}_{\mathrm{STO}}(L)$.
Hence, each ray $\mathfrak{r}$ in the space $\mathcal{L}$ of conserved quantities characterizes these manifolds (see also \textit{Example 5} below and Fig. \ref{fig:3D}(d-f)). 
To be more precise, for any $L,L' \in \mathfrak{r}$, $\mathcal{M}^{\mathcal{X}}_{\mathrm{STO}}(L) = \mathcal{M}^{\mathcal{X}}_{\mathrm{STO}}(L')$ holds. 
We denote the stoichiometric manifold corresponding to a ray $\mathfrak{r}$ as $\mathcal{M}^{\mathcal{X}}_{\mathrm{STO}}(L\in \mathfrak{r})$.

When $L = 0$, the stoichiometric manifold in $\mathcal{X}$ is written as 
\begin{align}
\mathcal{M}^{\mathcal{X}}_{\mathrm{STO}}(L = 0) =  \hat{\mathcal{M}}^{\mathcal{X}}_{\mathrm{STO}}\left( \hat{L} = 0 \right) \cap \mathcal{I}^{\mathcal{X}}(\tilde{\Pi}, \tilde{\mu}).\label{intersection_x0}
\end{align}
Since the dimension of $\alpha$ in Eq. (\ref{stoichiometric_manifold_x_isobaric_isochoric}) vanishes for $\mathcal{M}^{\mathcal{X}}_{\mathrm{STO}}(L = 0)$, we obtain $\dim \mathcal{M}^{\mathcal{X}}_{\mathrm{STO}}(L = 0) = \dim \mathcal{M}^{\mathcal{X}}_{\mathrm{STO}}(L \neq 0) - 1$.
In addition, for any given $L \neq 0$, we find that $\hat{\mathcal{M}}^{\mathcal{X}}_{\mathrm{STO}}\left( L/\alpha \right) \rightarrow \hat{\mathcal{M}}^{\mathcal{X}}_{\mathrm{STO}}( \hat{L} = 0 )$ when $\alpha \rightarrow \infty$. Thus, the stoichiometric manifold $\mathcal{M}^{\mathcal{X}}_{\mathrm{STO}}(L = 0)$ forms an interface of all the stoichiometric manifolds $\mathcal{M}^{\mathcal{X}}_{\mathrm{STO}}(L \neq 0)$.
These properties hold similarly in $\mathcal{Y}$ by replacing the superscript $\mathcal{X}$ with $\mathcal{Y}$.

The above statements are summarized as follows.

\begin{lemma}
\label{lemma1}
Let $\mathfrak{r}$ denote a ray in the space $\mathcal{L}$ of conserved quantities. 
For every $\mathfrak{r}$, the corresponding stoichiometric manifold $\mathcal{M}^{\mathcal{X}}_{\mathrm{STO}}(L \in \mathfrak{r})$ exists and the set of the stoichiometric manifolds partitions the isobaric manifold $\mathcal{I}^{\mathcal{X}}(\tilde{\Pi}, \tilde{\mu})$. In addition, their interface is given by the stoichiometric manifold $\mathcal{M}^{\mathcal{X}}_{\mathrm{STO}}(L = 0)$.
The isobaric manifold in the chemical potential space $\mathcal{Y}$ is similarly partitioned by stoichiometric manifolds (replace the superscript $\mathcal{X}$ with $\mathcal{Y}$).
\end{lemma}

\textit{Example 5:}
In Fig. \ref{fig:3D}(d-f), we schematically depict the partition of the isobaric manifold $\mathcal{I}^{\mathcal{X}}(\tilde{\Pi}, \tilde{\mu})$ by the stoichiometric manifolds $\mathcal{M}^{\mathcal{X}}_{\mathrm{STO}}(L \in \mathfrak{r})$ for the ideal gas case.
\rightline{$\blacksquare$}

\subsection{The equilibrium manifold}
\label{sec:thermodynamics}

In this subsection, we confirm that the candidates of the equilibrium states are given by Eq. (\ref{eq_manifold_y_outline}).  

The equilibrium states of the chemical reaction dynamics are given by the maxima of the total entropy with respect to the extent of reaction $\Xi$. 
From the differentiation of Eq. (\ref{Enttot_partialGP}) and using Eq. (\ref{volume}), we have the critical equation as 
\begin{align}
    \frac{\partial \Sigma^{\mathrm{tot}}}{\partial \Xi^r} &= -\frac{1}{\tilde{T}} \left\{ \partial_i \varphi \left( \frac{X}{\Omega(X)} \right) S^i_r + \tilde{\mu}_m O^m_r \right\} \notag \\
    &= 0,\label{enttot_critical_eq}
\end{align}
where $X = X_0 + S\Xi$ in Eq. (\ref{accessible_space_X}) (see Appendix \ref{appendix_criticaleq}).

Thus, at any equilibrium states, the density $x$ satisfies 
\begin{equation}
    \partial_i \varphi(x) S^i_r + \tilde{\mu}_m O^m_r = 0. \label{simultaneous_equation_outline_x}
\end{equation} 
By using the correspondence $y_i = \partial_i \varphi(x)$, it is represented as 
\begin{equation}
y_i S^i_r + \tilde{\mu}_m O^m_r = 0, \label{simultaneous_equation_outline}
\end{equation}
which confirms the balance of chemical potentials between reactants and products at the chemical equilibrium (see Eq. (\ref{simultaneous_equation_outline_new})). 

The set of solutions to Eq. (\ref{simultaneous_equation_outline_x}) represents the candidates of the equilibrium states in the density space $\mathcal{X}$, which we call an equilibrium manifold: 
\begin{equation}
    \mathcal{M}^{\mathcal{X}}_{\mathrm{EQ}}(\tilde{T}, \tilde{\mu}) := \Set{ x| \partial_i \varphi(x) S^i_r + \tilde{\mu}_m O^m_r = 0 }, \label{eq_manifold_x_outline}
\end{equation}
and in the chemical potential space $\mathcal{Y}$ as 
\begin{equation}
    \mathcal{M}^{\mathcal{Y}}_{\mathrm{EQ}}(\tilde{\mu}) := \Set{ y| y_i S^i_r + \tilde{\mu}_m O^m_r = 0 }. \label{eq_manifold_y_outline2}
\end{equation}
We note that the equilibrium manifold is an affine subspace in $\mathcal{Y}$, whereas it is generally curved in $\mathcal{X}$.
In addition, using the basis matrix $U$ of $\mathrm{Ker} [S^T]$ and solving Eq. (\ref{simultaneous_equation_outline}), the manifold in Eq. (\ref{eq_manifold_y_outline2}) is rewritten as 
\begin{equation}
    \mathcal{M}^{\mathcal{Y}}_{\mathrm{EQ}}(\tilde{\mu}) = \Set{ y| y_i = y_i^{\mathrm{P}} + h_l U^l_i, h \in \mathbb{R}^{\mathrm{dim}\mathrm{Ker} [S^T]}}, 
    \label{veq_element_outline}
\end{equation}
where $y^{\mathrm{P}}$ is a particular solution to Eq. (\ref{simultaneous_equation_outline}) and $h = \{ h_l \}$ represents a coordinate of $\mathrm{Ker} [S^T]$. 

\subsection{The equilibrium state as an intersection}
\label{sec:eq_as_intersection}

If the intersection $\mathcal{M}^{\mathcal{Y}}_{\mathrm{STO}}(L) \cap \mathcal{M}^{\mathcal{Y}}_{\mathrm{EQ}}(\tilde{\mu})$ is not empty, the system converges to the equilibrium state. 
We can compute the equilibrium state in $\mathcal{Y}$ by using Eqs. (\ref{stoichiometric_manifold_y}) and (\ref{eq_manifold_y_outline2}).
Similarly, we obtain the corresponding state in $\mathcal{X}$ as $\mathcal{M}^{\mathcal{X}}_{\mathrm{STO}}(L) \cap \mathcal{M}^{\mathcal{X}}_{\mathrm{EQ}}(\tilde{T}, \tilde{\mu})$ by using Eq. (\ref{stoichiometric_manifold_x}) and (\ref{eq_manifold_x_outline}).
In this subsection, we numerically confirm the convergence in $\mathcal{X}$ by the dynamics in Eq. (\ref{system_dynamics_chem}) using the specific setup in Sec. \ref{sec:numerical}. 

\begin{figure}
    \centering
    \includegraphics[width=0.5\textwidth]{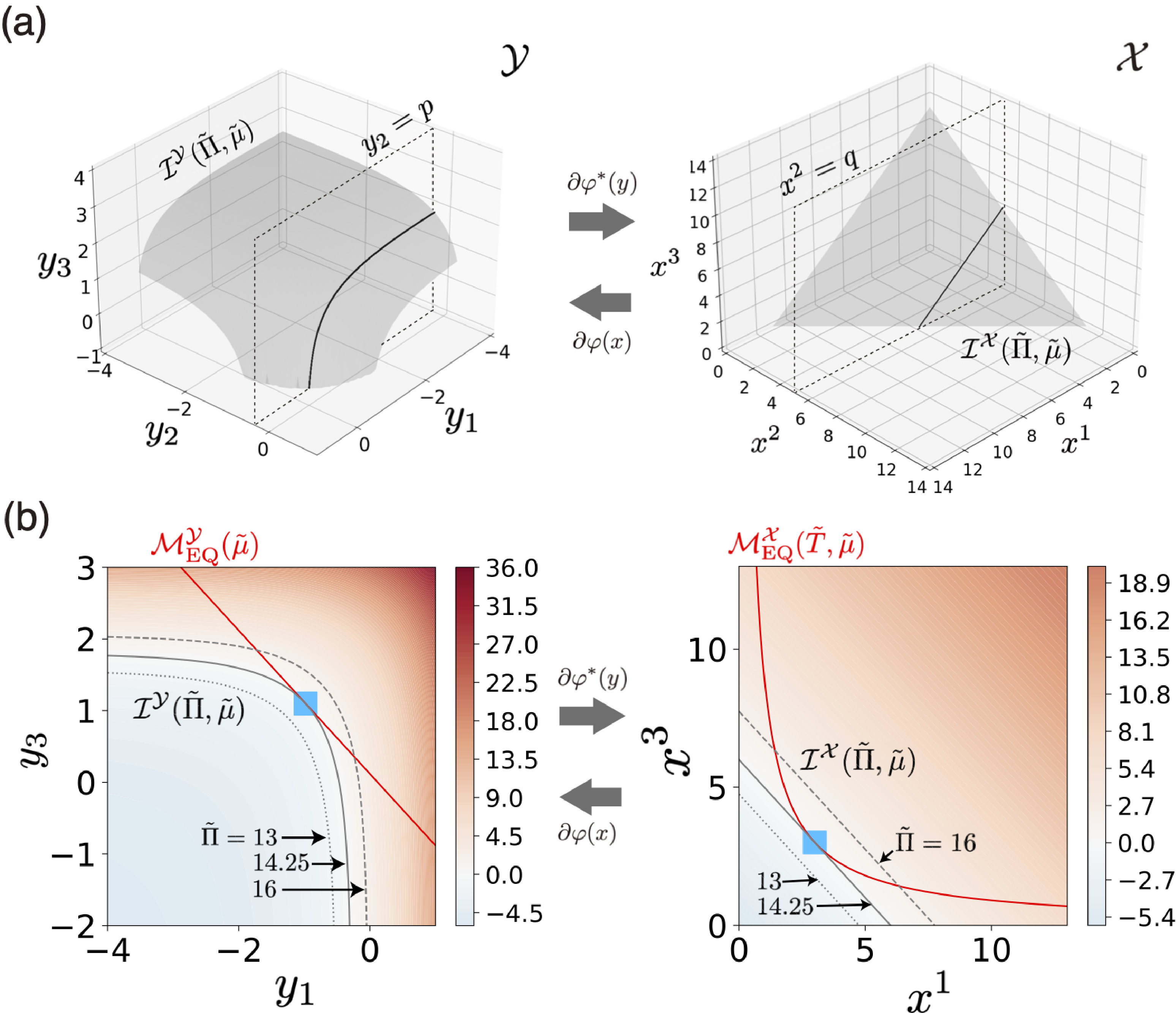}
    \caption{
    (a) For the ideal gas case, the isobaric manifold $\mathcal{I}^{\mathcal{Y}}(\tilde{\Pi},\tilde{\mu})$ in the chemical potential space $\mathcal{Y}$ is a level set for $\varphi^*(y)$ (left), whereas $\mathcal{I}^{\mathcal{X}}(\tilde{\Pi},\tilde{\mu})$ represents a simplex in the density space $\mathcal{X}$ (right). 
    We consider the planes $y_2 = p = \tilde{\mu}_1 - \tilde{\mu}_2$ in $\mathcal{Y}$ and $x^2 = q = \exp \{ (\tilde{\mu}_1 - \tilde{\mu}_2-\nu_2^o)/R\tilde{T}\}$ in $\mathcal{X}$, respectively. 
    The black curve in $\mathcal{Y}$ and the line in $\mathcal{X}$ indicate the cross section of the isobaric manifolds by the planes, respectively.
    (b) In the left panel, we plot $\mathcal{M}^{\mathcal{Y}}_{\mathrm{EQ}}(\tilde{\mu})$ (the red line) and $\mathcal{I}^{\mathcal{Y}}(\tilde{\Pi},\tilde{\mu})$ (gray curves) on the plane $y_2 = p$. In the right panel, $\mathcal{M}^{\mathcal{X}}_{\mathrm{EQ}}(\tilde{T}, \tilde{\mu})$ (the red curve) and $\mathcal{I}^{\mathcal{X}}(\tilde{\Pi},\tilde{\mu})$ (gray lines) are plotted on the plane $x^2 = q$. In $\mathcal{Y}$ (left), the three gray curves correspond to the isobaric manifolds $\mathcal{I}^{\mathcal{Y}}(\tilde{\Pi},\tilde{\mu})$ with the pressures $\tilde{\Pi} = 13$ (dotted), $14.25$ (solid) and $16$ (dashed), respectively. 
When $\tilde{\Pi} = 13$, the dotted curve does not have the intersection with $\mathcal{M}^{\mathcal{Y}}_{\mathrm{EQ}}(\tilde{\mu})$ (red line). When $\tilde{\Pi} = 14.25$, the solid curve has the unique intersecting point with the red line, denoted by the light blue square. It is located at $y^{\mathrm{min}} = (-0.981, -0.288, 1.099)$, where $y^{\mathrm{min}}$ has been obtained in Fig. \ref{Timeevolution}. When $\tilde{\Pi} = 16$, the dashed curve has two intersecting points with the red line. The heatmaps represent the values of $\varphi^*(y)-\varphi^*(y^{\mathrm{min}})$. In the right panel, we show the corresponding manifolds in $\mathcal{X}$. Here, the light blue square is located at $x_{\mathrm{min}} = (3.00,5.25,3.00)$, which is obtained by $x^i_{\mathrm{min}} = \partial^i \varphi^*(y^{\mathrm{min}}) = \exp\{(y^{\mathrm{min}}_i - \nu_i^o)/R\tilde{T}\}$.
    }
    \label{fig:xy_ideal}
\end{figure}

We first consider the equilibrium manifold.
The points in $\mathcal{M}^{\mathcal{Y}}_{\mathrm{EQ}}(\tilde{\mu})$ are written in Eq. (\ref{example_solution_y_new}). Here, the second component $y_2$ is constant $p := \tilde{\mu}_1 - \tilde{\mu}_2$. Thus, $\mathcal{M}^{\mathcal{Y}}_{\mathrm{EQ}}(\tilde{\mu})$ is a one-dimensional line located on the plane $y_2 = p$ in $\mathcal{Y}$ (see the left panel of Fig. \ref{fig:xy_ideal} (a) for the plane $y_2 = p$, and the red line in the left panel of Fig. \ref{fig:xy_ideal} (b) for $\mathcal{M}^{\mathcal{Y}}_{\mathrm{EQ}}(\tilde{\mu})$). 
By applying the mapping $\partial \varphi^*$, we can compute $\mathcal{M}^{\mathcal{X}}_{\mathrm{EQ}}(\tilde{T}, \tilde{\mu})$. 
For the ideal gas case in Eq. (\ref{fullGP_ideal}), the mapping is represented as $\partial^i \varphi^*(y) = \exp \{(y_i - \nu_i^o)/R\tilde{T}\}$.
Thus, the points in $\mathcal{M}^{\mathcal{X}}_{\mathrm{EQ}}(\tilde{T}, \tilde{\mu})$ are computed as 
\begin{align}
    \begin{pmatrix}
    x^1 \\
    x^2 \\
    x^3 
    \end{pmatrix}
    &= 
    \begin{pmatrix}
    \exp\{(-h-\nu_1^o)/R\tilde{T}\} \\
    \exp \{ (\tilde{\mu}_1 - \tilde{\mu}_2-\nu_2^o)/R\tilde{T}\} \\
    \exp \{ (\tilde{\mu}_1 - 2\tilde{\mu}_2 + h-\nu_3^o)/R\tilde{T} \} 
    \end{pmatrix},
    \label{example_solution_x}
\end{align}
where $h \in \mathbb{R}$ plays a role of the coordinate of $\mathcal{M}^{\mathcal{X}}_{\mathrm{EQ}}(\tilde{T}, \tilde{\mu})$. 
We note that the second component $x^2$ is also constant $q := \partial \varphi^*(p) = \exp \{ (\tilde{\mu}_1 - \tilde{\mu}_2-\nu_2^o)/R\tilde{T}\}$. Thus, $\mathcal{M}^{\mathcal{X}}_{\mathrm{EQ}}(\tilde{T}, \tilde{\mu})$ is a one-dimensional curve located on the plane $x^2 = q$.
(see the red curve in the right panel of Fig. \ref{fig:xy_ideal}(b)).

We next consider whether the intersection $\mathcal{I}^{\mathcal{X}}(\tilde{\Pi}, \tilde{\mu}) \cap \mathcal{M}^{\mathcal{X}}_{\mathrm{EQ}}(\tilde{T}, \tilde{\mu})$ is empty or not. 
As the pressure $\tilde{\Pi}$ increases, the number of intersecting points change as follows. 
When $\tilde{\Pi} = 13 < \varphi^*(y^{\mathrm{min}}) = 14.25$, 
the intersecting point $\mathcal{I}^{\mathcal{X}}(\tilde{\Pi},\tilde{\mu}) \cap \mathcal{M}^{\mathcal{X}}_{\mathrm{EQ}}(\tilde{T}, \tilde{\mu})$ does not exist (see the right panel of Fig. \ref{fig:xy_ideal}(b)). 
When $\tilde{\Pi} = \varphi^*(y^{\mathrm{min}})$, 
the intersection appears as the unique contact point corresponding to $y^{\mathrm{min}}$.
When $\tilde{\Pi} = 16 > \varphi^*(y^{\mathrm{min}})$, the intersection consists of two points. The above situation holds similarly in $\mathcal{Y}$ (see the left panel of Fig. \ref{fig:xy_ideal}(b)).

\begin{figure}
    \centering
    \includegraphics[width=0.5\textwidth]{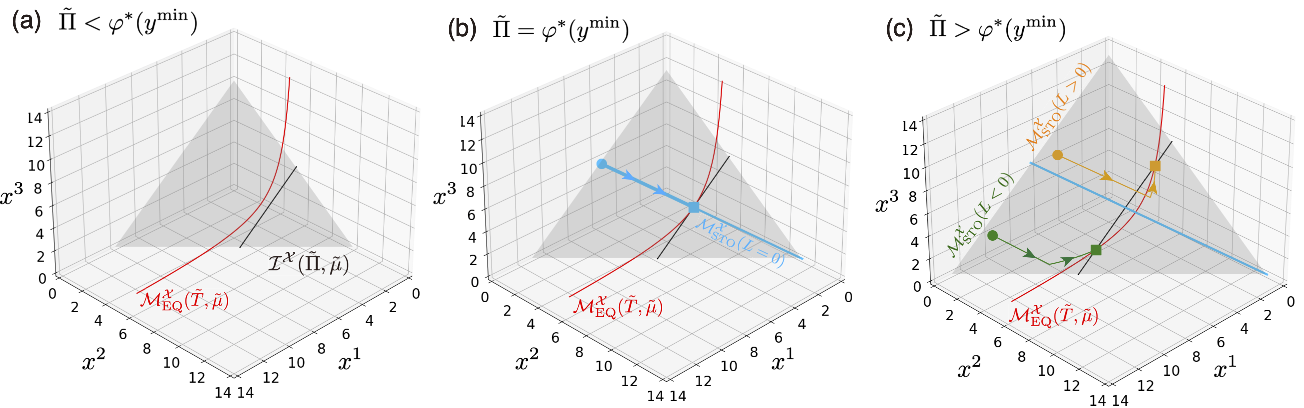}
    \caption{
    (a) When $\tilde{\Pi} = 13 < \varphi^*(y^{\mathrm{min}}) = 14.25$, $\mathcal{M}^{\mathcal{X}}_{\mathrm{EQ}}(\tilde{T}, \tilde{\mu})$ (red curve) and $\mathcal{I}^{\mathcal{X}}(\tilde{\Pi},\tilde{\mu})$ (gray simplex) do not have an intersecting point. The black line represents $\mathcal{I}^{\mathcal{X}}(\tilde{\Pi},\tilde{\mu})$ on the plane $x^2 = q$ (see also Fig. \ref{fig:xy_ideal}). (b) When $\tilde{\Pi} = 14.25 = \varphi^*(y^{\mathrm{min}})$, 
    the intersection consists of the unique point $x_{\mathrm{min}}$. It lies on the stoichiometric manifold $\mathcal{M}^{\mathcal{X}}_{\mathrm{STO}}(L=0)$ (the light blue line on the simplex). In numerical simulations, the time evolution of the system is also shown in a light blue line with arrows. From the initial condition (circle), it converges to the point $x_{\mathrm{min}}$ (square). (c) When $\tilde{\Pi} = 16 > \varphi^*(y^{\mathrm{min}})$, the intersection consists of two points (orange and green squares). They are located in the upper and the lower regions of the simplex, which correspond to $\mathcal{M}^{\mathcal{X}}_{\mathrm{STO}}(L)$ for $L>0$ and $L<0$, respectively (see Fig. \ref{fig:3D}(d)). We also show the time evolution of the system from the two initial conditions (circles). They respectively converge to the intersecting points (squares). In the numerical simulation, the initial conditions are given by 
    $(X^1_0, X^2_0, X^3_0) = (48, 1, 48)$ (light blue circle), $(148, 25, 200)$ (orange circle) and $(90, 5, 20)$ (green circle).} 
    \label{fig:intersection_numerical}
\end{figure}

We further depict the intersection $\mathcal{I}^{\mathcal{X}}(\tilde{\Pi}, \tilde{\mu}) \cap \mathcal{M}^{\mathcal{X}}_{\mathrm{EQ}}(\tilde{T}, \tilde{\mu})$ by the three dimensional plot in Fig. \ref{fig:intersection_numerical}. 
From Fig. \ref{fig:3D}(d), $\mathcal{I}^{\mathcal{X}}(\tilde{\Pi}, \tilde{\mu})$ is partitioned by the stoichiometric manifolds. 
When $\tilde{\Pi} = \varphi^*(y^{\mathrm{min}}) = 14.25$ (Fig. \ref{fig:intersection_numerical}(b)), the intersection lies on $\mathcal{M}^{\mathcal{X}}_{\mathrm{STO}}(L=0)$. When $\tilde{\Pi} = 16 > \varphi^*(y^{\mathrm{min}})$, the two intersecting points belong to each of $\mathcal{M}^{\mathcal{X}}_{\mathrm{STO}}(L)$ for $L > 0$ and $L < 0$, respectively (see Fig. \ref{fig:intersection_numerical}(c)). Thus, for a given $L$, the intersecting point is uniquely determined and it depends on a ray in the space $\mathcal{L}$ of the conserved quantities (see Fig. \ref{fig:3D}(e) and Lemma \ref{lemma1}).

In Fig. \ref{fig:intersection_numerical}(b,c), we also show the time evolution of the system in the numerical simulation. Indeed, the system converges to the intersection $\mathcal{M}^{\mathcal{X}}_{\mathrm{STO}}(L) \cap \mathcal{M}^{\mathcal{X}}_{\mathrm{EQ}}(\tilde{T}, \tilde{\mu})$. 

The above numerical result is an instance of the following theorem.
\begin{thm} The system converges to the intersecting point $\mathcal{M}^{\mathcal{Y}}_{\mathrm{STO}}(L) \cap \mathcal{M}^{\mathcal{Y}}_{\mathrm{EQ}}(\tilde{\mu})$ if it exists. Its existence is classified as follows: 
\label{thm1_new}
\begin{enumerate}
    \item $\varphi^{*}\left( y^{\mathrm{min}} \right)-\tilde{\Pi} > 0$ $\Rightarrow$ $\mathcal{M}^{\mathcal{Y}}_{\mathrm{STO}}(L) \cap \mathcal{M}^{\mathcal{Y}}_{\mathrm{EQ}}(\tilde{\mu}) = \emptyset$.
    \item $\varphi^{*}( y^{\mathrm{min}} ) - \tilde{\Pi} = 0$ $\Rightarrow$ 
    \begin{align*}
    \mathcal{M}^{\mathcal{Y}}_{\mathrm{STO}}(L) \cap \mathcal{M}^{\mathcal{Y}}_{\mathrm{EQ}}(\tilde{\mu}) = \begin{cases}
    \{ y^{\mathrm{min}} \} &\mbox{ for } L = 0, \notag \\ 
    \emptyset &\mbox{ for } L \neq 0.
    \end{cases}
    \end{align*}
    \item $\varphi^{*}( y^{\mathrm{min}} ) - \tilde{\Pi} < 0$ $\Rightarrow$ 
    \begin{align*}
    \mathcal{M}^{\mathcal{Y}}_{\mathrm{STO}}(L) \cap \mathcal{M}^{\mathcal{Y}}_{\mathrm{EQ}}(\tilde{\mu}) = \begin{cases}
    \emptyset &\mbox{ for } L = 0, \\
    \{ y^{\mathrm{EQ}} \} &\mbox{ for } L \neq 0,
    \end{cases}
    \end{align*}
    where the intersecting point $y^{\mathrm{EQ}}$ is uniquely determined for a given $L$, and it depends on a ray in the space $\mathcal{L}$ of conserved quantities.
\end{enumerate}
\end{thm}
A mathematical proof of this theorem will be shown in Sec. \ref{sec:proof}. 

From Theorem \ref{thm1_new}, we can derive claim \ref{thm2_new} as follows.
For the case 1 of claim \ref{thm2_new} in which $\varphi^*(y^{\mathrm{min}}) - \tilde{\Pi} = 0$ and $L=0$, Theorem \ref{thm1_new} states that the intersection $\mathcal{M}^{\mathcal{Y}}_{\mathrm{STO}}(L) \cap \mathcal{M}^{\mathcal{Y}}_{\mathrm{EQ}}(\tilde{\mu})$ consists of $y^{\mathrm{min}}$. From Eq. (\ref{ray_y}), this $y^{\mathrm{min}}$ is mapped to a ray $\mathfrak{r}^{\mathcal{Y}}(y^{\mathrm{min}})$ in the number space $\mathfrak{X}$. In addition, $\mathcal{M}^{\mathfrak{X}}_{\mathrm{STO}}(L=0)$ from Eq. (\ref{stoichiometric_manifold2}) indicates that $\alpha X \in \mathcal{M}^{\mathfrak{X}}_{\mathrm{STO}}(L=0)$ if $X \in \mathcal{M}^{\mathfrak{X}}_{\mathrm{STO}}(L=0)$ for arbitrary $\alpha > 0$. 
This means that all the points on the ray $\mathfrak{r}^{\mathcal{Y}}(y^{\mathrm{min}})$ is in $\mathcal{M}^{\mathfrak{X}}_{\mathrm{STO}}(L=0)$.
Thus, all the points on the ray are equilibrium states and the system converges to one of them. The state is not uniquely determined only by the entropy function and depends on the initial conditions and the functional form of $J(t)$. 
For the case 2 of claim \ref{thm2_new} in which $\varphi^*(y^{\mathrm{min}}) - \tilde{\Pi} < 0$ and $L\neq0$, Theorem \ref{thm1_new} states that the intersection $\mathcal{M}^{\mathcal{Y}}_{\mathrm{STO}}(L) \cap \mathcal{M}^{\mathcal{Y}}_{\mathrm{EQ}}(\tilde{\mu})$ exists as  $y^{\mathrm{EQ}}$. This also corresponds to a ray $\mathfrak{r}^{\mathcal{Y}}(y^{\mathrm{EQ}})$ in the number space $\mathfrak{X}$. 
Since $\mathcal{M}^{\mathfrak{X}}_{\mathrm{STO}}(L \neq 0)$ is an affine subspace which does not include the origin $X=0$, the intersection between $\mathcal{M}^{\mathfrak{X}}_{\mathrm{STO}}(L \neq 0)$ and an arbitrary ray is represented by a point. Thus, the equilibrium state is uniquely determined in this case. 

Theorem \ref{thm1_new} states the cases when the system converges to an equilibrium state in our claim \ref{thm1_new}.
In the next section, we will investigate the case when the system does not converge to an equilibrium state, that is  $\mathcal{M}^{\mathcal{Y}}_{\mathrm{STO}}(L) \cap \mathcal{M}^{\mathcal{Y}}_{\mathrm{EQ}}(\tilde{\mu}) = \emptyset$. 

\section{Form of the total entropy function}
\label{sec:entropy}

When the intersection $\mathcal{M}^{\mathcal{Y}}_{\mathrm{EQ}}(\tilde{\mu}) \cap \mathcal{M}^{\mathcal{Y}}_{\mathrm{STO}}(L)$ is empty, the system cannot converge to an equilibrium state, and it is expected that the system grows or shrinks. 
In this section, we observe if the system grows or shrinks by numerical plots of the total entropy function.

If $\mathcal{M}^{\mathfrak{X}}_{\mathrm{STO}}(L) \cap \mathcal{M}^{\mathfrak{X}}_{\mathrm{EQ}}(\tilde{T}, \tilde{\mu}) = \emptyset$, the total entropy function $\Sigma^{\mathrm{tot}}(\Xi)$ can not have the maximum inside of $\mathcal{M}^{\mathfrak{X}}_{\mathrm{STO}}(L)$ (see Sec. \ref{sec:thermodynamics}).
Thus, the functional form of $\Sigma^{\mathrm{tot}}(\Xi)$ has the following two possibilities: (1) $\Sigma^{\mathrm{tot}}(\Xi)$ is not bounded above. (2) $\Sigma^{\mathrm{tot}}(\Xi)$ is bounded above, but the supremum exists on the boundary of 
$\mathcal{M}^{\mathfrak{X}}_{\mathrm{STO}}(L)$. 
In the first case, we find $|\Xi(t)| \rightarrow \infty$ for $t \rightarrow \infty$, because the second law requires that the system must climb up the unbounded landscape of the total entropy function. 
Since we assume $\dim \mathrm{Ker}[S] = 0$, we get $|X| \rightarrow \infty$ for $|\Xi| \rightarrow \infty$. In addition, $\Omega(X) \rightarrow \infty$ for $|X| \rightarrow \infty$ (see Appendix \ref{appendix_volume}). Thus, if $\Sigma^{\mathrm{tot}}(\Xi)$ is not bounded above, the volume $\Omega$ grows. 
In the second case, we can further show that the supremum is possible only at the origin $X = 0$ (see Sec. \ref{sec:proof_theorem2}). Thus, the volume $\Omega$ is shrinking and finally vanishes. 

\begin{figure}
    \centering
\includegraphics[width=0.5\textwidth]{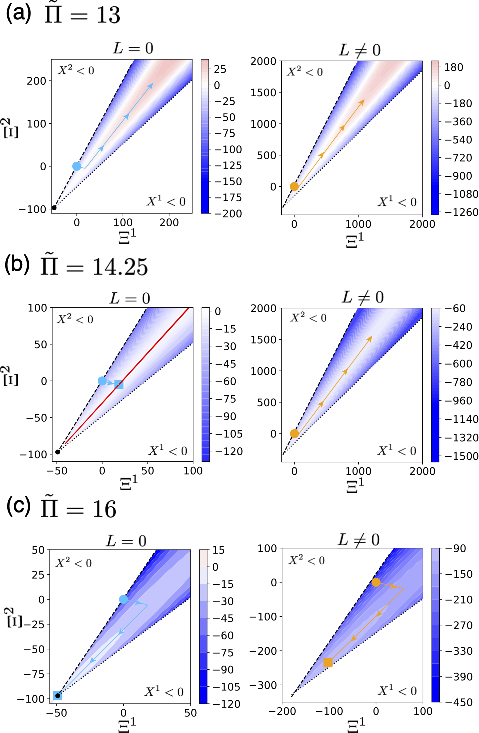}
    \caption{
    The entropy landscapes $\Sigma^{\mathrm{tot}}(\Xi)$ are shown for (a) $\tilde{\Pi} = 13 < \varphi^*(y^{\mathrm{min}}) = 14.25$, (b) $\tilde{\Pi} = \varphi^*(y^{\mathrm{min}})$ and (c) $\tilde{\Pi} = 16 > \varphi^*(y^{\mathrm{min}})$. The left and right panels show the cases $L=0$ and $L = 52 \neq 0$, respectively. The heatmaps represent the values of $\Sigma^{\mathrm{tot}}(\Xi)$. Here, we note that the domain of $(\Xi^1, \Xi^2)$ is restricted by $X \in \mathbb{R}^{\mathcal{N}_X}_{>0}$. The boundaries at which $X^2=0$ and $X^1=0$ are represented by dashed and dotted black lines, respectively. The black circles in the left panels denote the origin $X = 0$. The time evolutions of the system are shown by light blue and orange curves for the two initial conditions: (left) $X_0 = (48, 1, 48)$ with $L=0$ and (right) $X_0 = (148, 25, 200)$ with $L = 52$. 
    (a) The heatmaps in both panels indicate unbounded landscapes and the system goes to infinity. (b) For $L=0$, $\Sigma^{\mathrm{tot}}(\Xi)$ has maxima which form the red ray. The system converges to a point (the light blue square) on the ray. For $L \neq 0$, the heatmap shows unbounded landscape and the system goes to infinity. (c) For $L=0$, $\Sigma^{\mathrm{tot}}(\Xi)$ has the supremum at the origin $X=0$ and the system is approaching this point. For $L\neq 0$, $\Sigma^{\mathrm{tot}}(\Xi)$ has the unique maximum (the orange square) and the system converges to the point. 
    See the caption in Fig. \ref{Timeevolution} for specific values of the parameters.}
    \label{fig:entropy}
\end{figure}

In the following part of this section, we numerically plot the landscape of the total entropy function to determine the fate of the system.  
In Fig. \ref{fig:entropy}, we plot the total entropy function $\Sigma^{\mathrm{tot}}(\Xi)$ in Eq. (\ref{Enttot_partialGP}) for our example setup (see \textit{Example 1} and \textit{Example 2}). 
Following Theorem \ref{thm1_new}, the intersection $\mathcal{M}^{\mathcal{Y}}_{\mathrm{STO}}(L) \cap \mathcal{M}^{\mathcal{Y}}_{\mathrm{EQ}}(\tilde{\mu})$ is empty for the cases (1) $\varphi^*(y^{\mathrm{min}})-\tilde{\Pi} > 0$, (2) $\varphi^*(y^{\mathrm{min}})-\tilde{\Pi} = 0$ and  $L \neq 0$, and (3) $\varphi^*(y^{\mathrm{min}})-\tilde{\Pi} < 0$ and  $L = 0$,

In the case (1) $\tilde{\Pi} = 13 < \varphi^*(y^{\mathrm{min}}) = 14.25$, the total entropy function $\Sigma^{\mathrm{tot}}(\Xi)$ is not bounded above for both of $L = 0$ and $L\neq0$ (see Fig. \ref{fig:entropy}(a)). Indeed, $|\Xi(t)|\rightarrow \infty$ in the reaction dynamics of Eq. (\ref{system_dynamics_chem}). In the case (2) $\tilde{\Pi} = \varphi^*(y^{\mathrm{min}})$ and $L =52 \neq 0$, $\Sigma^{\mathrm{tot}}(\Xi)$ is still not bounded above and $|\Xi(t)|\rightarrow \infty$ (see the right panel of Fig. \ref{fig:entropy}(b)). In the case (3) $\tilde{\Pi} = 16 > \varphi^*(y^{\mathrm{min}})$ and $L = 0$, the supremum of $\Sigma^{\mathrm{tot}}(\Xi)$ is at the origin of the number space $\mathfrak{X}$. The system tends to approach the origin (see the left panel of Fig. \ref{fig:entropy}(c)).

The above numerical result demonstrates the following theorem.
\begin{thm} When $\mathcal{M}^{\mathcal{Y}}_{\mathrm{STO}}(L) \cap \mathcal{M}^{\mathcal{Y}}_{\mathrm{EQ}}(\tilde{\mu}) = \emptyset$, the landscape of $\Sigma^{\mathrm{tot}}(\Xi)$ is given as follows: 
\label{thm2_new}
\begin{enumerate}
    \item If $\varphi^{*}\left( y^{\mathrm{min}} \right)-\tilde{\Pi} > 0$, $\Sigma^{\mathrm{tot}}(\Xi)$ is not bounded above.
    \item If $\varphi^{*}( y^{\mathrm{min}} ) - \tilde{\Pi} = 0$ and $L\neq 0$, $\Sigma^{\mathrm{tot}}(\Xi)$ is not bounded above.
    \item If $\varphi^{*}( y^{\mathrm{min}} ) - \tilde{\Pi} < 0$ and $L=0$, the supremum of $\Sigma^{\mathrm{tot}}(\Xi)$ is at the origin of the number space $\mathfrak{X}$.
\end{enumerate}
\end{thm}
A mathematical proof of this theorem will be shown in Sec. \ref{sec:proof}. 

Theorem \ref{thm1_new} and \ref{thm2_new} lead to the claim \ref{claim1} and we find the fate of the system under isobaric condition.

Before closing this section, we confirm the claim \ref{claim2} in the left panel of Fig.\ref{fig:entropy}(b) and the right panel of Fig.\ref{fig:entropy}(c). In the former panel, the red ray represents the set of the maxima of $\Sigma^{\mathrm{tot}}(\Xi)$, which is $\mathfrak{r}^{\mathcal{Y}}(y^{\mathrm{min}})$.  We can see the system converges to a point on the ray. In the latter case, the maximum of $\Sigma^{\mathrm{tot}}(\Xi)$ is the unique point and the system converges to it. 

\section{Proof of the Theorems}
\label{sec:proof}

In this section, we prove Theorems \ref{thm1_new} and \ref{thm2_new}, and Corollary \ref{cor1}. 

\subsection{Proof of Theorem \ref{thm1_new}}
\label{sec:proof_theorem1}

\begin{figure}
    \centering
\includegraphics[width=0.5\textwidth]{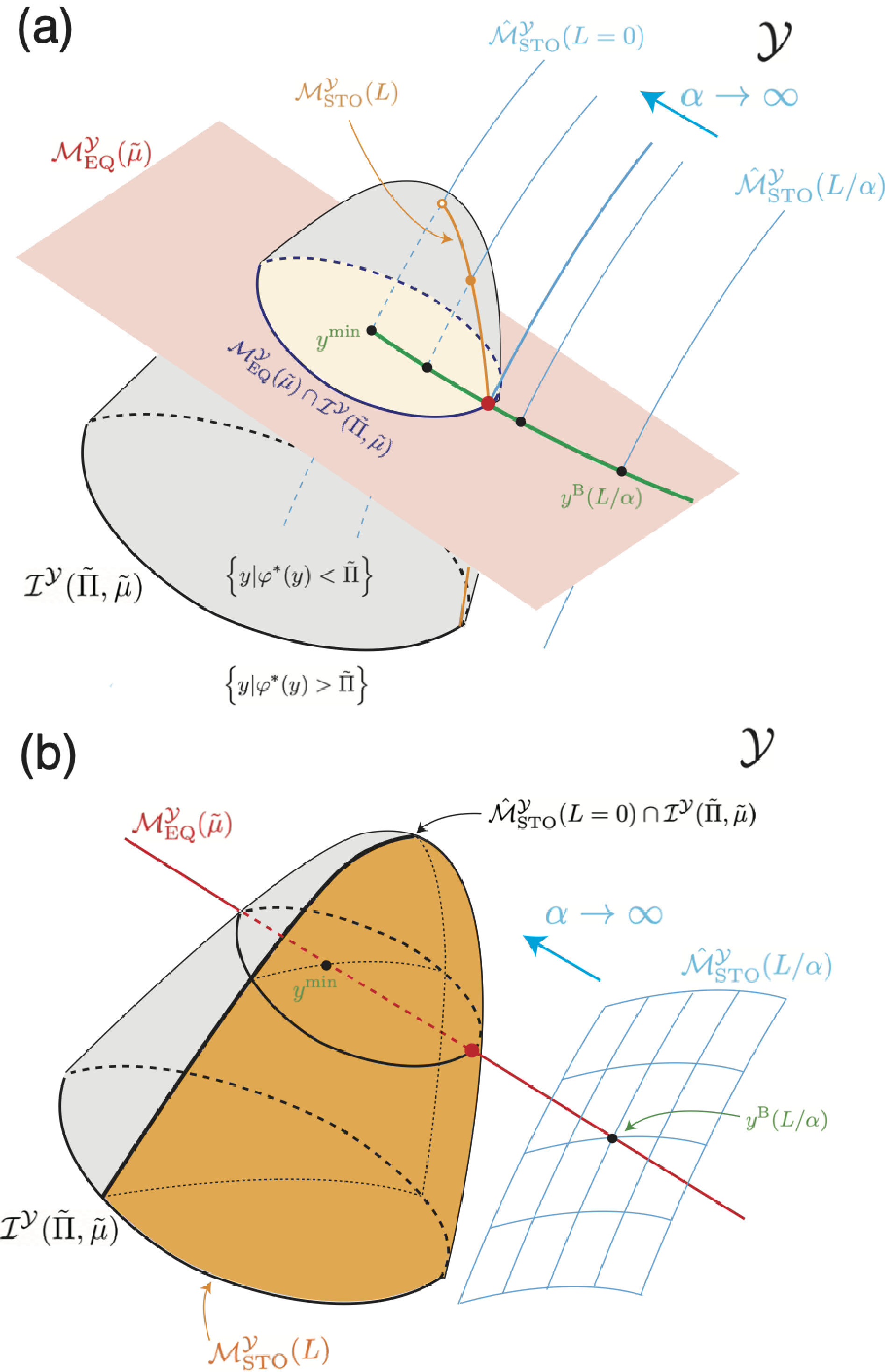}
\caption{Schematic illustration of the manifolds in $\mathcal{Y}$ for the case $\varphi^*(y^{\mathrm{min}}) - \tilde{\Pi} < 0$. In this illustration, we set $\dim \mathcal{Y} = \dim \mathrm{Ker}[S^T] + \dim \mathrm{Im}[S] = 3$. The conserved quantities $L$ are generally represented by $L = \{ L^l = U^l_i X^i_0 \}$ ($l = 1,...,\dim \mathrm{Ker}[S^T]$). The panels (a) and (b) represent the case for $\dim \mathrm{Ker}[S^T] = 2$ and $\dim \mathrm{Im}[S] = 1$, and the case for $\dim \mathrm{Ker}[S^T] = 1$ and $\dim \mathrm{Im}[S] = 2$, respectively. 
We note that $\dim \mathcal{M}^{\mathcal{Y}}_{\mathrm{EQ}}(\tilde{\mu}) = \dim \mathrm{Ker}[S^T]$ and $\dim  \hat{\mathcal{M}}^{\mathcal{Y}}_{\mathrm{STO}}(L/\alpha) = \dim \mathrm{Im}[S]$. 
The dimensions in the case (b) correspond to the illustrative example of chemical reactions in Eqs. (\ref{Reactions_example}) and (\ref{S_O_example}), whereas those in the case (a) correspond to the one in Sec. I of the Supplementary material.
(a) In this case, $\dim \mathcal{M}^{\mathcal{Y}}_{\mathrm{EQ}}(\tilde{\mu}) = 2$ and $\dim  \hat{\mathcal{M}}^{\mathcal{Y}}_{\mathrm{STO}}(L/\alpha) = 1$. Thus,  $\mathcal{M}^{\mathcal{Y}}_{\mathrm{EQ}}(\tilde{\mu})$ is denoted by the red plane, and the light blue curves represent $\hat{\mathcal{M}}^{\mathcal{Y}}_{\mathrm{STO}}(L/\alpha)$ for $\alpha \in (0, \infty)$. The gray surface is $\mathcal{I}^{\mathcal{Y}}(\tilde{\Pi}, \tilde{\mu})$. From Eq. (\ref{stoichiometric_manifold_y_isobaric_isochoric}), $\mathcal{M}^{\mathcal{Y}}_{\mathrm{STO}}(L)$ is given by the orange curve. Equation (\ref{birch_trajectory}) enables us to depict Birch's trajectory $\mathfrak{b}(L)$ as the green curve. From Eq. (\ref{birch_Lnon0}), the red circle denotes the intersecting point $\mathcal{M}^{\mathcal{Y}}_{\mathrm{STO}} (L) \cap \mathcal{M}^{\mathcal{Y}}_{\mathrm{EQ}} (\tilde{\mu})$ to which the system converges as the equilibrium state. 
(b) In this case, we consider $\dim \mathcal{M}^{\mathcal{Y}}_{\mathrm{EQ}}(\tilde{\mu}) = 1$ and $\dim \hat{\mathcal{M}}^{\mathcal{Y}}_{\mathrm{STO}}(L/\alpha) = 2$. Thus, $\mathcal{M}^{\mathcal{Y}}_{\mathrm{EQ}}(\tilde{\mu})$ is denoted by the red line, and $\hat{\mathcal{M}}^{\mathcal{Y}}_{\mathrm{STO}}(L/\alpha)$  represents the two dimensional manifold in light blue. Thus, $\mathcal{M}^{\mathcal{Y}}_{\mathrm{STO}}(L)$ is given by the orange surface. We note that Birch's trajectory $\mathfrak{b}(L)$ overlaps with the red line of $\mathcal{M}^{\mathcal{Y}}_{\mathrm{EQ}}(\tilde{\mu})$. The red circle represents the equilibrium state.}
    \label{fig:proof}
\end{figure}

We first provide a general proof valid for arbitrary dimensions. After concluding the proof, we schematically illustrate it for three dimensional space of $\mathcal{Y}$ in Fig. \ref{fig:proof}.

To investigate the existence of the intersection $\mathcal{M}^{\mathcal{Y}}_{\mathrm{STO}}(L) \cap \mathcal{M}^{\mathcal{Y}}_{\mathrm{EQ}}(\tilde{\mu})$, we introduce Birth's trajectory. 

Let $y^{\mathrm{B}}(\hat{L})$ represent the unique intersecting point between the \textit{isochoric} stoichiometric manifold $\hat{\mathcal{M}}^{\mathcal{Y}}_{\mathrm{STO}}(\hat{L})$ in Eq. (\ref{sto_isochoric_y}) and the equilibrium manifold $\mathcal{M}^{\mathcal{Y}}_{\mathrm{EQ}}(\tilde{\mu})$ in Eq. (\ref{eq_manifold_y_outline2}):
\begin{align}
\left\{ y^{\mathrm{B}}\left(\hat{L}\right) \right\} := \hat{\mathcal{M}}^{\mathcal{Y}}_{\mathrm{STO}}\left( \hat{L} \right) \cap \mathcal{M}^{\mathcal{Y}}_{\mathrm{EQ}}\left(\tilde{\mu}\right). \label{birch_point}
\end{align}
The uniqueness of $y^{\mathrm{B}}(\hat{L})$ is guaranteed by Birch's theorem and the point is known as Birth's point \cite{pachter2005algebraic, craciun2009toric, craciun2019, sughiyama2022hessian,kobayashi2022kinetic} (see Appendix \ref{appendix:birch}).
We define Birch's trajectory by a collection of Birch's point along a ray in the space of conserved quantities. 
For given $L \neq 0$, it is defined as  
\begin{align}
\mathfrak{b}(L) := \Set{ y^{\mathrm{B}}\left( \frac{L}{\alpha} \right) | \alpha > 0 }. \label{birch_trajectory}
\end{align}
We find that $y^{\mathrm{B}}(L/\alpha) \rightarrow y^{\mathrm{B}}(0)$ for $\alpha \rightarrow \infty$. Furthermore, from Eq. (\ref{Appendix_Birch_theorem_y_x0}) in Appendix \ref{appendix:birch}, we get 
\begin{align}
\{ y^{\mathrm{B}}(0) \} = \hat{\mathcal{M}}^{\mathcal{Y}}_{\mathrm{STO}}\left( \hat{L} = 0 \right) \cap \mathcal{M}^{\mathcal{Y}}_{\mathrm{EQ}}\left(\tilde{\mu}\right) = \{ y^{\mathrm{min}} \}.\label{birch_ymin}
\end{align}
Thus, the point $y^{\mathrm{min}}$ is located at one of the end points of Birch's trajectory, but is not included in the trajectory. 

For this Birch's trajectory, the following lemma is satisfied.
\begin{lemma}
\label{lemma2}
$\varphi^*(y)$ is a strictly increasing function along Birch's trajectory from the starting point $y^{\mathrm{min}}$.
\end{lemma}
A proof of this lemma is shown in Appendix \ref{appendix_lemma}.

By employing Birch's trajectory, we can represent the intersection for $L\neq0$ as 
\begin{align}
&\mathcal{M}^{\mathcal{Y}}_{\mathrm{STO}}(L) \cap \mathcal{M}^{\mathcal{Y}}_{\mathrm{EQ}}(\tilde{\mu}) \notag \\ 
& = \bigcup_{\alpha > 0} \hat{\mathcal{M}}^{\mathcal{Y}}_{\mathrm{STO}}\left( \frac{L}{\alpha} \right) \cap \mathcal{I}^{\mathcal{Y}}(\tilde{\Pi}, \tilde{\mu}) \cap \mathcal{M}^{\mathcal{Y}}_{\mathrm{EQ}}(\tilde{\mu}) \notag \\
& = \mathfrak{b}(L) \cap \mathcal{I}^{\mathcal{Y}}(\tilde{\Pi}, \tilde{\mu}),\label{birch_Lnon0}
\end{align}
where we use Eqs. (\ref{stoichiometric_manifold_y_isobaric_isochoric}), (\ref{birch_point}), and (\ref{birch_trajectory}). For $L=0$, the intersection is 
\begin{align}
&\mathcal{M}^{\mathcal{Y}}_{\mathrm{STO}}(L=0) \cap \mathcal{M}^{\mathcal{Y}}_{\mathrm{EQ}}(\tilde{\mu}) \notag \\ 
& = \hat{\mathcal{M}}^{\mathcal{Y}}_{\mathrm{STO}}(L=0) \cap \mathcal{I}^{\mathcal{Y}}(\tilde{\Pi}, \tilde{\mu}) \cap \mathcal{M}^{\mathcal{Y}}_{\mathrm{EQ}}(\tilde{\mu}) \notag \\
& = \{ y^{\mathrm{min}} \} \cap \mathcal{I}^{\mathcal{Y}}(\tilde{\Pi}, \tilde{\mu}),\label{birch_L0}
\end{align}
where we use Eqs. (\ref{intersection_x0}) and (\ref{birch_ymin}).

From Lemma \ref{lemma2}, we can determine the existence of the intersection $\mathcal{M}^{\mathcal{Y}}_{\mathrm{STO}}(L) \cap \mathcal{M}^{\mathcal{Y}}_{\mathrm{EQ}}(\tilde{\mu})$ by the position of the starting point $y^{\mathrm{min}}$.
When $\varphi^*(y^{\mathrm{min}}) - \tilde{\Pi} > 0$, the starting point $y^{\mathrm{min}}$ is located in the superlevel set: $\{ y | \varphi^*(y) > \tilde{\Pi} \}$. Taking into account the fact that $\mathcal{I}^{\mathcal{Y}}(\tilde{\Pi}, \tilde{\mu})$ is the level set, the intersecting point $\mathcal{M}^{\mathcal{Y}}_{\mathrm{STO}}(L) \cap \mathcal{M}^{\mathcal{Y}}_{\mathrm{EQ}}(\tilde{\mu})$ does not exist for any $L$. 
When $\varphi^*(y^{\mathrm{min}}) - \tilde{\Pi} = 0$, the starting point  $y^{\mathrm{min}}$ is on $\mathcal{I}^{\mathcal{Y}}(\tilde{\Pi}, \tilde{\mu})$. For $L=0$, the intersecting point uniquely exists as $y^{\mathrm{min}}$ because of Eq. (\ref{birch_L0}). For $L\neq 0$, Lemma \ref{lemma2} indicates $\varphi^*(y) > \varphi^*(y^{\mathrm{min}}) = \tilde{\Pi}$ at any point $y$ on Birch's trajectory $\mathfrak{b}(L)$. Thus, from Eq. (\ref{birch_Lnon0}), the intersecting point does not exist. 
When $\varphi^*(y^{\mathrm{min}}) - \tilde{\Pi} < 0$, the starting point $y^{\mathrm{min}}$ is located in the sublevel set: $\{ y | \varphi^*(y) < \tilde{\Pi} \}$. For $L=0$, the intersecting point does not exist. For $L\neq 0$, the intersecting point uniquely exists. 

This concludes the proof of Theorem $\ref{thm1_new}$. 

In Fig. \ref{fig:proof}, we provide schematic illustrations in three dimensional space of $\mathcal{Y}$ for the case $\varphi^*(y^{\mathrm{min}}) - \tilde{\Pi} < 0$. To demonstrate the generality of the proof, we visualize two cases with $\dim \mathrm{Ker}[S^T] = 2$ and $\dim \mathrm{Ker}[S^T] = 1$ in the panels (a) and (b), respectively.
From Eq. (\ref{conserved_quantity}), the case in (a) has two conserved quantities, whereas the case in (b) has one conserved quantity. 
We also have $\dim \mathrm{Im}[S] = 1$ and $\dim \mathrm{Im}[S] = 2$ for the two cases, respectively, because $\dim \mathcal{Y} = \dim \mathrm{Ker}[S^T] + \dim \mathrm{Im}[S]$ and $\dim \mathcal{Y} = 3$. The dimensions for the example in Eqs (\ref{Reactions_example}) and (\ref{S_O_example}) correspond to the case (b), whereas the case (a) corresponds to the one investigated in Sec. I of the Supplementary material. 

In the former case (Fig. \ref{fig:proof}(a)), the equilibrium manifold $\mathcal{M}^{\mathcal{Y}}_{\mathrm{EQ}}(\tilde{\mu})$ is represented by the two dimensional plane in red, whereas the \textit{isochoric} stoichiometric manifold $\hat{\mathcal{M}}^{\mathcal{Y}}_{\mathrm{STO}}(\hat{L})$
is given by a one-dimensional curve in light blue because $\dim \mathcal{M}^{\mathcal{Y}}_{\mathrm{EQ}}(\tilde{\mu}) = \dim \mathrm{Ker}[S^T]$ and $\dim  \hat{\mathcal{M}}^{\mathcal{Y}}_{\mathrm{STO}}(L/\alpha) = \dim \mathrm{Im}[S]$.
Accordingly, the black circles represent the intersecting points $y^{\mathrm{B}}(\hat{L})$ in Eq. (\ref{birch_point}) for $\hat{L} = L/\alpha$ with varying $\alpha$. The green curve indicates Birch's trajectory $\mathfrak{b}(L)$ in Eq. (\ref{birch_trajectory}), and one of the end points is located at $y^{\mathrm{min}}$ in Eq. (\ref{birch_ymin}). The isobaric manifold $\mathcal{I}^{\mathcal{Y}}(\tilde{\Pi}, \tilde{\mu})$ is represented by the gray surface, which is a level set for $\varphi^*(y)$. When $\varphi^*(y^{\mathrm{min}}) - \tilde{\Pi} < 0$, $y^{\mathrm{min}}$ is in the sublevel set $\{y|\varphi^{*}\left(y\right) < \tilde{\Pi} \}$. The stoichiometric manifold $\mathcal{M}^{\mathcal{Y}}_{\mathrm{STO}}(L)$ is given by the orange curve from Eq. (\ref{stoichiometric_manifold_y_isobaric_isochoric}). Thus, the intersection $\mathcal{M}^{\mathcal{Y}}_{\mathrm{STO}}(L) \cap \mathcal{M}^{\mathcal{Y}}_{\mathrm{EQ}}(\tilde{\mu})$ is represented by the red circle, which is also the intersecting point between $\mathfrak{b}(L)$ and $\mathcal{I}^{\mathcal{Y}}(\tilde{\Pi}, \tilde{\mu})$ in Eq. (\ref{birch_Lnon0}). 

In the latter case (Fig. \ref{fig:proof}(b)), $\mathcal{M}^{\mathcal{Y}}_{\mathrm{EQ}}(\tilde{\mu})$ is represented by the red line, whereas $\hat{\mathcal{M}}^{\mathcal{Y}}_{\mathrm{STO}}(\hat{L})$
is given by the two dimensional surface in light blue. The intersecting point $y^{\mathrm{B}}(\hat{L})$ is indicated by the black circle in Eq. (\ref{birch_point}) for $\hat{L} = L/\alpha$ with a given $\alpha$.
In this case, the Birch's trajectory $\mathfrak{b}(L)$ in Eq. (\ref{birch_trajectory}) overlaps with the red line, and one of the end points is located at $y^{\mathrm{min}}$ in Eq. (\ref{birch_ymin}). The stoichiometric manifold $\mathcal{M}^{\mathcal{Y}}_{\mathrm{STO}}(L)$ is given by the orange surface from Eq. (\ref{stoichiometric_manifold_y_isobaric_isochoric})\footnote{Similar to Fig. \ref{fig:3D}(d), the isobaric manifold $\mathcal{I}^{\mathcal{Y}}(\tilde{\Pi}, \tilde{\mu})$ is partitioned by $L > 0$ and $L < 0$ in this case.}. 
Thus, the intersection $\mathcal{M}^{\mathcal{Y}}_{\mathrm{STO}}(L) \cap \mathcal{M}^{\mathcal{Y}}_{\mathrm{EQ}}(\tilde{\mu})$ is represented by the red circle.

\subsection{Proof of Theorem \ref{thm2_new}}
\label{sec:proof_theorem2}

If $\mathcal{M}^{\mathfrak{X}}_{\mathrm{STO}}(L) \cap \mathcal{M}^{\mathfrak{X}}_{\mathrm{EQ}}(\tilde{T}, \tilde{\mu}) = \emptyset$, the system never relaxes to the equilibrium state and have the following two possibilities: 
(1) $\Sigma^{\mathrm{tot}}(\Xi)$ is not bounded above, and $|\Xi(t)| \rightarrow \infty$ with time. As we noted in Sec. \ref{sec:entropy}, the volume $\Omega$ grows in this setup. 
(2) $\Sigma^{\mathrm{tot}}(\Xi)$ is bounded above and the supremum exists on the boundary of 
$\mathcal{M}^{\mathfrak{X}}_{\mathrm{STO}}(L)$. 

First, we consider the case $L\neq 0$, where $\mathcal{M}^{\mathfrak{X}}_{\mathrm{STO}}(L \neq 0)$ does not contact with the origin. In this case, $\Sigma^{\mathrm{tot}}(\Xi)$ is not bounded above. To prove that, we will show that the supremum of $\Sigma^{\mathrm{tot}}(\Xi)$ does not exist on the boundary of $\mathcal{M}^{\mathfrak{X}}_{\mathrm{STO}}(L\neq0)$ as follows. 

The extent of reaction $\Xi$ provides a coordinate on the stoichiometric manifold $\mathcal{M}^{\mathfrak{X}}_{\mathrm{STO}}(L)$. Since $\mathfrak{X} = \mathbb{R}_{>0}^{\mathcal{N}_{X}}$, the domain of $\Xi$ has a boundary. We denote the coordinate of a point on a boundary by $\Xi_{B}$. Here, at least, one component of the vector $X_{B} := X_0 + S\Xi_{B}$ is zero.
We define the index set by $\mathfrak{I}_B(X_B) = \{ i | X^i_B = 0 \}$. 
Since the origin $X = 0$ is excluded, $\mathfrak{I}_B(X_B)$ never has all the indices. 
Also, the density $x_B = X_B/\Omega(X_B)$ has zero components as $x^i_B = 0$ for $i \in \mathfrak{I}_B(X_B)$.

To investigate the gradient of $\Sigma^{\mathrm{tot}}(\Xi)$ on the boundary, we calculate the thermodynamic force as 
\begin{align}
    f_r(\Xi) := \frac{\partial \Sigma^{\mathrm{tot}}}{\partial \Xi^r} = -\frac{1}{\tilde{T}} \left\{ \partial_i \varphi(x(X_0+S\Xi))S^i_r + \tilde{\mu}_m O^m_r \right\}.\label{force_proof}
\end{align}
If the supremum of $\Sigma^{\mathrm{tot}}(\Xi)$ does not exist on the boundary, we can find a direction from the boundary to the interior of $\mathcal{M}^{\mathfrak{X}}_{\mathrm{STO}}(L)$ such that $\Sigma^{\mathrm{tot}}(\Xi)$ increases.
To be more precise, by using $f_r(\Xi)$, this statement is mathematically phrased as follows:  
\textit{If the supremum of $\Sigma^{\mathrm{tot}}(\Xi)$ does not exist on the boundary, a vector $j = \{ j^r \}$ exists such that $f_r(\Xi_B) j^r > 0$ and $S^i_r j^r > 0$ for all the indices $i \in \mathfrak{I}_{B}$.}
In the next paragraph, we prove the above statement.

From the assumptions to $\varphi^*(y)$ in Sec. \ref{sec:proof1}, for every $i$, $y_i = \partial_i \varphi(x) \rightarrow -\infty$ when $x^i \rightarrow 0$. Thus, $\partial_i \varphi(x_B) \rightarrow -\infty$ for all $i\in \mathfrak{I}_{B}$. 
Choose a vector $j$ satisfying $S^i_r j^r > 0$ for $i \in \mathfrak{I}_{B}$. Then, we can show that, when $\Xi \rightarrow \Xi_B$,  $f_r (\Xi) j^r \rightarrow \infty$ from Eq. (\ref{force_proof}). 
This indicates that the supremum of $\Sigma^{\mathrm{tot}}(\Xi)$ does not exist on the boundary.

Second, we consider the case $L=0$. By employing rays on $\mathcal{M}^{\mathfrak{X}}_{\mathrm{STO}}(L=0)$, we investigate whether $\Sigma^{\mathrm{tot}}(X)$ is bounded above or not. 

Since we have assumed $\dim \mathrm{Ker}[S] = 0$, we can rewrite $\Sigma^{\mathrm{tot}}(\Xi)$ as a function of $X$. 
Using a particular solution $y^{\mathrm{P}}$ to Eq. (\ref{simultaneous_equation_outline}), we get
\begin{align}
     \Sigma^{\mathrm{tot}}(X) =& - \frac{1}{\tilde{T}} \biggl\{ \Omega(X) \varphi \left(\frac{X}{\Omega(X)} \right) + \tilde{\Pi} \Omega(X) \notag \\ &- y_i^{\mathrm{P}} \left( X^i - X^i_0 \right) \biggr\} + \mathrm{const}, \notag \\
    =& \frac{\Omega(X)}{\tilde{T}} K^{\mathcal{Y}}(y^{\mathrm{P}}; y(X)) - \frac{y^{\mathrm{P}}_i X^i_0}{\tilde{T}} + \mathrm{const},\label{enttot_form}
\end{align}
where 
\begin{equation}
    K^{\mathcal{Y}}(y^{\mathrm{P}}; y) := \varphi^*(y^{\mathrm{P}}) - \tilde{\Pi} - \mathcal{D}^{\mathcal{Y}} \left[ y^{\mathrm{P}} || y \right],  \label{kappa}
\end{equation}
(see Appendix \ref{appendix_psolution} for the details of the calculation).
Here, $\mathcal{D}^{\mathcal{Y}}[y^{\mathrm{P}}||y]$ is the Bregman divergence \cite{shima2007geometry,amari2000methods,bregman1967relaxation} induced by $\varphi^*(y)$, which is defined as 
\begin{equation}
    \mathcal{D}^{\mathcal{Y}}[y||y'] = \varphi^*(y)-\varphi^*(y')- \partial^i \varphi^*(y') \left\{ y_i - y'_i\right\}, \label{bregman}
\end{equation}
and $y(X) = \partial \varphi \circ \rho_{\mathcal{X}}(X)$.
Although Eq. (\ref{enttot_form}) is apparently a function of the particular solution $y^{\mathrm{P}}$, the value of $\Sigma^{\mathrm{tot}}$ does not depend on the choice of $y^{\mathrm{P}}$ as shown in Appendix \ref{appendix_ypchoice}. 
Since only the first term $\Omega(X) K^{\mathcal{Y}}(y^{\mathrm{P}}; y(X))/\tilde{T}$ depends on $X$, 
we consider its landscape. 

The stoichiometric manifold with $L=0$, $\mathcal{M}^{\mathfrak{X}}_{\mathrm{STO}}(L=0)$, can be described by the collection of rays in $\mathfrak{X}$ because it contacts with the origin $X=0$. 
Since a ray in the number space $\mathfrak{X}$ corresponds to a point in the chemical potential space $\mathcal{Y}$ as in Eq. (\ref{ray_y}), 
the value of $K^{\mathcal{Y}}(y^{\mathrm{P}}; y)$ is constant on each ray. 
In addition, $\Omega(X)$ is increasing along each ray from the origin.
Therefore, by considering the sign of $K^{\mathcal{Y}}(y^{\mathrm{P}}; y)$, we can determine if $\Sigma^{\mathrm{tot}}(X)$ increases along each ray in $\mathcal{M}^{\mathfrak{X}}_{\mathrm{STO}}(L=0)$. 
If $K^{\mathcal{Y}}(y^{\mathrm{P}}; y)$ is positive for a ray, $\Sigma^{\mathrm{tot}}(X)$ increases along the ray. 
If $K^{\mathcal{Y}}(y^{\mathrm{P}}; y)=0$, $\Sigma^{\mathrm{tot}}(X)$ is constant along the ray. 
If $K^{\mathcal{Y}}(y^{\mathrm{P}}; y)$ is negative for a ray, $\Sigma^{\mathrm{tot}}(X)$ decreases along the ray. 

When $\varphi^*(y^{\mathrm{min}}) - \tilde{\Pi} > 0$, we can find a point $y \in \mathcal{M}^{\mathcal{Y}}_{\mathrm{STO}}(L=0)$ such that $K^{\mathcal{Y}}(y^{\mathrm{P}}; y)$ is positive (see Appendix \ref{appendix_proof}). 
If such a point $y \in \mathcal{M}^{\mathcal{Y}}_{\mathrm{STO}}(L=0)$ exists, we can consider a corresponding ray in $\mathfrak{X}$ to the point $y$.
Along the ray, $\Sigma^{\mathrm{tot}}(X)$ increases and, thus, is not bounded above.
When $\varphi^*(y^{\mathrm{min}}) - \tilde{\Pi} = 0$, $K^{\mathcal{Y}}(y^{\mathrm{P}}; y)$ is negative for any $y \in \mathcal{M}^{\mathcal{Y}}_{\mathrm{STO}}(L=0)$ except $y = y^{\mathrm{min}}$, and $K^{\mathcal{Y}}(y^{\mathrm{P}}; y^{\mathrm{min}}) = 0$. Thus, the maxima of $\Sigma^{\mathrm{tot}}(X)$ forms the ray corresponding to $y^{\mathrm{min}}$, that is $\mathfrak{r}^{\mathcal{Y}}(y^{\mathrm{min}})$ in Eq. (\ref{ray_y}). 
When $\varphi^*({y^{\mathrm{min}}}) - \tilde{\Pi} < 0$, we can show that any $y\in \mathcal{M}^{\mathcal{Y}}_{\mathrm{STO}}(L=0)$ gives negative $K^{\mathcal{Y}}(y^{\mathrm{P}}; y)$ (see Appendix \ref{appendix_proof2}). 
Thus, $\Sigma^{\mathrm{tot}}(X)$ decreases along any ray and the supremum is located at the origin $X=0$. 

Summarizing the cases for $L\neq0$ and $L=0$, we obtain Theorem \ref{thm2_new}.

\subsection{Proof of Corollary \ref{cor1}}
\label{sec:proof_corollaries}

When $S$ is regular, the accessible region covers the whole space of $\mathfrak{X}$, that is $\mathcal{M}^{\mathfrak{X}}_{\mathrm{STO}}=\mathfrak{X}$. Thus, $\mathcal{M}^{\mathfrak{X}}_{\mathrm{STO}}$ can be described by the collection of rays in $\mathfrak{X}$. This situation coincides with the one for $L=0$ in the previous subsection. Therefore, by employing a similar proof, we can classify the fate of the system as Corollary $\ref{cor1}$ (see also \cite{sughiyama2022chemical}).

\section{Summary and Discussions}
\label{sec:summary}

In this work, we have considered chemical thermodynamics of growing CRSs with stoichiometric conservation laws in which the stoichiometric matrix $S$ has a nontrivial left kernel space. 
By clarifying the conditions for the fate of the system to grow, shrink, or equilibrate, we show that the existence of conservation laws qualitatively alter the fate of the system. It is emphasized again that our results are derived by general thermodynamic and stoichiometric structures without assuming any specific thermodynamic potentials or reaction kinetics; i.e., they are obtained based solely on the second law of thermodynamics and conservation laws.

Since we are mainly interested in the growth of the system, we have assumed the productivity of the stoichiometric matrix $S$: $\mathrm{Im}[S] \cap \mathbb{R}_{> 0}^{\mathcal{N}_{X}} \neq \emptyset$ (see Eq. (\ref{productiveS})).
This condition allows the time evolution of $X(t)$ so as to perpetually increase the number of all confined chemicals while satisfying the conservation laws.  
In the following two paragraphs, we comment on the fate of the system with the other classes of the stoichiometric matrix. Here, the complement of the productive $S$ is defined as $\mathrm{Im}[S] \cap \mathbb{R}_{> 0}^{\mathcal{N}_{X}} = \emptyset$. We divide it into two cases: (1) $\mathrm{Im}[S] \cap \mathbb{R}_{\geq 0}^{\mathcal{N}_{X}} = \emptyset$, and (2) $\mathrm{Im}[S] \cap \mathbb{R}_{\geq 0}^{\mathcal{N}_{X}} \neq \emptyset$ and $\mathrm{Im}[S] \cap \mathbb{R}_{> 0}^{\mathcal{N}_{X}} = \emptyset$ \footnote{Note the equality in $\geq 0$ of $\mathbb{R}_{\geq 0}^{\mathcal{N}_{X}}$.}.

In the first case, (1) $\mathrm{Im}[S] \cap \mathbb{R}_{\geq 0}^{\mathcal{N}_{X}} = \emptyset$,  
a vector $\mathbf{v}$ exists in $\mathrm{Ker}[S^T]$ such that all the components of $\mathbf{v}$ are positive. 
As a result, the conserved quantity $\mathbf{v}_i X^i = \mathbf{v}_i X^i_0$ implies that the total mass of the confined chemicals are conserved, and the system must converge to the equilibrium state, i.e., the growth of the system is not possible. We refer to this type of $S$ as non-productive.  A numerical example for this case is shown in Sec. III of the Supplementary material. 

In the second case, (2) $\mathrm{Im}[S] \cap \mathbb{R}_{\geq 0}^{\mathcal{N}_{X}} \neq \emptyset$ and $\mathrm{Im}[S] \cap \mathbb{R}_{> 0}^{\mathcal{N}_{X}} = \emptyset$, a vector $\mathbf{v}$ exists in $\mathrm{Ker}[S^T]$ such that all the components of $\mathbf{v}$ are non-negative, but $\mathrm{Ker}[S^T]$ does not have a vector $\mathbf{v}'$ such that all the components of $\mathbf{v}'$ are positive. 
This means that the mass conservation law exists for a proper subset of the confined chemicals. 
Thus, the volume growth of the system can still be allowed with the increase of the confined chemicals that do not participate in the mass conservation law. 
Although such a situation may not be biologically relevant, we refer to this type of $S$ as semi-productive.
A numerical example for this case is shown in Sec. IV of the Supplementary material. 
According to the example, it is expected that the claims $\ref{claim1}$ and $\ref{claim2}$ still hold for the semi-productive cases.

We can possibly correlate our theoretical findings with empirical observations. Our findings suggest that the fate changes with the initial conditions (conserved quantities $L$) even in the same environment. We especially find that, as shown in Table \ref{tab:table1}, a CRS does not vanish and is more likely to grow when it has non-zero $L$ than when it has $L=0$ or a regular $S$. If we suppose a population of the CRSs with varying initial conditions, and consider their fates in a slowly changing environment, surviving CRSs can be selective depending on $L$, where a CRS with non-zero $L$ might be advantageous. Identifying a possible mechanism to survive in a varying environment is biologically essential, and our findings may help explain the outcome.

Nevertheless, several extensions of our theory remain to be essential as future work to understand actual experimental situations of growing protocells or biological cells. 

The first extension is to consider the case when the system may relax to a state that continuously produces entropy with constant volume, namely, the conventional nonequilibrium steady state (NESS) \cite{rao2016nonequilibrium,polettini2014irreversible,ge2016nonequilibrium,qian2005nonequilibrium,craciun2019,perez2012chemical,craciun2009toric,horn1972general}. It is possible when the matrix $S$ has a nontrivial right kernel space, i.e., the assumption in Eq. (\ref{KerS0}) is not satisfied. It is a major challenge to understand how the growth of the system can be characterized and realized in such situations, and compatible with the NESS.

The second one is to consider a different hierarchy of the timescales in which the relaxation of the variables $(E, \Omega, N)$ may not be rapid compared to the timescale of chemical reactions. Although the present work assumes that the chemical reactions are slow (see the assumptions made above Eq. (\ref{Enttot_partialGP})), there may be cases in which the timescale of other processes is slow and/or comparable with that of the chemical reactions. A typical example is when we consider cell transport processes \cite{koch1997microbial}. In this case, the timescale of material exchanges between the system and the environment is relevant, and the processes may also couple with chemical reactions. In addition, the osmotic gradient (pressure difference) can play a driving force to change the volume, whereas isosmotic volume changes are also commonly known and widely investigated as a relevant process by alterations in intracellular solute content \cite{oneill1999physiological,strange2004cellular}. It is of interest to see how the conditions for the growth of the system change with relevant processes of the thermodynamic setup. 

The third one is to consider explicitly the effect of the membrane molecules. Although we neglect the tension of the membrane in this paper (see Fig. \ref{fig:1}), it could be relevant and changes with time because the membrane molecules themselves are produced and supplied by the CRSs in biological cells. It is important to investigate if and how the effect changes the results by extending our theoretical framework.

Alternatively, it would also be promising to apply our framework to fit into an experimental technique using membrane-free compartments such as droplets based on liquid-liquid phase separation \cite{alberti2019considerations}. 
Since its nature, a rigorous thermodynamic treatment would require an extended framework in which the entropy function is not strictly concave (see the assumption made below Eq. (\ref{ent_homogeneity})). 

In all the above extensions, the present work provides basic results to clarify the role of conservation laws to the growth of protocells and biological cells. 

\section{Acknowledgments}
The authors thank Dimitri Loutchko and Shuhei A. Horiguchi for discussions. This research is supported by JSPS KAKENHI Grant Numbers 19H05799, 21K21308, and 24K00542, and by JST CREST JPMJCR2011 and JPMJCR1927. Y. S. receives financial support from the Public\verb|\|Private R\&D Investment Strategic Expansion PrograM (PRISM) and programs for Bridging the gap between R\&D and the IDeal society (society 5.0) and Generating Economic and social value (BRIDGE) from Cabinet Office.


%

\clearpage

\appendix

\section{Calculation of the total entropy function}
\label{appendix_derivations}

The complete dynamics for the system is defined as
\begin{align}
    &\frac{dE}{dt} = J_E(t), \frac{d\Omega}{dt} = J_{\Omega}(t), \notag \\
    &\frac{dN^m}{dt} = O^m_r J^r(t) + J^m_D(t), \frac{dX^i}{dt} = S^i_r J^r(t),
    \label{system_dynamics}
\end{align}
where $J_E(t)$, $J_{\Omega}(t)$, $J_D(t) = \{ J^m_D(t) \}$ and $J(t) = \{ J^r(t) \}$ represent the energy, the volume, the chemical diffusion, and the chemical reaction fluxes, respectively (see Fig. \ref{fig:1}).
By contrast, the dynamics for the environment is given as 
\begin{equation}
    \frac{d\tilde{E}}{dt} = -J_E(t), \frac{d\tilde{\Omega}}{dt} = - J_{\Omega}(t), \frac{d\tilde{N}^m}{dt} = - J^m_D(t).
    \label{environment_dynamics}
\end{equation}

Since we have assumed that the time scale of the reactions is much slower than that of the others, we can analyze the dynamics, Eqs. (\ref{system_dynamics}) and (\ref{environment_dynamics}), by separating the slow one $J(t)$ from the fast ones $J_E(t), J_{\Omega}(t), J_D(t)$. 
Owing to this time-scale separation, the extensive variables $(E, \Omega, N)$ are rapidly relaxed to the values at the equilibrium state of the fast dynamics with a fixed $X$. They are computed by the thermodynamic variational problem:
\begin{align}
&\left(E(X), \Omega(X), N(X) \right) \notag \\ &= \arg \max_{E, \Omega, N} \left\{ \Sigma \left[ E, \Omega, N, X\right] - \frac{E}{\tilde{T}}
- \frac{\tilde{\Pi}}{\tilde{T}} \Omega + \frac{\tilde{\mu}_m}{\tilde{T}} N^m
\right\},
\end{align}
(see \cite{sughiyama2022hessian} and \cite{sughiyama2022chemical}).
If we use the densities $\sigma$, $\epsilon$ and $n$, this varational problem is equivalently written in
\begin{align}
&\left(\epsilon(X), \Omega(X), n(X) \right) \notag \\ &= \arg \max_{\epsilon, \Omega, n} \left\{ \Omega \left( \tilde{T}\sigma\left[ \epsilon, n, \frac{X}{\Omega} \right] - \epsilon - \tilde{\Pi} + \tilde{\mu}_m n^m \right)
\right\}.\label{appendix_evn}
\end{align}
By defining the partial grand potential density (see Eq. (\ref{varphi_x})), 
\begin{align}
\varphi(x) = \varphi \left[ \tilde{T}, \tilde{\mu}; x \right] := \min_{\epsilon,n} \left\{ \epsilon - \tilde{T}\sigma\left[ \epsilon, n, x \right] - \tilde{\mu}_m n^m \right\},\label{appendix_varphi}
\end{align}
and taking the maximization with respect to $\epsilon$ and $n$ in Eq. (\ref{appendix_evn}), we obtain
\begin{align}
\Omega(X) = \arg \min_{\Omega} \left\{ \Omega \varphi\left( \frac{X}{\Omega} \right) + \Omega \tilde{\Pi} \right\}.
\end{align}
This determines the volume $\Omega$ at $X$ (see Eq. (\ref{volume})).

By substituting $(\epsilon(X), \Omega(X), n(X))$ in Eq. (\ref{appendix_evn}) into Eqs. (\ref{system_dynamics}) and (\ref{environment_dynamics}), 
we get 
\begin{align}
    &\frac{dX^i}{dt} = S^i_r J^r(t), \frac{d\tilde{E}}{dt} = - \frac{d\Omega(X)\epsilon(X)}{dt},\notag \\
    &\frac{d\tilde{\Omega}}{dt} = - \frac{d\Omega(X)}{dt}, \frac{d\tilde{N}^m}{dt} = O^m_r J^r(t) - \frac{d\Omega(X) n^m(X)}{dt}. 
    \label{QEQ}
\end{align}
Integration of Eq. (\ref{QEQ}) leads to
\begin{align}
X_t &= X_0 + S \Xi_t, \notag \\
\tilde{E}(X_t) &= \tilde{E}(X_0) - \left\{ \Omega(X_t) \epsilon(X_t) - \Omega(X_0) \epsilon(X_0) \right\}, \notag \\
\tilde{\Omega}(X_t) &= \tilde{\Omega}(X_0) - \left\{ \Omega(X_t) - \Omega(X_0)\right\}, \notag \\
\tilde{N}(X_t) &= \tilde{N}(X_0) + O\Xi_t - \left\{ \Omega(X_t) n(X_t) - \Omega(X_0) n(X_0) \right\},\label{QEQ_integrate}
\end{align}
where $\Xi_t$ is the extent of reaction at time $t$.

The total entropy in Eq. (\ref{enttot}) is represented as 
\begin{align}
\Sigma^{\mathrm{tot}} =& \Omega(X_t) \sigma\left[ \epsilon(X_t), n(X_t), \frac{X_t}{\Omega(X_t)} \right] \notag \\ &+ \tilde{\Sigma}_{\tilde{T}, \tilde{\Pi}, \tilde{\mu}}\left[\tilde{E}(X_t), \tilde{\Omega}(X_t),\tilde{N}(X_t) \right].
\end{align}
By substituting Eq. (\ref{QEQ_integrate}) and employing the Taylor expansion for $\tilde{\Sigma}_{\tilde{T},\tilde{\Pi},\tilde{\mu}}$, we get
\begin{align}
&\Sigma^{\mathrm{tot}} = \Omega(X) \biggl[ \sigma\left[ \epsilon(X), n(X), \frac{X}{\Omega(X)} \right] \notag \\ & - \frac{\epsilon(X)}{\tilde{T}} - \frac{\tilde{\Pi}}{\tilde{T}} - \frac{\tilde{\mu}_m}{\tilde{T}} \left\{ O^m_r \frac{\Xi^r}{\Omega(X)} - n^m(X) \right\} \biggr] + \mathrm{const.},
\end{align}
where $X = X_0 + S\Xi$. Here, we used the thermodynamic relations: 
$\partial \tilde{\Sigma}_{\tilde{T},\tilde{\Pi},\tilde{\mu}}/\partial \tilde{E} = 1/\tilde{T}$, $\partial \tilde{\Sigma}_{\tilde{T},\tilde{\Pi},\tilde{\mu}}/\partial \tilde{\Omega} = \tilde{\Pi}/\tilde{T}$ and $\partial \tilde{\Sigma}_{\tilde{T},\tilde{\Pi},\tilde{\mu}}/\partial \tilde{N}^m = -\tilde{\mu}_m/\tilde{T}$.
Taking Eq. (\ref{appendix_varphi}) into account, we obtain
\begin{align}
    \Sigma^{\mathrm{tot}} = &- \frac{1}{\tilde{T}} \left\{ \Omega(X) \varphi \left( \frac{X}{\Omega(X)} \right) + \Omega(X) \tilde{\Pi} + \tilde{\mu}_m O^m_r \Xi^r \right\} \notag \\
    &+ \mathrm{const.}, 
\label{appendix_Enttot_partialGP}
\end{align}
which is Eq. (\ref{Enttot_partialGP}).

\section{Mass-action kinetics}
\label{appendix_numerical}

To numerically simulate the dynamics, Eq. (\ref{system_dynamics_chem}), we assume mass action kinetics for the flux function $J(t) = \{ J^1(t), J^2(t) \}$. 
For \textit{Example 1}, it is given as
\begin{align}
    J^1(t) &= w^1_+ \Omega \left( \frac{X^1}{\Omega} \right) \left( \frac{X^3}{\Omega}\right) \frac{N^1}{\Omega} - w^1_- \Omega \left( \frac{X^2}{\Omega} \right)^2, \notag \\ 
    J^2(t) &= w^2_+ \Omega \left( \frac{X^2}{\Omega}\right) - w^2_- \Omega \left(\frac{X^1}{\Omega}\right) \left(\frac{X^3}{\Omega}\right) \frac{N^2}{\Omega}, \label{massaction}
\end{align}
where $N = (N^1, N^2)$ denotes the number of $B = (B_1, B_2)$ in the system. The rate constants $w^r_+$ and $w^r_-$ satisfy the local detailed balance condition \cite{beard2008chemical,polettini2014irreversible,rao2016nonequilibrium,sughiyama2022hessian,kobayashi2022kinetic}: 
\begin{align}
    \log \frac{w^r_+}{w^r_-} = -\frac{1}{R\tilde{T}} \left\{ \nu_i^o(\tilde{T}) S_r^i + \mu_m^o(\tilde{T}) O^m_r \right\}.\label{localdetailedbalance}
\end{align}

For the ideal gas case, the density $N/\Omega$ of the open chemicals in the system is constant and fixed to that of the environment $\tilde{n}$.
Also, the volume $\Omega$ is given by the equation of state in Eq. (\ref{EqState}). Thus, we can rearrange Eq. (\ref{massaction}) as 
\begin{align}
    J^1(t) &= \hat{w}^1_+ \Omega \left( \frac{X^1}{\Omega} \right) \left(\frac{X^3}{\Omega}\right) - \hat{w}^1_- \Omega \left( \frac{X^2}{\Omega} \right)^2, \notag \\
    J^2(t) &= \hat{w}^2_+ \Omega \left( \frac{X^2}{\Omega} \right) 
    - \hat{w}^2_- \Omega \left( \frac{X^1}{\Omega}\right) \left(\frac{X^3}{\Omega}\right), \label{massaction2}
\end{align}
where we absorb the constant densities of the open chemicals into the rate constants as $\hat{w}^r_+$ and $\hat{w}^r_-$.
For these effective rate constants, the local detailed balance condition in Eq. (\ref{localdetailedbalance}) is written as 
\begin{align}
    \log \frac{\hat{w}^r_+}{\hat{w}^r_-} = -\frac{1}{R\tilde{T}} \left\{ \nu_i^o(\tilde{T}) S_r^i + \tilde{\mu}_m O^m_r \right\},\label{localdetailedbalance2}
\end{align}
where 
\begin{align}
\tilde{\mu}_m = \mu^o_m(\tilde{T}) + R\tilde{T} \log \tilde{n}^m,  
\end{align}
in Eq. (\ref{chemical_potential_idealgas}).
For our specific case, it is given as
\begin{align}
\frac{\hat{w}^1_+}{\hat{w}^1_-} = \frac{x^2_o x^2_o \tilde{n}^1}{x^1_o x^3_o n^1_o}, \hspace{5mm} \frac{\hat{w}^2_+}{\hat{w}^2_-} = \frac{x^1_o x^3_o n^2_o}{x^2_o \tilde{n}^2},\label{localdetailedbalance3}
\end{align}
where $x^i_o := e^{-\nu^o_i(\tilde{T})/R\tilde{T}}$ and $n^m_o := e^{-\mu^o_m(\tilde{T})/R\tilde{T}}$.

\section{The assumption for the pressure $\tilde{\Pi}$}
\label{appendix_min_pressure}

The assumption $\tilde{\Pi} > \Pi_{\mathrm{min}}$ is made to guarantee that, for a given $X$, the volume $\Omega(X)$ is uniquely determined and the system can relax to the equilibrium state of the fast dynamics. 

The volume $\Omega(X)$ is variationally determined in Eq. (\ref{volume}) as 
\begin{align}
    \Omega(X) = \arg \min_{\Omega} \left\{ \Omega \varphi\left( \frac{X}{\Omega} \right) + \tilde{\Pi} \Omega \right\}.\label{volume_variational_form_appendix}
\end{align}
For a given $X$, the critical equation is computed as 
\begin{align}
    h(\Omega) &:= \varphi \left(\frac{X}{\Omega} \right) - \frac{X^i}{\Omega} \partial_i \varphi\left( \frac{X}{\Omega} \right) + \tilde{\Pi} = 0.
    \label{volume_critical_eq_appendix}
\end{align}

The differentiation of $h(\Omega)$ is given as 
\begin{align}
    \frac{dh}{d\Omega} = \Omega^{-3} X^i \left[ \partial_i \partial_j \varphi \left( \frac{X}{\Omega} \right) \right] X^j.
\end{align}
Since $\varphi$ is strictly convex, its Hessian $\partial_i \partial_j \varphi$ is positive definite. Thus, the function $h(\Omega)$ is a strictly increasing function for $\Omega > 0$. 
In addition, $h(\Omega)$ is further calculated as 
\begin{align}
h(\Omega) = - \varphi^*\left( \partial \varphi \left( \frac{X}{\Omega} \right) \right) + \tilde{\Pi}.
\end{align}
If $\Omega \rightarrow 0$, $\partial \varphi(X/\Omega) \rightarrow \infty$ and 
$h(\Omega) \rightarrow -\infty$. By contrast, if $\Omega \rightarrow \infty$, $\partial \varphi(X/\Omega) \rightarrow -\infty$ and $h(\Omega) \rightarrow \tilde{\Pi} - \Pi_{\mathrm{min}}$, because $\varphi^*(y)$ is strictly increasing and its infimum is given by $\Pi_{\mathrm{min}}$. 
Thus, if $\tilde{\Pi} > \Pi_{\mathrm{min}}$, the critical equation has a unique solution with respect to $\Omega$, and, therefore, the system can relax to the equilibrium state of the fast dynamics. 

If $\tilde{\Pi} \leq \Pi_{\mathrm{min}}$, the critical equation $h(\Omega) = 0$ in Eq. (\ref{volume_critical_eq_appendix}) does not have a solution, and the volume $\Omega$ diverges from the variational form in Eq. (\ref{volume_variational_form_appendix}). 
This is consistent with our intuition; if the pressure $\tilde{\Pi}$ of the environment is smaller than the minimum pressure of the system, the pressures cannot be balanced and the volume of the system would diverge. 

\section{Derivation of Eq. (\ref{enttot_critical_eq})}
\label{appendix_criticaleq}

The total entropy function is given in Eq. (\ref{Enttot_partialGP}) as
\begin{align}
    \Sigma^{\mathrm{tot}}(\Xi) = &- \frac{1}{\tilde{T}} \left\{ \Omega(X) \varphi \left( \frac{X}{\Omega(X)} \right) + \Omega(X) \tilde{\Pi} + \tilde{\mu}_m O^m_r \Xi^r \right\} \notag \\
    &+ \mathrm{const},
    \label{Enttot_partialGP_appendix}
\end{align}
where $X = X_0 + S\Xi$. 
It is rewritten as a function of $\Omega$, $X$ and $\Xi$:
\begin{align}
    \bar{\Sigma}^{\mathrm{tot}}(\Omega, X, \Xi) = &- \frac{1}{\tilde{T}} \left\{ \Omega \varphi \left( \frac{X}{\Omega} \right) + \Omega \tilde{\Pi} + \tilde{\mu}_m O^m_r \Xi^r \right\} \notag \\
    &+ \mathrm{const}.
    \label{Enttot_partialGP_appendix2}
\end{align}
Its total derivative is calculated as
\begin{align}
d\bar{\Sigma}^{\mathrm{tot}} &= \frac{\partial \bar{\Sigma}^{\mathrm{tot}}}{\partial \Omega} d\Omega + \frac{\partial \bar{\Sigma}^{\mathrm{tot}}}{\partial X^i} dX^i + \frac{\partial \bar{\Sigma}^{\mathrm{tot}}}{\partial \Xi^r} d\Xi^r.
\label{Enttot_partialGP_appendix3}
\end{align}
The first coefficient $\partial \bar{\Sigma}^{\mathrm{tot}}/\partial \Omega$ vanishes at $\Omega = \Omega(X)$ because of Eq. (\ref{volume}). The second and third coefficients are computed as $\partial \bar{\Sigma}^{\mathrm{tot}}/\partial X^i = - \partial_i \varphi(X/\Omega)/\tilde{T}$ and $\partial \bar{\Sigma}^{\mathrm{tot}}/\partial \Xi^r = - \tilde{\mu}_m O^m_r/\tilde{T}$.
Since $dX^i = S^i_r d\Xi^r$, we obtain
\begin{align}
d\bar{\Sigma}^{\mathrm{tot}} &= - \frac{1}{\tilde{T}} \left\{ \partial_i \varphi\left( \frac{X}{\Omega(X)} \right)S^i_r + \tilde{\mu}_m O^m_r \right\}d\Xi^r.
\end{align}
This corresponds to Eq. (\ref{enttot_critical_eq}).

\section{Proof of $\Omega(X) \rightarrow \infty$ when $|X| \rightarrow \infty$}
\label{appendix_volume}

In this Appendix, we show that the volume $\Omega(X) \rightarrow \infty$ when $|X| \rightarrow \infty$. 
For a given $X$, the volume $\Omega(X)$ is variationally determined by Eq. (\ref{volume}).  

In the number space $\mathfrak{X}$, we can show that the volume function $\Omega(X)$ satisfies the homogeneity: $\Omega(\alpha X) = \alpha \Omega(X)$ for $\alpha > 0$. 
Thus, for a given $X_{\mathfrak{r}}$, we have 
\begin{align}
    \frac{\partial \Omega(\alpha X_{\mathfrak{r}})}{\partial \alpha} = \frac{\partial}{\partial \alpha} \left\{ \alpha \Omega(X_{\mathfrak{r}}) \right\} = \Omega(X_{\mathfrak{r}}).
\end{align}
This implies that the volume function $\Omega(X)$ increases with the rate $\Omega(X_{\mathfrak{r}})$ along the ray including $X_{\mathfrak{r}}$. 
Therefore, the volume $\Omega(X)$ goes to infinity for $|X| \rightarrow \infty$.

\section{Birch's theorem}
\label{appendix:birch}

In this Appendix, we introduce Birch's theorem and its extension.

Consider the following variational problem:
\begin{align}
\arg \min_y \Set{ \varphi^*(y) - y_i x^i_0 | y \in \mathcal{M}^{\mathcal{Y}}_{\mathrm{EQ}}(\tilde{\mu})}, \label{Appendix_Birch_theorem_y}
\end{align}
where $x_0 \in \mathcal{X} = \mathbb{R}_{>0}^{\mathcal{N}_{X}}$ is an arbitrary constant such that $Ux_0 = \hat{L}$ for a given $\hat{L}$; it corresponds to the initial condition in the main text.
Also, $\varphi^*(y)$ is a strictly increasing function as introduced in Sec. \ref{sec:proof1}.
From Eq. (\ref{veq_element_outline}), the variational problem is rewritten as 
\begin{align}
\arg \min_y \Set{ \varphi^*(y) - y_i x^i_0 | y \in y^{\mathrm{P}}+\mathrm{Im}[U^T]} = \{ y^{\mathrm{B}}(\hat{L}) \}.
\label{Appendix_Birch_theorem_y2}
\end{align}
The point $y^{\mathrm{B}}(\hat{L})$ to attain the minimum uniquely exists, because $\varphi^*(y)- y_i x^i_0$ is strictly convex and $x_0 > 0$. To be more precise, we have $\varphi^*(y)- y_i x^i_0 \rightarrow \infty$ as $|y|\rightarrow \infty$.
By the directional derivative, the left hand side of Eq. (\ref{Appendix_Birch_theorem_y2}) can be represented as 
\begin{align}
&\arg \min_y \Set{ \varphi^*(y) - y_i x^i_0 | y \in y^{\mathrm{P}}+\mathrm{Im}[U^T] } \notag \\
&= \Set{ y | \partial \varphi^*(y) \in x_0 + \mathrm{Ker} [U], y \in y^{\mathrm{P}} + \mathrm{Im}[U^T] } \notag \\
&= \Set{ y | U^l_i \partial^i \varphi^*(y) = \hat{L}^l, y \in y^{\mathrm{P}} + \mathrm{Im}[U^T] }.
\label{Appendix_Birch_theorem_y3}
\end{align}
Taking Eq. (\ref{sto_isochoric_y}) into account, we obtain 
\begin{align}
\hat{\mathcal{M}}^{\mathcal{Y}}_{\mathrm{STO}}(\hat{L}) \cap \mathcal{M}^{\mathcal{Y}}_{\mathrm{EQ}}(\tilde{\mu}) = \{ y^{\mathrm{B}}(\hat{L}) \},
\end{align}
which is called Birch's theorem. 

Next, we extend Birch's theorem to the case $x_0 \rightarrow 0$. 
In this case, Eq. (\ref{Appendix_Birch_theorem_y}) is formally rewritten as 
\begin{align}
\{ y^{\mathrm{min}} \} &= \arg \min_y \Set{ \varphi^*(y) | y \in \mathcal{M}^{\mathcal{Y}}_{\mathrm{EQ}}(\tilde{\mu})} \nonumber \\
&=\hat{\mathcal{M}}^{\mathcal{Y}}_{\mathrm{STO}}(\hat{L}=0) \cap \mathcal{M}^{\mathcal{Y}}_{\mathrm{EQ}}(\tilde{\mu}) = \{ y^{\mathrm{B}}(0) \}.
\label{Appendix_Birch_theorem_y_x0}
\end{align}
However, the existence of the point $y^{\mathrm{min}}$ is not trivial because $\varphi^*(y)$ is strictly increasing and $y^{\mathrm{min}}$ may goes to infinity. In the following, we show that the productivity of $S$ guarantees its existence.

When we write $\{ U_i^l \} = (\textbf{U}^1, \textbf{U}^2, ...)^T$, the productivity of $S$ guarantees that any vector $\textbf{U}^l$ has both positive and negative components.
The directional derivative on $\mathcal{M}^{\mathcal{Y}}_{\mathrm{EQ}}(\tilde{\mu})$ leads to 
\begin{align}
    \frac{\partial \varphi^*}{\partial \eta_l} = U^l_i \partial^i \varphi^*(y^{\mathrm{P}}+\eta U),\label{appendix_directional_derivative}
\end{align}
where $\eta$ is a coordinate of $\mathrm{Im}[U^T]$.
For a given $l$, as $\eta_l \rightarrow \infty$, the following holds: 
\begin{align*}
    \partial^i \varphi^*(y^{\mathrm{P}}+\eta U) \rightarrow \begin{cases}
    \infty &\mbox{ for } i \mbox{ } s.t. \mbox{ } U^l_i > 0,  \notag \\ 
    0 &\mbox{ for } i \mbox{ } s.t. \mbox{ } U^l_i < 0, \notag \\
    \partial^i \varphi^*(y^{\mathrm{P}}) &\mbox{ for } i \mbox{ } s.t. \mbox{ } U^l_i = 0.
    \end{cases}
\end{align*}
Therefore, by taking Eq. (\ref{appendix_directional_derivative}) and the productivity of $S$ into account, we conclude that, for any $l$, $\partial \varphi^*/\partial \eta_l \rightarrow \infty$ as $\eta_l \rightarrow \infty$. 
In a similar manner, we can show that, for any $l$, $\partial \varphi^*/\partial \eta_l \rightarrow - \infty$ as $\eta_l \rightarrow -\infty$. 
As a result, $\varphi^*(y)$ restricted on $\mathcal{M}^{\mathcal{Y}}_{\mathrm{EQ}}(\tilde{\mu})$ has a unique minimum. Note that, if $S$ is not productive, the above statement is not satisfied.

\section{Proof of Lemma \ref{lemma2}}
\label{appendix_lemma}
The starting point $y^{\mathrm{min}}$ of Birch's trajectory gives the infimum of $\varphi^*(y)$ in the trajectory because $y^{\mathrm{min}}$ gives the minimum $\varphi^*(y)$ in the equilibrium manifold $\mathcal{M}^{\mathcal{Y}}_{\mathrm{EQ}}(\tilde{\mu})$ (see Eq. (\ref{y_min_outline})) and any points in the trajectory is in the manifold $\mathcal{M}^{\mathcal{Y}}_{\mathrm{EQ}}(\tilde{\mu})$. Thus, the statement of the Lemma \ref{lemma2} can be proved if we show that $\varphi^*(y)$ is a strictly monotonic function along Birch's trajectory. 
We prove that by contradiction. 

For a given $L\neq 0$, we pick arbitrary two different Birch's points on Birch's trajectory: $y^{\mathrm{B1}}, y^{\mathrm{B2}} \in \mathfrak{b}(L)$.
We can find $\alpha_1$ and $\alpha_2$
such that $y^{\mathrm{B1}} \in \hat{\mathcal{M}}^{\mathcal{Y}}_{\mathrm{STO}}(L/\alpha_1)$ and $y^{\mathrm{B2}} \in \hat{\mathcal{M}}^{\mathcal{Y}}_{\mathrm{STO}}(L/\alpha_2)$ where $\alpha_1 \neq \alpha_2$ and $\alpha_1, \alpha_2 > 0$.
From Eq. (\ref{sto_isochoric_y}), we have
\begin{align}
    U^l_i \partial^i \varphi^*(y^{\mathrm{B1}}) = \frac{L^l}{\alpha_1}, \mbox{ } U^l_i \partial^i \varphi^*(y^{\mathrm{B2}}) = \frac{L^l}{\alpha_2}.
\end{align}
Since $U$ is a basis matrix of $\mathrm{Ker} [S^T]$: $\{ U_i^l \}:= (\textbf{U}^1, \textbf{U}^2, ...)^T$, and $y^{\mathrm{B1}}, y^{\mathrm{B2}} \in \mathcal{M}^{\mathcal{Y}}_{\mathrm{EQ}}(\tilde{\mu})$, we can choose $\textbf{U}^1$ as $y^{\mathrm{B2}}-y^{\mathrm{B1}}$. Then, we obtain 
\begin{align}
    U^1_i \partial^i \varphi^*(y^{\mathrm{B1}}) = \frac{\alpha_1}{\alpha_2}U^1_i \partial^i \varphi^*(y^{\mathrm{B2}}),\label{appendix_sign}
\end{align}
which implies that the signs of $U^1_i \partial^i \varphi^*(y)$ are the same at $y^{\mathrm{B1}}$ and $y^{\mathrm{B2}}$.

If $\varphi^*(y)$ is not strictly monotonic along Birch's trajectory, we can find $y^{\mathrm{B1}} \neq y^{\mathrm{B2}}$ such that $\varphi^*(y^{\mathrm{B1}}) = \varphi^*(y^{\mathrm{B2}})$.
Since the vector $\textbf{U}^1$ points in the direction from $y^{\mathrm{B1}}$ to $y^{\mathrm{B2}}$, and $\varphi^*(y)$ is strictly convex, 
the signs of $U^1_i \partial^i \varphi^*(y)$
are different at $y^{\mathrm{B1}}$ and $y^{\mathrm{B2}}$.
It contradicts with Eq. (\ref{appendix_sign}), and 
therefore, $\varphi^*(y)$ should be strictly monotonic along Birch's trajectory.

\section{The form of the total entropy function using a particular solution $y^{\mathrm{P}}$}
\label{appendix_psolution}

Substituting a particular solution $y^{\mathrm{P}}$ to Eq. (\ref{simultaneous_equation_outline}) into   
Eq. (\ref{Enttot_partialGP}) (also Eq. (\ref{appendix_Enttot_partialGP})), we get 
\begin{align}
    \Sigma^{\mathrm{tot}} = &- \frac{1}{\tilde{T}} \left\{ \Omega(X) \varphi \left( \frac{X}{\Omega(X)} \right) + \Omega(X) \tilde{\Pi} - y_i^{\mathrm{P}} S^i_r \Xi^r \right\} \notag \\ 
    &+ \mathrm{const}.
\end{align}
Since $S^i_r \Xi^r = X^i - X^i_0$ from Eq. (\ref{accessible_space_X}), it can be represented as a function of the number of the confined chemicals $X$: 
\begin{align}
    \Sigma^{\mathrm{tot}} = &- \frac{1}{\tilde{T}} \left\{ \Omega(X) \varphi \left( \frac{X}{\Omega(X)} \right) + \Omega(X) \tilde{\Pi} - y_i^{\mathrm{P}} \left( X^i - X^i_0 \right) \right\} \notag \\
    &+ \mathrm{const}.
    \label{Enttot_X}
\end{align}
This is further calculated as 
\begin{align}
\Sigma^{\mathrm{tot}} =& - \frac{\Omega(X)}{\tilde{T}} \left\{ \varphi \left( \frac{X}{\Omega(X)} \right) - y^{\mathrm{P}}_i \frac{X^i}{\Omega(X)} + \tilde{\Pi} \right\} - \frac{y^{\mathrm{P}}_i X^i_0}{\tilde{T}} \notag \\
=& - \frac{\Omega(X)}{\tilde{T}} \biggl[ - \varphi^* \left( \partial \varphi \left( \frac{X}{\Omega(X)} \right) \right) \notag \\ &- \left\{ y^{\mathrm{P}}_i - \partial_i \varphi \left( \frac{X}{\Omega(X)} \right) \right\} \frac{X^i}{\Omega(X)} + \tilde{\Pi} \biggr] - \frac{y^{\mathrm{P}}_i X^i_0}{\tilde{T}},
\end{align}
where we omit the constant term in Eq. (\ref{Enttot_X}) for simplicity.
In addition, by defining $y(X) := \partial \varphi(X/\Omega(X)) = \partial \varphi \circ \rho_{\mathcal{X}}(X)$, we get 
\begin{align}
&\Sigma^{\mathrm{tot}} = - \frac{\Omega(X)}{\tilde{T}} \biggl[ 
\varphi^*(y^{\mathrm{P}}) - \varphi^*(y(X)) \notag \\ &- \partial^i \varphi^*(y(X)) \left\{ y_i^{\mathrm{P}} - y_i(X) \right\} + \tilde{\Pi} - \varphi^*(y^{\mathrm{P}})
\biggr] - \frac{y^{\mathrm{P}}_i X^i_0}{\tilde{T}} \notag \\
&= \frac{\Omega(X)}{\tilde{T}} \biggl[ \varphi^*(y^{\mathrm{P}}) - \tilde{\Pi} - \mathcal{D}^{\mathcal{Y}} \left[ y^{\mathrm{P}} || y \right] \biggr] - \frac{y^{\mathrm{P}}_i X^i_0}{\tilde{T}}.
\end{align}
This corresponds to Eq. (\ref{enttot_form}). 

\section{The entropy function does not depend on the choice of a particular solution $y^{\mathrm{P}}$}
\label{appendix_ypchoice}

We show that the value of the total entropy function in Eq. (\ref{enttot_form}) (i.e., Eq. (\ref{Enttot_X})) does not depend on the particular choice of $y_i^{\mathrm{P}}$. 
The general solution to Eq. (\ref{simultaneous_equation_outline}) is represented by using the particular solution $y^{\mathrm{P}}$ as 
\begin{equation}
    y_i = y^{\mathrm{P}}_i + h_l U^l_i. \label{ymanifold_generalsolution}
\end{equation}

Then, the term including $y^{\mathrm{P}}_i$ in Eq. (\ref{Enttot_X}) is written as 
\begin{align}
    y_i^{\mathrm{P}} \left( X^i - X_0^i \right) &= y_i \left(X^i - X_0^i \right) - h_l \left( U^l_i X^i - U^l_i X^i_0 \right) \notag\\
    &=  y_i \left(X^i - X_0^i \right). \label{equality_choice_of_yp}
\end{align}
Here, the second equality holds because the quantities $U^l_i X^i$ is conserved in the time evolution, Eq. (\ref{system_dynamics_chem}). 
Thus, the value in Eq. (\ref{enttot_form}) does not depend on the choice of $y^{\mathrm{P}}_i$.

\section{Proof of the existence of a point $y \in \mathcal{M}^{\mathcal{Y}}_{\mathrm{STO}}(L=0)$ with positive $K^{\mathcal{Y}}(y^{\mathrm{P}}; y)$ when $\varphi^*(y^{\mathrm{min}})- \tilde{\Pi} > 0$ }
\label{appendix_proof}

We show that, when $\varphi^*(y^{\mathrm{min}}) - \tilde{\Pi} > 0$, we can find $y \in \mathcal{M}^{\mathcal{Y}}_{\mathrm{STO}}(L=0)$ such that $K^{\mathcal{Y}}(y^{\mathrm{P}}; y)$ is positive.

When the stoichiometric matrix $S$ is productive (see Eq. (\ref{productiveS})), we first show that $\mathcal{M}^{\mathcal{Y}}_{\mathrm{STO}}(L=0) \neq \emptyset$, i.e., there exists $y \in \mathcal{I}^{\mathcal{Y}}(\tilde{\Pi}, \tilde{\mu})$ such that $U^l_i \partial^i \varphi^*(y) = 0$. From the productivity of $S$, we can construct a vector $\bm{v} = \{ v^i \}$ such that its components are positive and satisfies $U^l_i v^i = 0$. In addition, we can find $y \in \mathcal{I}^{\mathcal{Y}}(\tilde{\Pi}, \tilde{\mu})$ such that the vector $\partial \varphi^*(y)$ has the same direction with $\bm{v}$ because the range of $\partial^i \varphi^*(y)$ covers the positive orthant. This point $y$ is on the stoichiometric manifold $\mathcal{M}^{\mathcal{Y}}_{\mathrm{STO}}(L=0)$. Thus, we get $\mathcal{M}^{\mathcal{Y}}_{\mathrm{STO}}(L=0) \neq \emptyset$.

\begin{figure}[h]
    \centering
    \includegraphics[width=0.5\textwidth]{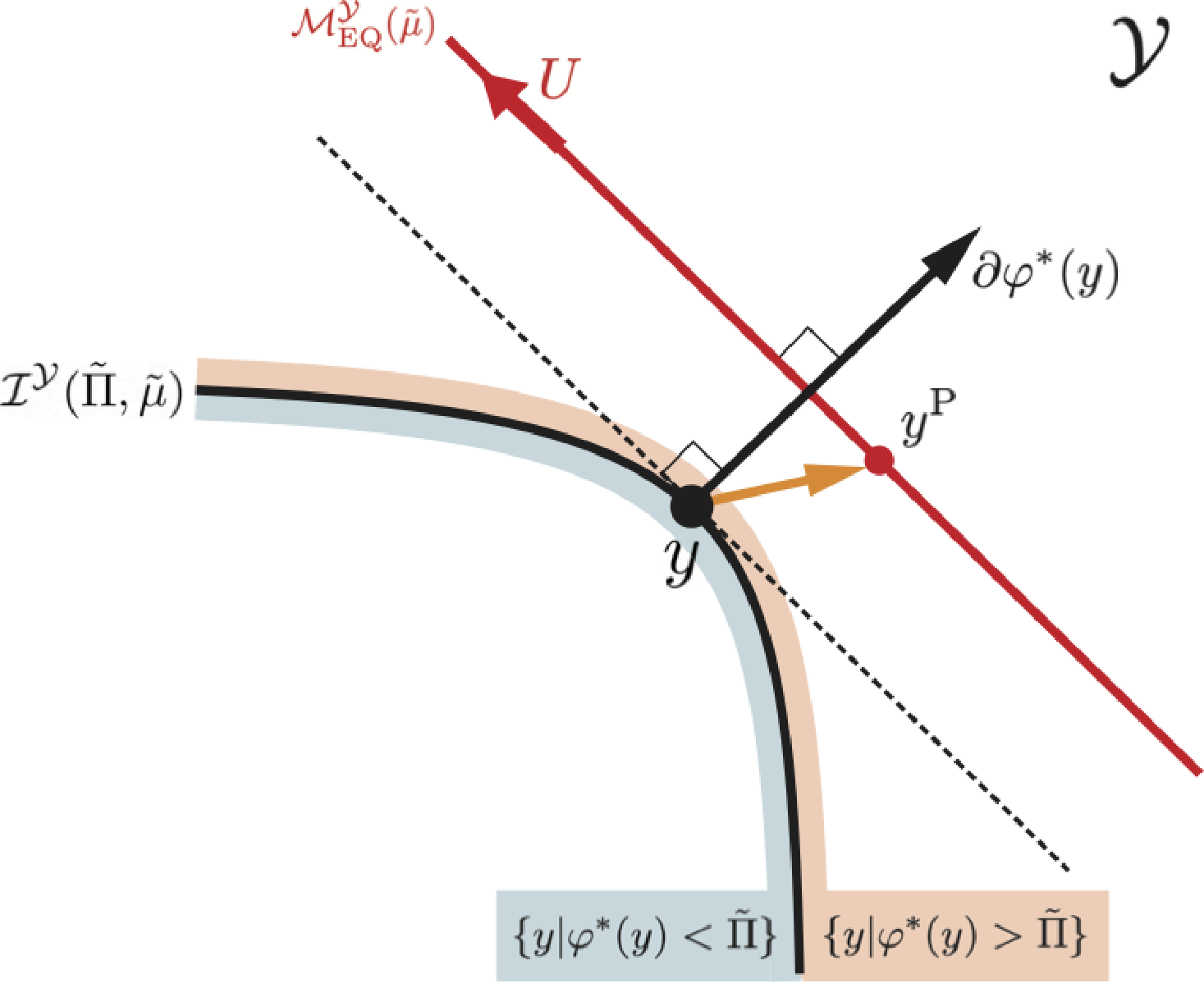}
    \caption{Illustration of the proof for the existence of $y\in \mathcal{M}^{\mathcal{Y}}_{\mathrm{STO}}(L=0)$ such that $K^{\mathcal{Y}}(y^{\mathrm{P}}; y) > 0$ when $\varphi^*(y^{\mathrm{min}}) - \tilde{\Pi} > 0$. The isobaric manifold $\mathcal{I}^{\mathcal{Y}}(\tilde{\Pi}, \tilde{\mu})$ (solid black curve) represents the level set $\{ y | \varphi^*(y) = \tilde{\Pi} \}$, which divides the space $\mathcal{Y}$ into the sublevel set (lower left) and the superlevel set (upper right). The equilibrium manifold $\mathcal{M}^{\mathcal{Y}}_{\mathrm{EQ}}(\tilde{\mu})$ (red line) is located in the superlevel set. The black vector is the normal vector $\partial \varphi^*(y)$ of the level set. Since the range of $\partial \varphi^*(y) = x$ covers the positive orthant, $y \in \mathcal{I}^{\mathcal{Y}}(\tilde{\Pi}, \tilde{\mu})$ exists such that it satisfies $U \partial \varphi^*(y) = 0$, shown by the black circle. This point is on the stoichiometric manifold $\mathcal{M}^{\mathcal{Y}}_{\mathrm{STO}}(L=0)$. The orange vector represents $(y^{\mathrm{P}}-y)$, and the dashed line expresses the tangent plane of the level set at the point $y$. One can find $y\in \mathcal{M}^{\mathcal{Y}}_{\mathrm{STO}}(L=0)$ such that any $y^{\mathrm{P}} \in \mathcal{M}^{\mathcal{Y}}_{\mathrm{EQ}}(\tilde{\mu})$  is located on the same side of the tangent plane with the black vector $\partial \varphi^*(y)$. For such $y\in \mathcal{M}^{\mathcal{Y}}_{\mathrm{STO}}(L=0)$, the inner product $\partial \varphi^*(y) (y^{\mathrm{P}}-y)$ is positive.}
    \label{fig:appendix}
\end{figure}

For $y \in \mathcal{M}^{\mathcal{Y}}_{\mathrm{STO}}(L=0)$, we have $\varphi^*(y) = \tilde{\Pi}$ because $\mathcal{M}^{\mathcal{Y}}_{\mathrm{STO}}(L=0) \subset \mathcal{I}^{\mathcal{Y}}(\tilde{\Pi}, \tilde{\mu})$.
Thus, the expression in Eq. (\ref{kappa}) is rearranged for $y \in \mathcal{M}^{\mathcal{Y}}_{\mathrm{STO}}(L=0)$ as 
\begin{align}
    K^{\mathcal{Y}}(y^{\mathrm{P}}; y) &= \varphi^*(y^{\mathrm{P}}) - \tilde{\Pi} - \mathcal{D}^{\mathcal{Y}} \left[ y^{\mathrm{P}} || y \right] \notag \\
    &= \partial^i \varphi^*(y) \left(y^{\mathrm{P}}_i - y_i\right). \label{appendix_kappa}
\end{align}
The vector $\partial \varphi^*(y) = \{ \partial^i \varphi^*(y) \}$ represents a gradient of the convex function $\varphi^*(y)$, which is a normal vector at a point $y$ of the level set $\{ y | \varphi^*(y) = \tilde{\Pi} \}$. Note that the orientation of the normal vector points to the superlevel set (see Fig. \ref{fig:appendix}). Also, $(y^{\mathrm{P}}-y)$ is a vector from the point $y$ on the level set to the point $y^{\mathrm{P}} \in \mathcal{M}^{\mathcal{Y}}_{\mathrm{EQ}}(\tilde{\mu})$. $K^{\mathcal{Y}}(y^{\mathrm{P}}; y)$ represents the inner product of the two vectors.

To investigate the sign of the inner product, we consider the tangent plane of the level set at the point $y \in \mathcal{M}^{\mathcal{Y}}_{\mathrm{STO}}(L=0)$ (see the dashed line in Fig. \ref{fig:appendix}). 
This plane divides the space of $\mathcal{Y}$ into two regions. If the orientations of the two vectors, $\partial \varphi^*(y)$ and $(y^{\mathrm{P}}-y)$, point to the same side of the tangent plane, their inner product is positive. If they point to the opposite sides, the inner product is negative. 

When $\varphi^*(y^{\mathrm{min}}) - \tilde{\Pi} > 0$, the equilibrium manifold $\mathcal{M}^{\mathcal{Y}}_{\mathrm{EQ}}(\tilde{\mu})$ is located in the superlevel set.  
In addition, the tangent plane at the point $y \in \mathcal{M}^{\mathcal{Y}}_{\mathrm{STO}}(L=0)$ is parallel to $\mathcal{M}^{\mathcal{Y}}_{\mathrm{EQ}}(\tilde{\mu})$. 
Then, we can find $y\in \mathcal{M}^{\mathcal{Y}}_{\mathrm{STO}}(L=0)$ such that $\mathcal{M}^{\mathcal{Y}}_{\mathrm{EQ}}(\tilde{\mu})$ is located on the same side of the tangent plane with the orientation of the normal vector $\partial \varphi^*(y)$ at the point $y$ (see Fig. \ref{fig:appendix}). In such a case, the inner product between $\partial \varphi^*(y)$ and $(y^{\mathrm{P}}_i-y_i)$ is positive for $y^{\mathrm{P}} \in \mathcal{M}^{\mathcal{Y}}_{\mathrm{EQ}}(\tilde{\mu})$. 
Thus, $y\in \mathcal{M}^{\mathcal{Y}}_{\mathrm{STO}}(L=0)$ exists such that $K^{\mathcal{Y}}(y^{\mathrm{P}}; y) > 0$. 

\section{Proof of negative $K^{\mathcal{Y}}(y^{\mathrm{P}}; y)$ for any $y \in \mathcal{M}^{\mathcal{Y}}_{\mathrm{STO}}(L=0)$ when $\varphi^*(y^{\mathrm{min}})-\tilde{\Pi} < 0$}
\label{appendix_proof2}

We show that, when $\varphi^*(y^{\mathrm{min}}) - \tilde{\Pi} < 0$, $K^{\mathcal{Y}}(y^{\mathrm{P}}; y)$ is negative for $y \in \mathcal{M}^{\mathcal{Y}}_{\mathrm{STO}}(L=0)$. Here, we assume that the stoichiometric matrix $S$ is productive, therefore, $\mathcal{M}^{\mathcal{Y}}_{\mathrm{STO}}(L=0) \neq \emptyset$ (see Appendix \ref{appendix_proof}). 

\begin{figure}[h]
    \centering
    \includegraphics[width=0.5\textwidth]{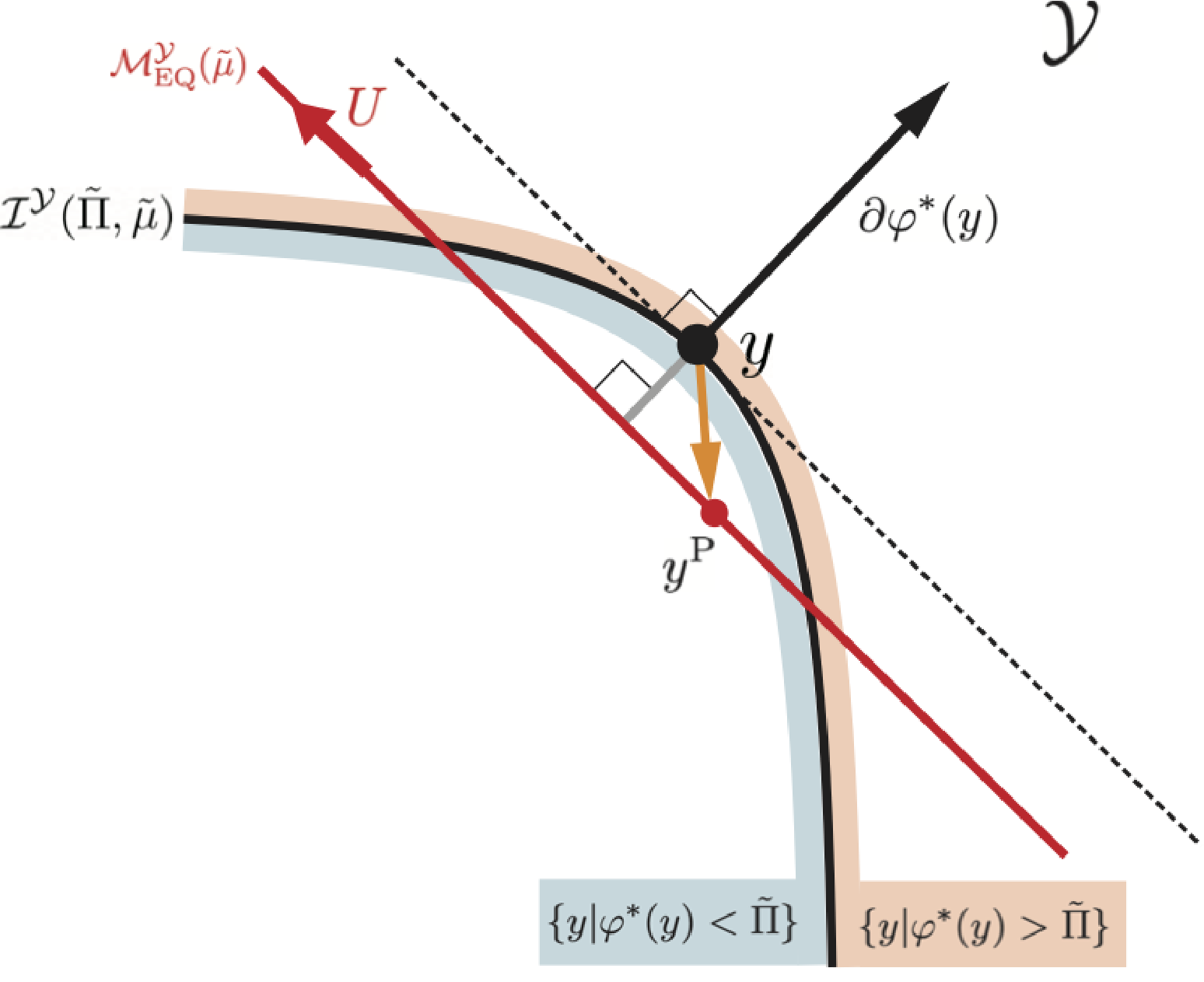}
    \caption{Illustration of the proof that $K^{\mathcal{Y}}(y^{\mathrm{P}}; y) = \partial \varphi^*(y) (y^{\mathrm{P}}-y)$ is negative for $y\in \mathcal{M}^{\mathcal{Y}}_{\mathrm{STO}}(L=0)$ when $\varphi^*(y^{\mathrm{min}}) - \tilde{\Pi} < 0$. The isobaric manifold $\mathcal{I}^{\mathcal{Y}}(\tilde{\Pi}, \tilde{\mu})$ (solid black curve) represents the level set $\{ y | \varphi^*(y) = \tilde{\Pi} \}$, which divides the space $\mathcal{Y}$ into the sublevel set (lower left) and the superlevel set (upper right). The equilibrium manifold $\mathcal{M}^{\mathcal{Y}}_{\mathrm{EQ}}(\tilde{\mu})$ (red line) is located both in the sublevel and the superlevel sets. The black vector is the normal vector $\partial \varphi^*(y)$ of the level set, which satisfies $U \partial \varphi^*(y) = 0$. Thus, the point $y$ shown by the black circle is on the stoichiometric manifold $\mathcal{M}^{\mathcal{Y}}_{\mathrm{STO}}(L=0)$. The orange vector represents $(y^{\mathrm{P}}-y)$ and the dashed line expresses the tangent plane of the level set at the point $y$.
    Any $y^{\mathrm{P}} \in \mathcal{M}^{\mathcal{Y}}_{\mathrm{EQ}}(\tilde{\mu})$ is located on the opposite side of the tangent plane to the orientation of the black vector. Thus, the inner product $\partial \varphi^*(y) (y^{\mathrm{P}}-y)$ is negative. }
    \label{fig:appendix2}
\end{figure}

Consider any point $y\in \mathcal{M}^{\mathcal{Y}}_{\mathrm{STO}}(L=0)$. At the point $y$, we can consider the tangent plane of the level set $\{ y | \varphi^*(y) = \tilde{\Pi} \}$, which divides the space of $\mathcal{Y}$ into two regions. 
Since $K^{\mathcal{Y}}(y^{\mathrm{P}}; y)$ represents the inner product of the two vectors $\partial \varphi^*(y)$ and $(y^{\mathrm{P}}-y)$ in Eq. (\ref{appendix_kappa}), the sign of $K^{\mathcal{Y}}(y^{\mathrm{P}}; y)$ is determined from the sides of the tangent plane to which the two vectors point.
We also note that any point $y^{\mathrm{P}} \in \mathcal{M}^{\mathcal{Y}}_{\mathrm{EQ}}(\tilde{\mu})$ is on the one side of the tangent plane because the plane is parallel to the equilibrium manifold $\mathcal{M}^{\mathcal{Y}}_{\mathrm{EQ}}(\tilde{\mu})$ (see Fig. \ref{fig:appendix2}). 

When $\varphi^*(y^{\mathrm{min}}) - \tilde{\Pi} < 0$, the equilibrium manifold $\mathcal{M}^{\mathcal{Y}}_{\mathrm{EQ}}(\tilde{\mu})$ (the red line in Fig. \ref{fig:appendix2}) is located both in the sublevel and the superlevel sets.
From the convexity of the function $\varphi^*(y)$, any point in the sublevel set is located on the opposite side of the tangent plane to the orientation of the normal vector $\partial \varphi^*(y)$ at the point $y$ (see Fig. \ref{fig:appendix2}). 
Since any point $y \in \mathcal{M}^{\mathcal{Y}}_{\mathrm{EQ}}(\tilde{\mu})$ is on the one side of the tangent plane, we can show that the particular solution $y^{\mathrm{P}} \in \mathcal{M}^{\mathcal{Y}}_{\mathrm{EQ}}(\tilde{\mu})$ is located on the opposite side of the tangent plane to the orientation of $\partial \varphi^*(y)$. 
Thus, for $y^{\mathrm{P}} \in \mathcal{M}^{\mathcal{Y}}_{\mathrm{EQ}}(\tilde{\mu})$, the inner product between $\partial \varphi^*(y)$ and $(y^{\mathrm{P}}_i-y_i)$ is negative. 
Therefore, for $y\in \mathcal{M}^{\mathcal{Y}}_{\mathrm{STO}}(L=0)$, $K^{\mathcal{Y}}(y^{\mathrm{P}}; y)$ is always negative. 

\section{Symbols and Notations}
\label{appendix_notations}

\begin{table*}[h]
\caption{\label{tab:table_appendix}%
List of symbols and notations for open chemical reaction systems (CRSs)
}
\begin{tabularx}{\textwidth}{lll}
\hline
Symbol & Description/definition & First appearance\\
\hline
$X = \{ X^i \}$ & The number of confined chemicals ($i$ = $1,\ldots, \mathcal{N}_{X}$). & Sec. \ref{sec:setup}\\
$X_0 = \{ X^i_0 \}$ & The initial condition of $X$ & Sec. \ref{sec:setup}/Eq. (\ref{accessible_space_X})\\
$N = \{ N^m \}$ & The number of open chemicals ($m$ = $1,\ldots, \mathcal{N}_{N}$) & Sec. \ref{sec:setup}\\
$\mathcal{N}_{X}$ & The number of species of confined chemicals & Sec. \ref{sec:setup}\\
$\mathcal{N}_{N}$ & The number of species of open chemicals & Sec. \ref{sec:setup}\\
$E$ & The energy of the CRS & Sec. \ref{sec:setup}\\
$\Omega$ & The volume of the CRS & Sec. \ref{sec:setup}\\
$\tilde{\mu} = \{ \tilde{\mu}_m \}$ & The chemical potential of open chemicals in the environment & Sec. \ref{sec:setup}\\
$\tilde{T}$ & The temperature of the environment & Sec. \ref{sec:setup}\\
$\tilde{\Pi}$ & The pressure of the environment & Sec. \ref{sec:setup}\\
$\tilde{N} = \{ \tilde{N}^m \}$ & The number of open chemicals in the environment & Sec. \ref{sec:setup}\\
$\tilde{E}$ & The energy of the environment & Sec. \ref{sec:setup}\\
$\tilde{\Omega}$ & The volume of the environment & Sec. \ref{sec:setup}\\
$\Sigma[E, \Omega, N, X]$ & The entropy function for the CRS & Sec. \ref{sec:setup}\\
$\tilde{\Sigma}_{\tilde{T},\tilde{\Pi},\tilde{\mu}}[\tilde{E},\tilde{\Omega},\tilde{N}]$ & The entropy function for the environment & Sec. \ref{sec:setup}\\
$\Sigma^{\mathrm{tot}}$ & The total entropy function & Sec. \ref{sec:setup}/Eq. (\ref{enttot})  \\
$\sigma[\epsilon,n,x]$ & The entropy density & Sec. \ref{sec:setup}/Eq. (\ref{ent_homogeneity})\\
$(\epsilon,n,x)$ & $(E/\Omega,N/\Omega,X/\Omega)$ & Sec. \ref{sec:setup}\\
$S = \{ S^i_r\}$ & The stoichiometric matrix for confined chemicals & Sec. \ref{sec:setup}/Fig. \ref{fig:1}\\
$O = \{ O^i_r\}$ & The stoichiometric matrix for open chemicals & Sec. \ref{sec:setup}/Fig. \ref{fig:1}\\ 
$J(t) = \{ J^r(t) \}$ & The chemical reaction flux ($r$ = $1, \ldots, \mathcal{N}_{R}$). & Sec. \ref{sec:setup}/Eq. (\ref{system_dynamics_chem})\\
$\mathcal{N}_{R}$ & The number of reactions & Sec. \ref{sec:setup} \\
$J_E(t)$ & The energy exchange rate with the environment & Sec. \ref{sec:setup}/Fig. \ref{fig:1}\\
$J_{\Omega}(t)$ & The volume expansion rate & Sec. \ref{sec:setup}/Fig. \ref{fig:1}\\
$J_D(t) = \{ J^m_D(t) \}$ & The diffusion flux for open chemicals & Sec. \ref{sec:setup}/Fig. \ref{fig:1}\\
$\Xi(t) = \{ \Xi^r(t) \}$ & The extent of reaction & Sec. \ref{sec:setup}/Eq. (\ref{accessible_space_X})\\
$L = \{L^l = U^l_i X^i_0\}$ & The conserved quantities ($l$ = $1,\ldots, \dim \mathrm{Ker}[S^T]$). & Sec. \ref{sec:setup}/Eq. (\ref{conserved_quantity})\\ 
$U = \{U^l_i\}$ & A basis matrix whose row vectors form the basis of $\mathrm{Ker}[S^T]$ & Sec. \ref{sec:setup}/Eq. (\ref{conserved_quantity})\\
$\hat{L}$ & The conserved quantities under the $isochoric$ condition & Sec. \ref{sec:conservation_law} \\ 
$\varphi(x)$ & The partial grand potential density & Sec. \ref{sec:setup}/Eq. (\ref{varphi_x})\\
$y$ & The Legendre dual variable of $x$ & Sec. \ref{sec:outline_eq} \\
$\varphi^*(y)$ & The full grand potential density & Sec. \ref{sec:outline_eq}/Eq. (\ref{varphi_y})\\
$\Pi_{\mathrm{min}}$ & The minimum pressure that the CRS can take  & Sec. \ref{sec:proof1}/Eq. (\ref{min_pressure}) \\
$f_r(\Xi)$ & The thermodynamic force & Sec. \ref{sec:proof_theorem2}/Eq. (\ref{force_proof})\\ 
\hline
\end{tabularx}
\end{table*}

\begin{table*}[h]
\caption{\label{tab:table_appendix2}%
List of symbols and notations for geometric descriptions and manifolds
}
\begin{tabularx}{\textwidth}{lll}
\hline
Symbol & Description/definition & First appearance \\
\hline
$\mathfrak{X}$ & The number space $\mathbb{R}_{> 0}^{\mathcal{N}_{X}}$ & Sec. \ref{sec:setup} \\
$\mathcal{X}$ & The density space $\mathbb{R}_{>0}^{\mathcal{N}_{X}}$ & Sec. \ref{sec:setup}\\
$\mathcal{Y}$ & The chemical potential space $\mathbb{R}^{\mathcal{N}_{X}}$ & Sec. \ref{sec:outline_eq}\\
$\mathcal{L}$ & The space of conserved quantities $\mathbb{R}^{\dim \mathrm{Ker}[S^T]}$ & Sec. \ref{sec:setup} \\
$\mathcal{M}^{\mathcal{Y}}_{\mathrm{EQ}}(\tilde{\mu})$ & The equilibrium manifold in $\mathcal{Y}$ & Sec. \ref{sec:outline_eq}/Eq. (\ref{eq_manifold_y_outline}) \\
$\mathcal{M}^{\mathcal{X}}_{\mathrm{EQ}}(\tilde{T}, \tilde{\mu})$ & The equilibrium manifold in $\mathcal{X}$ & Sec. \ref{sec:thermodynamics}/Eq.(\ref{eq_manifold_x_outline}) \\
$\mathcal{I}^{\#}(\tilde{\Pi},\tilde{\mu})$ & The isobaric manifold in $\# = \mathcal{X}$ or $\mathcal{Y}$ & Sec. \ref{sec:isobaric}/Eq. (\ref{isobaric_manifold_x}), Eq. (\ref{isobaric_manifold_y})\\
$\mathcal{M}^{\#}_{\mathrm{STO}}(L)$ & The stoichiometric manifold/compatibility class in $\# = \mathfrak{X}$, $\mathcal{X}$, or $\mathcal{Y}$ & Sec. \ref{sec:eq_as_intersection_outline}/Eq. (\ref{stoichiometric_manifold})-(\ref{stoichiometric_y_claim}) \\
$\hat{\mathcal{M}}^{\#}_{\mathrm{STO}}(\hat{L})$ & The stoichiometric manifold in $\# = \mathcal{X}$ or $\mathcal{Y}$ under the $isochoric$ condition & Sec. \ref{sec:conservation_law}/Eq. (\ref{sto_isochoric_x}), Eq. (\ref{sto_isochoric_y})\\
$\rho_{\mathcal{X}}(X)$ & A nonlinear function of $X$ defined as $\rho_{\mathcal{X}}(X) = X/\Omega(X)$ & Sec. \ref{sec:setup}/Eq. (\ref{map_density_of_x}) \\
$\mathfrak{r}^{\mathcal{X}}(x)$ & The corresponding ray to $x$ in the space $\mathfrak{X}$ & Sec. \ref{sec:setup}/Eq. (\ref{ray_x}) \\
$\mathfrak{r}^{\mathcal{Y}}(y)$ & The corresponding ray to $y$ in the space $\mathfrak{X}$ & Sec. \ref{sec:outline_eq}/Eq. (\ref{ray_y}) \\
$y^{\mathrm{min}}$ & Eq. (\ref{y_min_outline}) & Sec. \ref{sec:claims}\\
$x_{\mathrm{min}}$ & $\partial \varphi^*(y^{\mathrm{min}})$ & Sec. \ref{sec:numerical}/Fig. \ref{Timeevolution} \\
$y^{\mathrm{P}}$ & A particular solution to Eq. (\ref{simultaneous_equation_outline}) & Sec. \ref{sec:thermodynamics}/Eq. (\ref{veq_element_outline}) \\
$h = \{ h_l \}$ & A coordinate of $\mathrm{Ker}[S^T]$ & Sec. \ref{sec:thermodynamics}/Eq. (\ref{veq_element_outline}) \\ 
$y^{\mathrm{EQ}}$ & The intersecting point $\mathcal{M}^{\mathcal{Y}}_{\mathrm{STO}}(L) \cap \mathcal{M}^{\mathcal{Y}}_{\mathrm{EQ}}(\tilde{\mu})$ & Sec. \ref{sec:eq_as_intersection}/Theorem \ref{thm1_new}\\
& For regular $S$, $y^{\mathrm{EQ}}_i = -\tilde{\mu}_m O^m_r (S^{-1})^r_i$. & Sec. \ref{sec:productivity} \\
$y^{\mathrm{B}}(\hat{L})$ & Birch's point & Sec. \ref{sec:proof_theorem1}/Eq. (\ref{birch_point}) \\
$\mathfrak{b}(L)$ & Birch's trajectory & Sec. \ref{sec:proof_theorem1}/Eq. (\ref{birch_trajectory}) \\
$K^{\mathcal{Y}}(y^{\mathrm{P}}; y)$ & Eq. (\ref{kappa}) & Sec. \ref{sec:proof_theorem2} \\
$\mathcal{D}^{\mathcal{Y}}[y'||y]$ & The Bregman divergence & Sec. \ref{sec:proof_theorem2}/Eq. (\ref{bregman}) \\ 
$\Xi_{B}$ & The extent of reaction $\Xi$ at the boundary of the domain & Sec. \ref{sec:proof_theorem2}\\
$X_{B}$ & $X_0 + S\Xi_{B}$ & Sec. \ref{sec:proof_theorem2}\\
$\mathfrak{I}_B(X_B)$ & The index set such that $X^i_B = 0$ & Sec. \ref{sec:proof_theorem2}\\
\hline
\end{tabularx}
\end{table*}

\begin{table*}[h]
\caption{\label{tab:table_appendix3}%
List of symbols and notations used for describing the ideal gas and mass action kinetics
}
\begin{tabularx}{\textwidth}{lll}
\hline
Symbol & Description/definition & First appearance \\
\hline
$R$ & The gas constant & Sec.\ref{sec:setup} \\
$\nu^o(\tilde{T}) = \{ \nu^o_i(\tilde{T}) \}$ & The standard chemical potentials of confined chemicals & Sec.\ref{sec:setup} \\
$\mu^o(\tilde{T}) = \{ \mu^o_m(\tilde{T}) \}$ & The standard chemical potentials of open chemicals & Sec.\ref{sec:setup} \\
$\tilde{n}= \{ \tilde{n}^m \}$ & The density of open chemicals in the environment & Sec.\ref{sec:setup}/Eq. (\ref{chemical_potential_idealgas}) \\
$x^i_o$ & $\exp[-\nu^o_i(\tilde{T})/R\tilde{T}]$ & Fig. \ref{Timeevolution}/Appendix \ref{appendix_numerical} \\
$n^m_o$ & $\exp[-\mu^o_m(\tilde{T})/R\tilde{T}]$ & Fig. \ref{Timeevolution}/Appendix \ref{appendix_numerical} \\
$w^r_+$, $w^r_-$ & The rate constants of $r$-th forward and backward reactions & Appendix \ref{appendix_numerical} \\
$\hat{w}^r_+$, $\hat{w}^r_-$ & The rate constants where we absorb the constant densities of open chemicals & Appendix \ref{appendix_numerical}\\
\hline
\end{tabularx}
\end{table*}

\end{document}